\documentclass[aps,prd,12pt,citeautoscript,superscriptaddress,showpacs,scrartcl,amsmath,amssymb]{revtex4-2}
\usepackage[utf8x]{inputenc}

\usepackage{xcolor}
\usepackage[colorlinks=true, pdfstartview=FitV,linkcolor=blue, citecolor=blue, urlcolor=blue]{hyperref}

\usepackage[bottom]{footmisc}

\usepackage{graphicx}
\usepackage[figuresleft]{rotating}

\makeatletter
\newcommand*\bigcdot{\mathpalette\bigcdot@{.5}}
\newcommand*\bigcdot@[2]{\mathbin{\vcenter{\hbox{\scalebox{#2}{$\m@th#1\bullet$}}}}}
\makeatother

\newcommand{\be}{\begin{eqnarray}}
\newcommand{\ee}{\end{eqnarray}}
\newcommand{\bea}{\begin{eqnarray}}
\newcommand{\eea}{\end{eqnarray}}

\newcommand{\nn}{\nonumber}

\newcommand{\thetaw}{\theta_{\mbox{\tiny W}}}
\newcommand{\mz}{m_{\mbox{\tiny Z}}}
\newcommand{\mh}{m_{\mbox{\tiny H}}}
\newcommand{\mw}{m_{\mbox{\tiny W}}}

\newcommand{\F}{{B}}
\newcommand{\WW}{{\rm W}}

\usepackage{natbib}
\usepackage{graphicx}
\usepackage{amsmath}
\usepackage{amssymb}
\usepackage{mathtools}
\usepackage{tensor}
\usepackage{accents}
\usepackage{cancel}
\usepackage{titlesec}
\usepackage{subfigure}
\usepackage[a4paper]{geometry}
\usepackage{psfrag}
\usepackage{latexsym,array,theorem,mathrsfs,subfigure,bm,float}
\usepackage{amsfonts}
\usepackage{amssymb}

\renewcommand{\theequation}{\arabic{section}.\arabic{equation}}

\newcommand{\A}{{B}}

\newcommand{\T}{{\rm T}}

\titleformat{\paragraph}
{\normalfont\normalsize\bfseries}{\theparagraph}{1em}{}
\titlespacing*{\paragraph}
{0pt}{3.25ex plus 1ex minus .2ex}{1.5ex plus .2ex}

\begin{document}

\title{Electroweak multi-monopoles}

\author{Romain Gervalle}
\email{romain.gervalle@univ-tours.fr}
\affiliation{
Institut Denis Poisson, UMR - CNRS 7013, \\ 
Universit\'{e} de Tours, Parc de Grandmont, 37200 Tours, France}

\author{Mikhail~S.~Volkov}
\email{volkov@lmpt.univ-tours.fr}
\affiliation{
Institut Denis Poisson, UMR - CNRS 7013, \\ 
Universit\'{e} de Tours, Parc de Grandmont, 37200 Tours, France}

\ 

\vspace{1 cm}

\begin{abstract}

\vspace{1 cm}

We construct the multi-charge generalizations for the electroweak magnetic monopole solution of Cho and Maison
within  a wide range of values of the magnetic charge. 
We use the same ansatz for the axially symmetric fields as the one previously 
employed  to construct the electroweak sphalerons and compare the internal structure of 
monopoles with that of sphalerons. 
The monopoles  have zero dipole moment but a finite quadrupole momentum that rapidly increases with growing magnetic charge.  
For large charges, the monopole configurations  are strongly squashed  and show inside a bubble of symmetric phase 
filled with a U(1) hypercharge field produced by a pointlike magnetic charge at the origin, strong enough to 
suppress all other fields and restore the full gauge symmetry.  The bubble is surrounded by a large belt of   broken phase containing a 
magnetically charged ring filled with 
a nonlinear W-condensate, squeezed   between two superconducting rings of opposite electric currents. 
In the far field region there remains only the magnetic field supported by the total magnetic charge contained 
at the origin and in the magnetic ring. 
The axially symmetric monopoles are  probably just a special  case of more general monopole solutions 
not possessing  any continuous symmetries. The Cho-Maison monopole  is stable but the stability 
of its multi-charge generalizations  is not yet confirmed. All electroweak monopoles have infinite energy due to the 
pointlike U(1) charge at the origin, but the energy is expected to  
become finite after taking gravity into account, which should provide 
a cutoff via creating an event horizon to shield the U(1) charge.

\end{abstract}


\maketitle
\newpage

\tableofcontents

\section{INTRODUCTION}

The magnetic monopole in the U(1) electrodynamics is described
by the Coulombian magnetic field, 
$
\vec{\boldsymbol{\mathcal{B}}}=\bm{P}\vec{\bm r}/\bm{r}^3.
$
As was noticed by Dirac \cite{Dirac:1931kp} 
(see also \cite{Wu:1976ge}), although one cannot find a globally regular vector potential $\vec{\boldsymbol{\mathcal{A}}}$
such that $\vec{\boldsymbol{\mathcal{B}}}=\vec{\nabla}\wedge \vec{\boldsymbol{\mathcal{A}}}$, one can use two locally regular potentials 
related to each other via a gauge transformation in a transition region. This imposes 
the  quantization condition for the magnetic charge,
\be                 \label{P}
\bm{P}=\frac{\bm{\hbar c}}{2\bm e}\times n,~~~~n=\pm 1,\pm 2,\ldots
\ee
Extending the gauge group to SU(2) and adding a Higgs field in the adjoint representation, allows one to obtain monopoles 
described by a globally  regular potential 
and without the central singularity,
as was noticed by t'Hooft  \cite{tHooft:1974kcl} and by Polyakov \cite{Polyakov:1974ek}. 
These monopoles  have a finite energy and 
contain massive fields in the central region, while at large distances only the massless U(1) gauge field 
survives and approaches that of the Dirac monopole. 
This discovery  triggered a large number of theoretical studies 
(see \cite{Goddard:1977da,Coleman:1982cx,Konishi:2007dn,Manton:2004tk,Shnir:2005vvi} for reviews
and, e.g., \cite{Chamseddine:1997nm,Forgacs:2003yh} for particular aspects of monopoles), but the experimental 
search for magnetic monopoles has always been giving negative results 
(see   \cite{Rajantie:2016paj,Mitsou:2019mrs,Mavromatos:2020gwk} 
for recent reviews).  One of the explanation for this is the fact  that the  t'Hooft-Polyakov monopoles are not described by the Standard Model, 
because the latter contains in the electroweak sector the Higgs field in the fundamental and not adjoint representation. 
As a result, the standard topological arguments \cite{Manton:2004tk} 
for  the existence and stability  of monopoles do not apply.

One may wonder then if there are any magnetic monopoles in the electroweak theory at all ? 
The answer is  of course positive because the Dirac monopoles should be solutions of the theory 
containing the U(1) electrodynamics as a special limit.  Another type of electroweak monopoles 
was described by Nambu \cite{Nambu:1977ag}, who noticed that the electroweak theory contains vortex 
solutions similar to the Abrikosov-Nielsen-Olesen vortices in the Abelian Higgs model 
\cite{Abrikosov:1956sx,Nielsen:1973cs}. Unlike the latter, however, the electroweak vortices can terminate, 
and then the magnetic flux trapped inside the vortex comes out through the termination point and spreads out all over the space, 
which imitates the magnetic monopole. To describe this, Nambu used the ``isospinor" form for the 
Higgs field, 
\be               \label{Higgs0}
\Phi_{\rm mon}=\phi\, 
 \begin{pmatrix}
\sin\frac{\vartheta}{2}\,e^{-i{\varphi}}\ \\
-\cos\frac{\vartheta}{2}
\end{pmatrix}, 
\ee
which is ill-defined at the negative part of the $z$-axis since it has no limit for $\vartheta\to\pi$. To
cure this, Nambu assumed that the amplitude $\phi$ vanishes at $\vartheta=\pi$,  thereby producing a semi-infinite 
vortex extending along the negative part of the $z$-axis and terminating at the monopole at $z=0$. Analyzing the 
fields inside the vortex and those spreading out to infinity through the vortex termination, Nambu  arrived at the 
following expression for the magnetic charge, 
\be                \label{Nam}
\bm{P}=\frac{\bm{\hbar c}}{\bm e}\times \sin^2\thetaw\,, 
\ee
where $\thetaw$ is the weak mixing angle. This corresponds to the Dirac value \eqref{P} for $n=2$ but with the additional 
factor of $\sin^2\thetaw$ 
(in general the charge can be an integer multiple of \eqref{Nam}). 
If the vortex is semi-infinite, then the resulting system  has an infinite energy and 
cannot be static since the vortex will be pulling the monopole. However, 
the vortex may  have a finite length and terminate
some distance away  on an antimonopole, then the resulting  monopole-antimonopole pair 
will have a finite energy and 
will be spinning around the common center of mass 
\cite{Urrestilla:2001dd}. 

Yet one more possibility to introduce monopoles into the electroweak theory was found by 
 Cho and Maison (CM) \cite{Cho:1996qd}, who used  the same form for  the Higgs field 
 as for the Nambu monopole \eqref{Higgs0},  but assumed  that its singularity at $\vartheta=\pi$ is a gauge 
 artefact and can be handled  by using two local gauges, as for the Dirac monopole. In other words, one assumes that 
 $\Phi_{\rm mon}$ in \eqref{Higgs0} should  be used only in the upper hemisphere where it is regular, while in the lower
hemisphere one uses its gauge-transformed version $\tilde{\Phi}_{\rm mon}=e^{i\varphi}\Phi_{\rm mon}$ 
which is regular for $\vartheta\to \pi$. 
The U(1) gauge transformation $e^{i\varphi}$ relating the two gauges is regular in the equatorial transition region. This 
provides a globally regular description for  a static and spherically symmetric monopole whose 
magnetic charge is the same  as  for the Dirac  monopole \eqref{P} with $n=2$. 

The CM monopole solution contains a regular non-Abelian part  which is similar to the t'Hooft-Polyakov monopole, 
but  it   contains  also a Coulombian U(1) hypercharge field which diverges at the origin thus rendering  the energy infinite \cite{Cho:1996qd}.
The latter feature is  not very appealing and there have been attempts  to regularize  the monopole energy  in some way, 
 but they require to modify the Lagrangian of the theory \cite{Cho:2013vba,Pak:2013jaa,Blaschke:2017pym,Ellis:2020bpy,Hung:2020vuo}. 
At the same time, since the Standard Model describes the real world extremely well, it seems to be more logical to consider 
the CM monopole as it is, with infinite energy.  In any case, its energy 
certainly becomes finite  when gravity is taken into account \cite{Bai:2020ezy}. 

In a recent analysis, the stability of the CM monopole was studied  and it was found 
that it is stable with respect to arbitrary (small) perturbations \cite{Gervalle:2022npx}. 
At the same time, all Dirac monopoles with $|n|>1$ are unstable with respect to perturbations 
in the sector with the angular momentum $j=|n|/2-1$. In particular, the Dirac monopole with 
$|n|=2$ is unstable only in the $j=0$ sector while the  CM monopole is stable and also has $|n|=2$. 
This suggests  that the CM monopole may be viewed as a stable remnant of the decay of the Abelian monopole. 
One may similarly conjecture that stable remnants exist also for monopoles with $|n|>2$, hence the CM monopole is just the 
first member of a  sequence of non-Abelian monopole solutions labeled by their magnetic charge $n$. 
 Only the CM monopole is spherically symmetric, while the non-Abelian monopoles with $|n|>2$ are not rotationally invariant. 

In what follows, we confirm this conjecture by explicitly constructing generalizations of the Cho-Maison monopole 
for higher values of the magnetic charge in the simplest case of axial symmetry. 
At the same time,  we could not yet  check their stability. 
We construct the solutions numerically for various values of the charge, compute their regularized energy, the 
quadrupole momentum, and study their inner  structure.  
It turns out that the elementary Cho-Maison monopoles inside the multi-charge 
monopole merge together  to form a magnetically charged toroidal condensate,
 accompanied by circular  electric currents. 

Monopoles have zero dipole moment but a 
finite quadrupole momentum that rapidly increases with growing magnetic charge.  
For large values of the charge, the monopoles are strongly squashed  and 
develop in the center a bubble of symmetric phase containing 
the U(1) hypercharge field created by a pointlike magnetic charge at the center. This field 
is strong enough to suppress all other fields and restore the full electroweak gauge symmetry in the bubble. 
The bubble is encircled by a  belt of broken phase containing the W-condensate in the form of a magnetically charged ring squeezed 
between two superconducting rings of oppositely directed electric currents. 
The total magnetic charge of the monopole splits 
into the pointlike U(1) part at the origin and the SU(2) part smoothly distributed over the ring volume. 
The pointlike charge at the origin makes an infinite contribution to the energy, 
 but the energy is expected to  
become finite after taking gravity into account, which will provide 
a cutoff via creating an event horizon to shield the U(1) charge.

We use the same ansatz for the axially symmetric fields as the one previously employed   to construct the 
electroweak sphalerons \cite{Kleihaus:1991ks,Kunz:1992uh}.  The sphalerons are static and 
spherically symmetric if $\thetaw=0$ 
\cite{Dashen:1974ck,Yaffe:1989ms}, while for $\thetaw\neq 0$ they are
axially symmetric \cite{Kleihaus:1991ks,Kunz:1992uh,James:1992re}.  Sphalerons are quite different physically 
from monopoles  --  they are neutral and  unstable  \cite{Klinkhamer:1984di}, but 
from the technical viewpoint they are similar to monopoles, 
and we were able to obtain solutions of both types by simply changing 
the boundary conditions in the  equations. This provides a good consistency check for our numerical scheme. 
Besides, sphalerons contain inside  monopoles and antimonopoles of Nambu \cite{Hindmarsh:1993aw}, and 
we find that these Nambu monopoles  and our monopoles, after subtracting their 
divergent U(1) part, are very similar to each other -- 
they have the same quantization condition for the magnetic charge, 
a similar ring distribution of the charge for $|n|>2$, 
and almost the same energy for $|n|=2$.

The rest of the text is organized as follows. Equations of the classical electroweak theory are presented in Section II,
and the axially symmetric fields are described in  Section III. This section also shows 
 the desingularization procedure for removing the line singularities in the fields. The spherically symmetric  monopole and sphaleron
 are described in Section V. 
The main results -- the non-Abelian multi-monopole 
solutions and their various properties --  are presented in Section V.  
The comparison with the sphalerons is discussed in Section VI, and
concluding remarks are given  in Section VII. The two Appendices contain  technical details, such as 
solutions in the asymptotic region, solutions close to the origin, and properties of the gauge conditions. 

In our analysis we used the FreeFem++  numerical solver based on the finite element method  \cite{MR3043640}. 
Each of us run his own numerical code and we compared our results till reaching the agreement.

\section{ELECTROWEAK THEORY}
\setcounter{equation}{0}

The dimensionful action of the bosonic part of the electroweak theory of Weinberg and Salam (WS) can be represented in the form 
\be                                     \label{00}
{\bf S}=\frac{1}{\bm{ c\,g}_0^2} \int {\cal L}_{\rm WS}\,\sqrt{-{\rm g}}\, d^4 x\,,
\ee
with the Lagrangian 
\be                             \label{L}
{\cal L}_{\rm WS}=
-\frac{1}{4g^2}\,\WW^a_{\mu\nu}\WW^{a\mu\nu}
-\frac{1}{4g^{\prime 2}}\,{\F}_{\mu\nu}{\F}^{\mu\nu}
-(D_\mu\Phi)^\dagger D^\mu\Phi
-\frac{\beta}{8}\left(\Phi^\dagger\Phi-1\right)^2,
\ee
where  all fields and couplings  as well as the spacetime coordinates $x^\mu$ and metric ${\rm g}_{\mu\nu}$ 
are rendered dimensionless by rescaling. 
The Abelian U(1) and non-Abelian SU(2) field strengths are 
\begin{align}
{\F}_{\mu\nu}=\partial_\mu{\A}_\nu
-\partial_\nu{\A}_\mu\, ,~~~~~~
\WW^a_{\mu\nu}=\partial_\mu\WW^a_\nu
-\partial_\nu \WW^a_\mu
+\epsilon_{abc}\WW^b_\mu\WW^c_\nu\, ,~~~~~
\end{align}
while  the Higgs field $\Phi$
is in the fundamental 
representation of SU(2) with the covariant derivative 
\begin{align}
D_\mu\Phi
&=\left(\partial_\mu-\frac{i}{2}\,{\A}_\mu
-\frac{i}{2}\,\tau^a \WW^a_\mu\right)\Phi\,, \label{unbold} 
\end{align}
where $\tau^a$ are the Pauli matrices.
The two coupling constants are 
$g=\cos\thetaw$ and
$g^\prime=\sin\thetaw$ where the physical value of the Weinberg angle is such that 
$
g^{\prime 2}=\sin^2\thetaw=0.23. 
$

The dimensionful parameters (we denote all dimensionful quantities  boldfaced)  
in the action \eqref{00} are the speed of light ${\bm c}$ and also ${\bm g}_0$ 
related to the electron charge ${\bm e}$, 
\be                              \label{e}
\frac{\bm e^2}{4\pi\bm \hbar \bm c}=\frac{\bm \hbar {\bm c}}{4\pi}\,{\left(gg^\prime{\bm g}_0\right)^2}
\approx \frac{1}{137}~~~~~\Rightarrow~~~~~~{\bm e}={\bm \hbar \bm c\bm g}_0 e~~~~~\text{with}~~~~~e\equiv gg^\prime \,.
\ee
The dimensionful fields often used in the literature   are 
${\bm \A}_\mu=({\mbox{\boldmath $\Phi$}_0}/g^\prime)\A_\mu$,  
${\bm W}^a_\mu=({\mbox{\boldmath $\Phi$}_0}/g)\WW^a_\mu$ and 
${\mbox{\boldmath $\Phi$}}={\mbox{\boldmath $\Phi$}_0}\Phi$ where 
$\mbox{\boldmath $\Phi$}_0=246$ GeV 
is the Higgs field vacuum expectation value. The dimensionful coordinates are 
${\bm x}^\mu={\bm L}_{\rm WS}\,x^\mu$  with  the electroweak  
length scale  ${\bm L}_{\rm WS}=1/({\bm g}_0{\bm \Phi}_0)=1.52\times 10^{-16}$~cm.  

The theory  is invariant under 
SU(2)$\times$U(1) gauge transformations
\be                               \label{gauge}
\Phi\to {\rm U}\Phi,~~~~~~~~
{\cal W}\to {\rm U}{\cal W}{\rm U}^{-1}
+i{\rm U}\partial_\mu {\rm U}^{-1}dx^\mu\,,
\ee
with 
\be                            \label{U}
{\cal W}=
\frac12\, (B_\mu+\tau^a\WW^a_\mu)\, dx^\mu\,,~~~~~~~~~
{\rm U}=\exp\left(\frac{i}{2}\,\Sigma+\frac{i}{2}\,\tau^a\theta^a\right),
\ee
where $\Sigma$ and $\theta^a$ are functions of $x^\mu$. 
Varying the action 
gives the equations,
\begin{align}
\nabla^\mu {B}_{\mu\nu}&=g^{\prime 2}\,\frac{i}{2}\,
(\Phi^\dagger D_\nu\Phi -(D_\nu\Phi)^\dagger\Phi
)\equiv g^{\prime 2}J_\nu,\nn 
\\
{\cal D}^\mu \WW^a_{\mu\nu}
&=g^{2}\,\frac{i}{2}\,
(
\Phi^\dagger\tau^a D_\nu\Phi
-(D_\nu\Phi)^\dagger\tau^a\Phi
)
\equiv g^{2} J^{a}_{\nu}, \nn 
\\
D_\mu D^\mu\Phi&-\frac{\beta}{4}\,(\Phi^\dagger\Phi-1)\Phi=0,      \label{P2}
\end{align}
with ${\cal D}_\mu\WW^a_{\alpha\beta}=\nabla_\mu \WW^a_{\alpha\beta}
+\epsilon_{abc}\WW^b_\mu\WW^c_{\alpha\beta}$ where $\nabla_\mu$ is the geometrical covariant derivative with respect 
to the spacetime metric. 
Varying the action with respect to the latter determines the energy-momentum tensor, 
\be                      \label{TT}
T_{\mu\nu}=
\frac{1}{g^2}\,\WW^a_{~\mu\sigma}\WW^{a~\sigma}_{~\nu}
+\frac{1}{g^{\prime\,2}}B_{\mu\sigma}B_\nu^{~\sigma}
+(D_\mu\Phi)^\dagger D_\nu\Phi
+(D_\nu\Phi)^\dagger D_\mu\Phi+g_{\mu\nu}\mathcal{L}_{\rm WS}\,.
\ee

The vacuum is defined as the configuration with $T_{\mu\nu}=0$. Modulo gauge transformations, it can be chosen as 
\be
\WW^a_\mu=\A_\mu=0,~~~~~~ 
 \Phi=\begin{pmatrix}
0  \\
1
\end{pmatrix}.
\ee 
Allowing for small fluctuations around the vacuum and 
linearizing the field equations 
with respect to the fluctuations gives the perturbative mass spectrum 
containing the massless photon 
and the massive Z, W and Higgs bosons with dimensionless masses
\be                                   \label{masses}
\mz=\frac{1}{\sqrt{2}},~~~~~~
\mw=g\,\mz,~~~~~~
\mh=\sqrt{\beta}\,\mz.
\ee
 Multiplying  these by 
${\bm e}\mbox{\boldmath $\Phi$}_0/(gg^\prime)$
gives the dimensionful masses, 
for example one has
$
{\bm m}_{\mbox{\tiny Z}}{\bm c}^2=
{\bm e}\mbox{\boldmath $\Phi$}_0/(\sqrt{2}gg^\prime)
\approx 91~ {\rm GeV}. 
$
Using the Higgs  mass 
${\bm m}_{\mbox{\tiny H}}{\bm c}^2\approx 125$ GeV
yields the value $\beta\approx 1.88$. 

Summarizing, the dimensionless parameters in the equations are 
\be
g^{\prime 2}=0.23,~~~~g^2=1-g^{\prime 2},~~~~\beta=1.88. 
\ee
We shall adopt  the definition  of Nambu for the 
electromagnetic and Z fields \cite{Nambu:1977ag}, 
\be                                  \label{Nambu}
F_{\mu\nu}=\frac{g}{g^\prime}\,  
\A_{\mu\nu}-\frac{g^{\prime}}{g}\,N^a\WW^a_{\mu\nu}\,,~~~~~~
{Z}_{\mu\nu}=\A_{\mu\nu}+N^a\WW^a_{\mu\nu}\,,
\ee
where 
$
N^a=\Phi^\dagger\tau^a\Phi/(\Phi^\dagger\Phi). 
$
The magnetic part of $F_{\mu\nu}$ will be denoted by the calligraphic symbol, ${\cal B}^i=\frac12\, \epsilon^{ijk} F_{jk}$, 
not to be confused with the hypercharge field $B=B_\mu dx^\mu$.

Using the electromagnetic tensors $F_{\mu\nu}$ and its dual, 
 \be
 \tilde{F}^{\mu\nu}=\frac{1}{2\sqrt{-\rm g}}\,\epsilon^{\mu\nu\alpha\beta}F_{\alpha\beta}\,,~~~~~
 \ee
 one can define the conserved electric and magnetic currents,
 \be             \label{cur}
 J^\mu =\frac{1}{4\pi}\,\frac{1}{\sqrt{-\rm g}}\,\partial_\nu \left(\sqrt{-\rm g}\,F^{\mu\nu}\right),~~~~~~~
 \tilde{J}^\mu =\frac{1}{4\pi}\,\frac{1}{\sqrt{-\rm g}}\,\partial_\nu \left(\sqrt{-\rm g}\,\tilde{F}^{\mu\nu}\right). 
 \ee
  Since $F_{\mu\nu}$ consists of two parts, 
 both $J^\mu$ and $\tilde{J}^\mu$ split into a sum of two 
 separately conserved currents --  the U(1) current determined by the contribution of $B_{\mu\nu}$ and the SU(2) current 
 determined by $W^a_{\mu\nu}$. 
 We shall be considering purely magnetic systems for which the non-vanishing components are the electric current $J^k$ and the 
 magnetic charge density $\tilde{J}^0$. The magnetic charge and its density then split into the U(1) and SU(2) parts, 
 \be       \label{rhom}
 \tilde{J}^0=\frac{1}{4\pi}\,\vec{\nabla}\cdot \vec{\cal B}\equiv \rho_{\rm U(1)}+\rho_{\rm SU(2)}\,,
 \ee
 and 
 \be                   \label{PT}
 P=\left.\left.\int \right(\rho_{\rm U(1)}+\rho_{\rm SU(2)}\right) \sqrt{-g}\,d^3 x\equiv P_{\rm U(1)}+P_{\rm SU(2)}, 
 \ee
 where $P_{\rm U(1)}$ and $P_{\rm SU(2)}$  are separately conserved. 
 Since the $B$ field is Abelian, one has 
 \be               \label{PB}
 P_{\rm U(1)}=\frac{g}{g^\prime}\,\frac{1}{4\pi} \oint_{S^2} dB\,,
 \ee
 where the integration is performed over a two-sphere at infinity. This integral vanishes unless $B$ is topologically non-trivial,
 in which  case the value of the integral is determined by the topology and 
 does not depend on the radius of the sphere.

\section{AXIAL SYMMETRY \label{II}}
\setcounter{equation}{0}
To describe axially symmetric fields, 
it is convenient to express the spacetime  metric in spherical coordinates, 
\be                   \label{metr}
{\rm g}_{\mu\nu}dx^\mu dx^\nu =-dt^2+dr^2+r^2\left(d\vartheta^2+\sin^2\vartheta d\varphi^2\right).
\ee
Let $\T_a=\frac12\, \tau_a$ be the SU(2) gauge group generators such that $[\T_a,\T_b]=i\epsilon_{abc}\T_c$.
The SU(2) gauge field, the U(1) hypercharge field and the Higgs field are 
\be               \label{RR}
W\equiv \T_a {\rm W}^a_\mu dx^\mu&=&\T_2\left( F_1\,dr+F_2\,d\vartheta \right)+\nu\left( \T_3\,F_3-\T_1 F_4\,\right)d\varphi\,, \nn \\
B\equiv B_\mu dx^\mu&=&\nu\, Yd\varphi\,,~~~~~~
\Phi=\begin{pmatrix}
\phi_1 \\
\phi_2
\end{pmatrix}\,,
\ee
where $F_1,F_2,F_3,F_4,Y,\phi_1,\phi_2$ are 7 real-valued functions of $r,\vartheta$ and $\nu$ is a real parameter. 
The SU(2) field here corresponds to the purely magnetic ansatz of Rebbi and Rossi \cite{Rebbi:1980yi}. The ansatz keeps its form under gauge 
transformations \eqref{gauge} generated by ${\rm U}=\exp\left\{i\chi(r,\vartheta)\T_2\right\}$, whose effect  is 
\be               \label{res}
F_1\to F_1+\partial_r \chi,~~~~~~~~~
F_2\to F_2+\partial_\vartheta \chi,~~~~~~~~Y\to Y,\nn \\
F_3\to F_3\,\cos\chi-F_4\,\sin\chi\,,~~~~~~
F_4\to F_4\,\cos\chi+F_3\,\sin\chi\,,~~~~~~\nn \\
\phi_1\to \phi_1\,\cos(\chi/2)+ \phi_2\,\sin(\chi/2),~~~~~~~
\phi_2\to \phi_2\,\cos(\chi/2)- \phi_1\,\sin(\chi/2).
\ee

Inserting this to \eqref{TT} defines the energy,
\be                \label{EN}
E=\int T_{00}\, \sqrt{-\rm g}\, d^3 x=2\pi\int_0^\infty dr \int_0^\pi d\vartheta\left(
\frac{{\cal E}_W}{2g^2}+\frac{{\cal E}_B}{2g^{\prime 2}}+{\cal E}_\Phi+V\right),
\ee
where 
\be                 \label{EN1}
{\cal E}_W&=&\left(\partial_\vartheta F_1-\partial_r F_2\right)^2\,\sin\vartheta \nn \\
&&+\left(
(\partial_r F_3+F_1F_4)^2
+(\partial_r F_4-F_1F_3)^2
\left)\,\frac{\nu^2}{\sin\vartheta}\right.\right.
\nn \\
&&+\left(
(\partial_\vartheta F_3+F_2F_4)^2
+(\partial_\vartheta F_4-F_2F_3)^2
\left)\,\frac{\nu^2}{r^2\,\sin\vartheta}\right.\right.\,, \nn \\
{\cal E}_B&=&\left(
(\partial_r Y)^2
+\frac{1}{r^2}\,(\partial_\vartheta Y)^2
\left)\,\frac{\nu^2}{\sin\vartheta}\right.\right.\,,
\nn \\
{\cal E}_\Phi&=&r^2\left(
\left(\partial_r\phi_1-\frac{F_1}{2}\,\phi_2\right)^2
+
\left(\partial_r\phi_2+\frac{F_1}{2}\,\phi_1\right)^2
\right)\,\sin\vartheta \nn \\
&&+
\left(
\left(\partial_\vartheta\phi_1-\frac{F_2}{2}\,\phi_2\right)^2
+
\left(\partial_\vartheta\phi_2+\frac{F_2}{2}\,\phi_1\right)^2
\right)\,\sin\vartheta \nn \\
&&+
\left(
\left((F_3+Y)\phi_1-F_4\phi_2\right)^2
+
\left((F_3-Y)\phi_2+F_4\phi_1\right)^2
\left)\,\frac{\nu^2}{4\sin\vartheta}\right.\right.\,,
 \nn \\
 V&=&\left.\left.\frac{\beta r^2}{8}\right(\phi_1^2+\phi_2^2-1\right)^2\sin\vartheta\,.
\ee
The energy is gauge invariant. 
Modulo gauge transformations \eqref{res}, 
the zero energy configuration is 
\be
F_1=F_2=F_4=\phi_1=0,~~~\phi_2=1,~~~~~F_3=Y=const.\equiv Y_\infty.
\ee
This vacuum keeps its form under gauge transformations generated by ${\rm U}=\exp\left\{i\,{\cal C}\,\nu\varphi (1+\tau_3)/2 \right\}$
with a constant ${\cal C}$, whose effect is $Y_\infty\to Y_\infty+{\cal C}$.   

The above formulas apply to describe both monopoles and sphalerons. The difference between the two cases is in the boundary 
conditions for the field amplitudes. Specifically, 
let us require the energy  to be invariant under the reflection in the equatorial plane, $\vartheta\to\pi-\vartheta$. 
This implies that certain fields amplitudes do not change so that they are ``even"  while the others change sign under the reflection
hence  they are  ``odd". 
Assuming that $\phi_2\to 1$ at infinity, 
the direct inspection of Eqs.\eqref{EN},\eqref{EN1} 
shows two possible options that  we call ``monopole case" and ``sphaleron case":
\be
\text{\underline{monopole case:}}~~ 
\text{odd}~~~F_1,F_3,Y,\phi_1~~\text{and even}~~
F_2,F_4,\phi_2; \nn \\
\text{\underline{sphaleron case:}}~~ 
\text{odd}~~~F_1,F_4,\phi_1~~\text{and even}~~
F_2,F_3,Y,\phi_2.
\ee
Let us redefine the gauge  field amplitudes as follows,
\be                \label{var}
F_1=-\frac{H_1(r,\vartheta)}{r},~~~~F_2=H_2(r,\vartheta),~~~~F_3=\Theta(\vartheta)+H_3(r,\vartheta)\sin\vartheta\,, \nn \\
F_4=H_4(r,\vartheta)\sin\vartheta\,,~~~~~~~~~~~~Y=\Theta(\vartheta)+y(r,\vartheta)\sin\vartheta\,,
\ee
where the function $\Theta(\vartheta)$ and the behaviour under  $\vartheta\to\pi-\vartheta$ are as follows:
\be         \label{eq}
\text{\underline{monopole case:}}&&~~~~~\Theta(\vartheta)=\cos\vartheta,~~~~~~~\text{odd}~~ H_1,H_3,y,\phi_1~~\text{and even}~~~
H_2,H_4,\phi_2; \nn \\
\text{\underline{sphaleron case:}}&&~~~~~\Theta(\vartheta)=1,~~~~~~~~~~~~\text{odd}~~ H_1,H_4,\phi_1~~\text{and even}~~~
H_2,H_3,y,\phi_2\,.
\ee
The energy density will be finite at the polar axis if only 
all coefficients in front of the $1/\sin\vartheta$ terms in \eqref{EN1} vanish, which requires that 
\be        \label{axis}
H_1=H_3=y=\phi_1=0,~~~~~~H_2=H_4~~~~~\text{for}~~~~~\vartheta=0,\pi.
\ee
These conditions  guarantee that the fields  can be transformed to a regular gauge. Specifically, 
the $\varphi$-components of the gauge fields in \eqref{RR} do not vanish for $\vartheta=0,\pi$, 
which implies a line singularity of the Dirac string type along the symmetry axis. 
However,  this singularity can be gauged away, but if only the parameter $\nu$ in \eqref{RR} is integer.
The regularizing gauge transformation for monopoles is not the same as for sphalerons. 

\subsection{Removing string singularity in the monopole case}

Setting $\Theta(\vartheta)=\cos\vartheta$ in \eqref{var}, 
the gauge transformation that removes the singularity in \eqref{RR} 
is generated by
\be                   \label{Ureg} 
 {\rm U}_\pm =
 e^{- i\nu\varphi \T_3 }
 e^{- i\vartheta \T_2}
 e^{\pm i\nu\varphi/2}
 = 
 e^{\pm i\nu\varphi/2}\,
\begin{pmatrix}
\cos\frac{\vartheta}{2}\,e^{-i{\nu \varphi/2}}  & ~-\sin\frac{\vartheta}{2}\,e^{-i{\nu \varphi/2}}  \\
\sin\frac{\vartheta}{2}\,e^{i{\nu \varphi/2}}  ~& ~\cos\frac{\vartheta}{2}\, e^{i{\nu \varphi/2}} 
\end{pmatrix}, 
 \ee
which brings the SU(2) field to the form 
 \be           \label{gauge2}
W&=&\left.\left.\T_\varphi\left(-\frac{H_1}{r}\,dr+(H_2-1)\,d\vartheta \right)+\nu\,\right( \T_r\,H_3+\T_\vartheta\,(1-H_4) \right)\sin\vartheta\, d\varphi\,.
\ee
This form of the field (and the notation) is often used in the literature; see, e.g., \cite{Kleihaus:1997mn}. 
Here the angle-dependent generators, 
\be                \label{win1}
\T_r =n^a \T_a, ~~~~~\T_\vartheta=\partial_\vartheta \T_r,~~~~~~\T_\varphi=\frac{1}{\nu\sin\vartheta}\,\partial_\varphi \T_r,
\ee
are expressed in terms of the unit vector 
\be   \label{win2}
n^a=\left[\,\sin\vartheta\cos(\nu\varphi),\sin\vartheta\sin(\nu\varphi),\cos\vartheta \,\right]. 
\ee
They  satisfy the standard commutation relations, for example $[\T_r,\T_\vartheta]=i\T_\varphi$.
It is clear that the parameter $\nu$ should be integer  since otherwise the vector $n^a$ is not single-valued. 
Now, \eqref{axis} implies that in the vicinity of the symmetry axis 
 $W=(\T_1\, dx^2-\T_2\, dx^1)(1-H_2)+\ldots$ where $x^a=r n^a$ are the Cartesian coordinates and the dots denote terms that vanish 
 at the axis. This field is regular at the axis and the Dirac string is gone.

The ``$+$" and ``$-$"  sign choices in \eqref{Ureg} determine two locally regular gauges for $B,\Phi$:
\be          \label{gauge2a}
B_\pm &=&\nu\,(\cos\vartheta\pm 1+\,y\,\sin\vartheta)\, d\varphi\,,~~~
\Phi_\pm=
e^{\pm i\nu\varphi/2}\,
\begin{pmatrix}
\left.\left. \,\right(\phi_1\,\cos\frac{\vartheta}{2}-\phi_2\,\sin\frac{\vartheta}{2}\right) e^{-i\nu\varphi/2} \\
\left.\left. \,\right(\phi_1\,\sin\frac{\vartheta}{2}+\phi_2\,\cos\frac{\vartheta}{2}\right) e^{+i\nu\varphi/2}
\end{pmatrix}.~~~~
\ee
Here $B_{-}$ and $\Phi_{-}$ are  regular for $\vartheta=0$, but $B_{-}$ shows the Dirac string singularity along 
the negative $z$-axis at $\vartheta=\pi$, whereas $\Phi_{-}$ has no limit there. 
Therefore, this gauge can be used only in the upper part of the sphere, for $\vartheta\in[0,\pi-\epsilon)$. 
On the other hand, $B_{+}$ and $\Phi_{+}$ are  regular for $\vartheta=\pi$ and can be used in the lower hemisphere, for  $\vartheta\in(\epsilon,\pi]$.  
Therefore,  $B$ and $\Phi$  will be completely regular if 
one uses two  local gauges: $B_{-},\Phi_{-}$  in the upper hemisphere and 
$B_{+},\Phi_{+}$ in the lower hemisphere. The transition from one local gauge to the other 
is performed in the equatorial region, $\epsilon<\vartheta<\pi-\epsilon$,
and provided by ${\rm U}=\exp(i\nu\varphi)$, which is 
single-valued if $\nu\in \mathbb{Z}$. This provides a regular description for all fields.

The U(1) part of the magnetic charge in \eqref{PB} is defined by the integral 
\be                 \label{flux}
\frac{1}{4\pi} \oint_{S^2} dB=\frac{1}{4\pi} \oint_{S^1}(B_{-}-B_{+})=-\frac{\nu}{2\pi}\oint_{S^1}d\varphi
=-\nu\,,
\ee
where $S^1$ is a circle around  the equatorial region of $S^2$ where both $B_{+}$ and $B_{-}$ are regular. 
The winding number $\nu$ is the topological index -- the first Chern class of the U(1) bundle over $S^2$. 
The U(1) part of the magnetic charge and 
the corresponding charge density are 
\be             \label{PA}
P_{\rm U(1)}=-\frac{g}{g^\prime}\,\nu\,,~~~~~\rho_{\rm U(1)}=P_{\rm U(1)}\delta^3(\vec{x}),
\ee
so that the charge is pointlike and located at the origin.

\subsection{Removing string singularity in the sphaleron case}

Setting $\Theta(\vartheta)=1$ in  \eqref{var}, 
the gauge transformation that removes the singularity in \eqref{RR} 
is generated by
${\rm U}=\exp\left\{-i\nu(1+\tau_3)\,\varphi/2 \right\}$. This brings the fields to the form 
\be           \label{gauge1}
W&=&\T_\varphi\left(-\frac{H_1}{r}\,dr+H_2\,d\vartheta \right)+\nu\left( \T_3\,H_3-\T_\rho\,H_4 \right)\sin\vartheta\, d\varphi\,,\nn \\
B&=&\nu\,y\,\sin\vartheta\, d\varphi\,,~~~~~~~~~~
\Phi=\begin{pmatrix}
 e^{-i\nu\varphi} \,\phi_1\\
\phi_2
\end{pmatrix}\,,
\ee
where $\T_\rho=\cos(\nu\varphi)\T_1+\sin(\nu\varphi)\T_2$. Here 
$B$ and $\Phi$ are regular at the symmetry axis and  one has close to the axis 
$W=(\T_1\, dx^2-\T_2\, dx^1)H_2+\ldots$ which is also regular. 

Defining 
$\tilde{H}_2=1+H_2$, 
$\tilde{H}_3=H_3\,\cos\vartheta -H_4\,\sin\vartheta$, 
$\tilde{H}_1=1+H_3\,\sin\vartheta + H_4\,\cos\vartheta $, the field 
 $W$ in  \eqref{gauge1} can be represented 
exactly in the same form as $W$ in \eqref{gauge2}, 
 \be           \label{gauge22}
W&=&\T_\varphi\left(-\frac{H_1}{r}\,dr+(\tilde{H}_2-1)\,d\vartheta \right)+\nu\left( \T_r\,\tilde{H}_3+\T_\vartheta\,(1-\tilde{H}_4) \right)\sin\vartheta\, d\varphi\,,
\ee
which form is often used in the literature \cite{Kleihaus:2008gn,Kleihaus:2008cv,Ibadov:2010ei}. 
This does not mean that sphalerons and monopoles
can be related by simply redefining the field amplitudes, since  the fields $B,\Phi$ in the monopole case  given by \eqref{gauge2a} 
are not the same as  those in in the sphaleron case given by \eqref{gauge1}. 

The $B$ field in the sphaleron case case is topologically trivial, hence the U(1) magnetic charge density  vanishes. 
The SU(2) part of the charge density,  $\rho_{\rm SU(2)}$, does not necessarily vanish, but the total magnetic charge is zero, as we shall see below.

Summarizing, the fields \eqref{RR} can be transformed to a regular gauge only if $\nu$ is integer.
This is an important conclusion, since the field equations can formally be considered for any real $\nu$ giving  perfectly smooth 
solutions for the 7 field amplitudes $H_1,\ldots ,\phi_2$. However, unless $\nu$ is integer, the fields will contain unremovable 
string singularities along the symmetry axis. The only exception is the special case when $H_1=H_2=H_3=H_4=0$ when the SU(2) 
field becomes Abelian.  As will be shown below, $\nu$ can then assume also half-integer values. 

\subsection{Fixing the gauge}

The field equations can be obtained by injecting \eqref{var} to the energy \eqref{EN} and varying with respect to $H_1,H_2,H_3,H_4,y,\phi_1,\phi_2$. 
These equations admit pure gauge solutions due to the residual gauge invariance \eqref{res}, and such zero modes should be removed 
by fixing the gauge, since otherwise the differential operators in the equations will not be invertible. The gauge can be fixed setting to zero 
the divergence of the two-vector  $F_1\, dr+F_2\,d\vartheta$ in \eqref{RR}, which requires that \cite{Kleihaus:1991ks,Kunz:1992uh}
\be                \label{fix}
r\partial_r H_1=\partial_\vartheta H_2. 
\ee
The advantage of this gauge condition is that it is simple, globally defined and yields a good numerical convergence. The disadvantage,
as will be shown in Appendix A, is that it gives rise to a spurious long-range mode contained in solutions at large $r$. This 
spurious mode can be removed by passing to the unitary gauge, but the latter turns out to be singular at the origin, as will be shown in Appendix B. 
Therefore, the gauge condition \eqref{fix} seems to be preferable.

Using this condition, all equations assume a manifestly elliptic form with the standard differential operator
\be
\Delta=
\partial^2_{rr}+\frac{2}{r}\,\partial_r+\frac{1}{r^2}\left(\partial^2_{\vartheta\vartheta}+\cot\vartheta\,\partial_\vartheta\right).
\ee
The equations should be  solved in the domain $r\in[0,\infty)$, $\vartheta\in [0,\pi/2]$, and 
it is also convenient to use  the compact radial variable $x\in[0,1]$ related to $r$ via 
\be            \label{comp}
r=\frac{x}{1-x}. 
\ee
The boundary conditions at $\vartheta=0,\pi/2$ have been described above, while those at the origin $r=0$ and at infinity $r=\infty$
will be described below. 

\section{SPHERICALLY SYMMETRIC SOLUTIONS}
\setcounter{equation}{0}

Solutions of the field equations can be spherically symmetric in exceptional cases, and such solutions can be 
magnetically charged (monopoles) or neutral (sphalerons).

\subsection{Monopoles}

Choosing in \eqref{var} 
\be                          \label{ssm}
\Theta(\vartheta)=\cos\vartheta,~~~~~H_1=H_3=y=\phi_1=0,~~~~~H_2=H_4=f(r),~~~~~\phi_2=\phi(r),
\ee
the angular variables decouple and the equations reduce to 
\be            \label{eq-sph}
f^{\prime\prime}&=&\nu^2\,\frac{f(f^2-1)}{r^2}+\frac{g^2}{2}\,\phi^2 f\,, \nn \\
(r^2\phi^\prime)^\prime &=&\frac14\,(\nu^2+1) f^2\phi+\frac{\beta r^2}{4}\,(\phi^2-1)\phi\,, \nn \\
(\nu^2-1) f^\prime&=&(\nu^2-1) f\phi=0.
\ee
\subsubsection{Abelian monopoles of Dirac}
The simplest solution of these equations exists for any value of $\nu$,
\be
f=0,~~~~\phi=1.
\ee
This describes the Dirac magnetic monopole embedded into the electroweak theory. 
Returning for a moment to the original 
parameterization \eqref{RR} yields 
\be
B=\nu\cos\vartheta\,d\varphi,~~~~W=T_3\,B,~~~~~
\Phi=
\begin{pmatrix}
0 \\
1
\end{pmatrix},~~~~
\ee
and after the gauge transformation 
generated by 
$
{\rm U}_\pm=\exp\left(\pm i\nu\varphi(1+\tau_3)/2\right)
$
this becomes 
\be             \label{Bpm}
B_\pm=\nu(\cos\vartheta\pm 1)\,d\varphi,~~~~W_\pm=T_3\,B_\pm,~~~~~
\Phi=
\begin{pmatrix}
0 \\
1
\end{pmatrix}. 
\ee
Here $W_{-}, B_{-}$ are regular at $\vartheta=0$ and can be used in the northern hemisphere, while 
$W_{+}, B_{+}$ are regular at $\vartheta=\pi$ and can be used in the southern hemisphere. Using these two local gauges 
provides a completely regular description. The transition from   $W_{-}, B_{-}$ to $W_{+}, B_{+}$ is provided by the
gauge transformation in the equatorial region with 
\be             \label{W2c}
{\rm U}=\exp\left(i\nu\,\varphi(1+\tau_3)\right)= 
\begin{pmatrix}
\exp\left( 2i\,\nu\varphi\right)  & 0  \\
0 ~& ~1
\end{pmatrix},
\ee
which is single-valued if $\nu$ is integer or {\it  half-integer}. The latter 
is an important conclusion since generically  $\nu$ 
should be integer, but we see that half-integer values of $\nu$ are also allowed in the particular case when the field configuration is {Abelian}.

Computing the electromagnetic field $F_{\mu\nu}$ in \eqref{Nambu} shows that it admits a potential, 
$F=d{\cal A}$ with 
\be                       \label{ADir}
{\cal A}={\cal A}_\mu d{x^\mu}&=& \left(\frac{g}{g^\prime}\,B_\mu+\frac{g^\prime }{g}\, W^3_\mu\right) 
dx^\mu=
\frac{g^2+g^{\prime 2}}{gg^\prime}B_\pm =
\frac{\nu}{e}\,(\cos\vartheta\pm 1)\,d\varphi\,, 
\ee
which is the potential of the Dirac monopole  
\be              \label{Dir}
\vec{\cal B}=\vec{\nabla}\wedge \vec{\cal A}=\frac{{P}\,\vec{r}}{r^3}\,,
\ee
with the magnetic charge
\be            \label{Pc}
{P}=-\frac{\nu}{e}\,.
\ee
Here $e=gg^\prime$ is the dimensionless electron charge defined in \eqref{e}. 
It will be commonly  assumed  below that $\nu>1$, hence the magnetic charge $P$ defined by \eqref{Pc} is {\it negative}
(the opposite sign convention for the charge was made in \cite{Gervalle:2022npx}). 
Since $\nu$ in \eqref{Pc} can be integer or half-integer, it follows that 
\be
n\equiv -2\nu
\ee
is integer (notice the minus sign here), hence  the magnetic charge  fulfills 
the standard Dirac quantization condition, 
\be
e P=\frac{n}{2}~~~~\text{with}~~~~n\in \mathbb{Z}. 
\ee

The magnetic charge can  be split into  two parts according to \eqref{PT}, 
corresponding to the Abelian $B_\mu$ and non-Abelian $W^3_\mu$ contributions to \eqref{ADir},
\be           \label{SU0}
{P}={P}_{\rm U(1)}+{P}_{\rm SU(2)}~~~~~~~~~~\text{with}~~~~~
{P}_{\rm U(1)}=g^2{P}\,,~~
{P}_{\rm SU(2)}=g^{\prime 2}{P}\,,
\ee
and it is worth noting that the non-Abelian part, 
\be
{P}_{\rm SU(2)}=\sin^2\thetaw\times \frac{n}{2e}=\frac{g^\prime}{g}\,\frac{n}{2},
\ee 
is quantized as in  the Nambu formula \eqref{Nam}. 
The U(1) and SU(2) parts of the 
magnetic charge density \eqref{rhom} 
are
\be            \label{SU}
\rho_{\rm U(1)}=g^2{P}\delta^3(\vec{x}),~~~~~~~~~
\rho_{\rm SU(2)}=g^{\prime 2}{P}\delta^3(\vec{x}). 
\ee
Both parts of the magnetic charge make singular contributions to the energy 
\be                \label{01}
 E=\frac{2\pi\nu^2}{g'^2}\int_0^\infty{\frac{dr}{r^2}}+\frac{2\pi\nu^2}{g^2}\int_0^\infty{\frac{dr}{r^2}}\equiv E_{\rm U(1)}+E_{\rm SU(2)}.
\ee

Summarizing, the Dirac monopole can be viewed as a superposition of two pointlike magnetic charges $P_{\rm U(1)}$ and 
$P_{\rm SU(2)}$ located at the origin, both making an infinite contribution to the energy. Below we shall be considering other, 
more general solutions approaching the Dirac monopole configuration in the far field region. 
Their  $P_{\rm U(1)}$  charge is still pointlike, 
but the $P_{\rm SU(2)}$ charge is smoothly distributed over a finite volume and its contribution to the total energy is finite. 
The simplest solution of this type is the Cho-Maison monopole.

\subsubsection{The non-Abelian monopole of Cho and Maison  \label{CMM}}

\begin{figure}
    \centering

       	\includegraphics[width=5 cm,angle =-90 ]{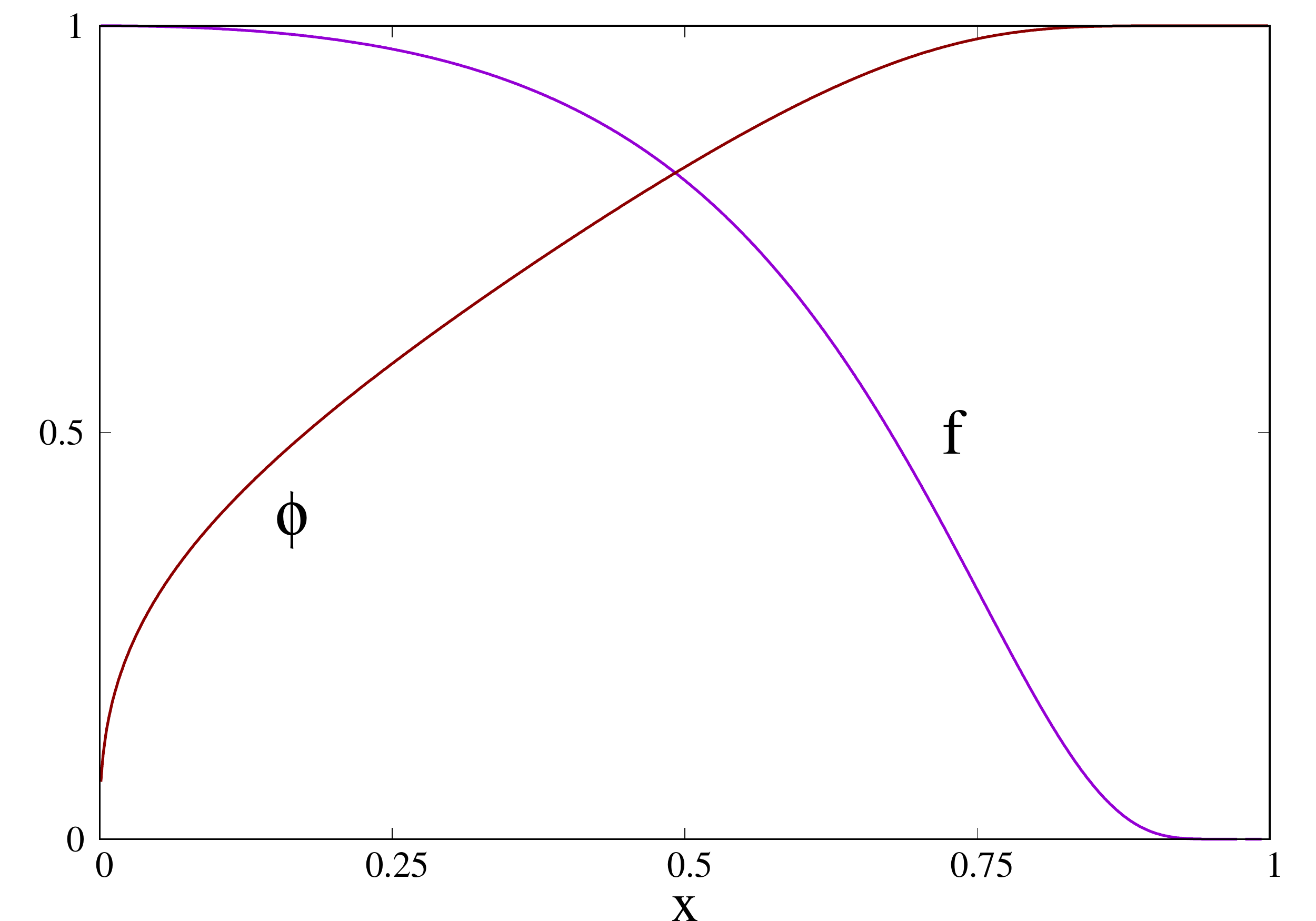}
		\includegraphics[width=5 cm,angle =-90 ]{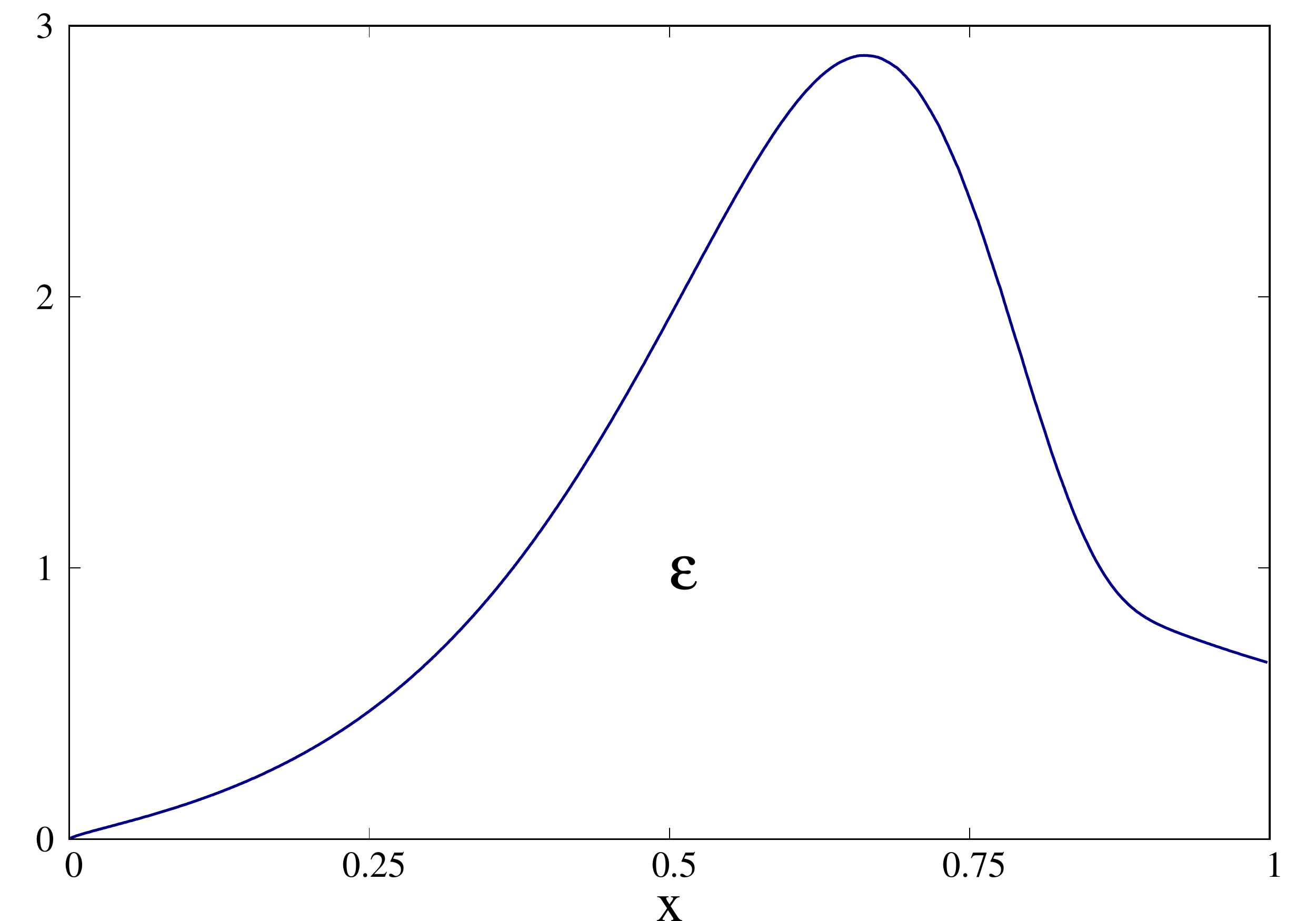}

    \caption{Profiles of the gauge field and Higgs amplitudes $f$ and $\phi$ (left) and the energy density (right) against the compact radial coordinate $x$ for the
    spherically symmetric CM monopole with $g^{\prime 2}=0.23$.}
    \label{cho}
\end{figure}

For $\nu=\pm 1$ (hence for $n=\pm 2$), 
Eqs.\eqref{eq-sph} admit a smooth non-Abelian solution for which the amplitudes $f,\phi$ interpolate between 
the following asymptotic values:
$f=1+{\cal O}(r^2)$, $\phi={\cal O}(r^\delta)$ for $r\to 0$, 
where $\delta=(\sqrt{3}-1)/2$, and 
$f={\cal O}\left(e^{-\mw r} \right)$, $\phi=1+{\cal O}\left(e^{-\mh r} \right)$ for $r\to \infty$; see Fig.\ref{cho}. 
This solution was  found numerically by Cho and Maison (CM) \cite{Cho:1996qd}, 
and its existence was proven by Yang \cite{yang2014solitons}. 
At infinity the fields approach those for the Dirac monopole with $n=\pm 2$, while at the 
origin the non-Abelian field is regular and its contribution to the energy is finite.
However, the U(1) contribution to the energy is still infinite due to the ${\cal E}_B$ term in  \eqref{EN1},
since for $Y=\cos\vartheta$ and $\nu=\pm 1$ one has ${\cal E}_B=\sin\vartheta/r^2$ whose contribution to the energy is 
the same as $E_{\rm U(1)}$ with $\nu^2=1$ in \eqref{01}.

The total energy is $E=E_{\rm U(1)}+E_{\rm SU(2)}$ where 
   \be               \label{Energy}
 E_{\rm SU(2)}&=&4\pi\int_0^\infty  \left(\frac{1}{g^2}\left(\frac{\nu^2+1}{2}f'^2+\nu^2\frac{ (f^2-1)^2}{2r^2}\right)\nn\right. \\
 &&+\left.(r\phi')^2+\frac{\nu^2+1}{4}(f\phi)^2+\frac{r^2\beta}{8}(\phi^2-1)^2\right)dr. 
 ~~~~~~~
   \ee
Unless otherwise stated  (the only exception will be made in Section~V.E), 
it will always be assumed in this formula  that $\nu^2=1$, since only in this case the spherical symmetry can be maintained on-shell.
 Equations \eqref{eq-sph} then can be obtained by varying $ E_{\rm SU(2)}$
with respect to $f,\phi$.   

 For the Dirac monopole with $\nu^2=1$ one has $f=0$ hence $E_{\rm SU(2)}=\infty$, 
 but for the CM monopole one obtains a finite value 
$
E_{\rm CM}\equiv E_{\rm SU(2)}=15.759,
$
assuming that $g^{\prime 2}=1-g^2=0.23$. 
Therefore, even though the total energy  is infinite due to the U(1) field,
this solution is less energetic than the Dirac monopole. It is convenient to use the compact coordinate $x$ defined in \eqref{comp} 
to represent the energy as
\be            \label{ESU2}
 E_{\rm SU(2)}=4\pi \int_0^1 \varepsilon(x)\, dx\,,
 \ee
 where the energy density $\varepsilon$ is the integrand in \eqref{Energy} multiplied by ${dr}/{dx}$. Due to the longe-range 
 magnetic field of the monopole, the integrand in  \eqref{Energy}  decays at large $r$ as $1/r^2$,  while ${dr}/{dx}\sim r^2$, 
 hence $\varepsilon(x)$ approaches at infinity a constant value $\varepsilon_\infty=1/(2g^2)$, as seen in Fig.\ref{cho}.
 As a result, the non-Abelian part of the energy is smoothly distributed in space.

 The magnetic charge density is defined in \eqref{rhom}.  Its U(1) part is given by the general formula \eqref{PA} which applies to all monopoles, 
 while the SU(2) part is 
 \be
 \rho_{\rm SU(2)}=\frac{1}{4\pi}\,\frac{g^\prime}{g\,r^2}\,(f^2)^\prime \,.
 \ee
This determines  the SU(2) part of the magnetic charge, 
 \be           \label{char}
 {P}_{\rm SU(2)} =\int \rho_{\rm SU(2)} \sqrt{-\rm g}\, d^3x=\nu\,\frac{g^\prime}{g}\int_0^\infty (f^2)^\prime dr
 =-\nu\, \frac{g^\prime}{g}=-g^{\prime 2}\,\frac{\nu}{e}=g^{\prime 2}P. 
 \ee
This is the same as  ${P}_{\rm SU(2)}$ with $\nu^2=1$ in \eqref{SU0}, and
the U(1) part of the charge is the same ${P}_{\rm U(1)}$ in \eqref{SU0}. 
Therefore, the SU(2) part 
 of the magnetic charge is distributed all over the space while its U(1) part is 
 concentrated at the origin 
 as for the Dirac monopole.

\begin{figure}
    \centering

       	\includegraphics[width=5 cm,angle =-90 ]{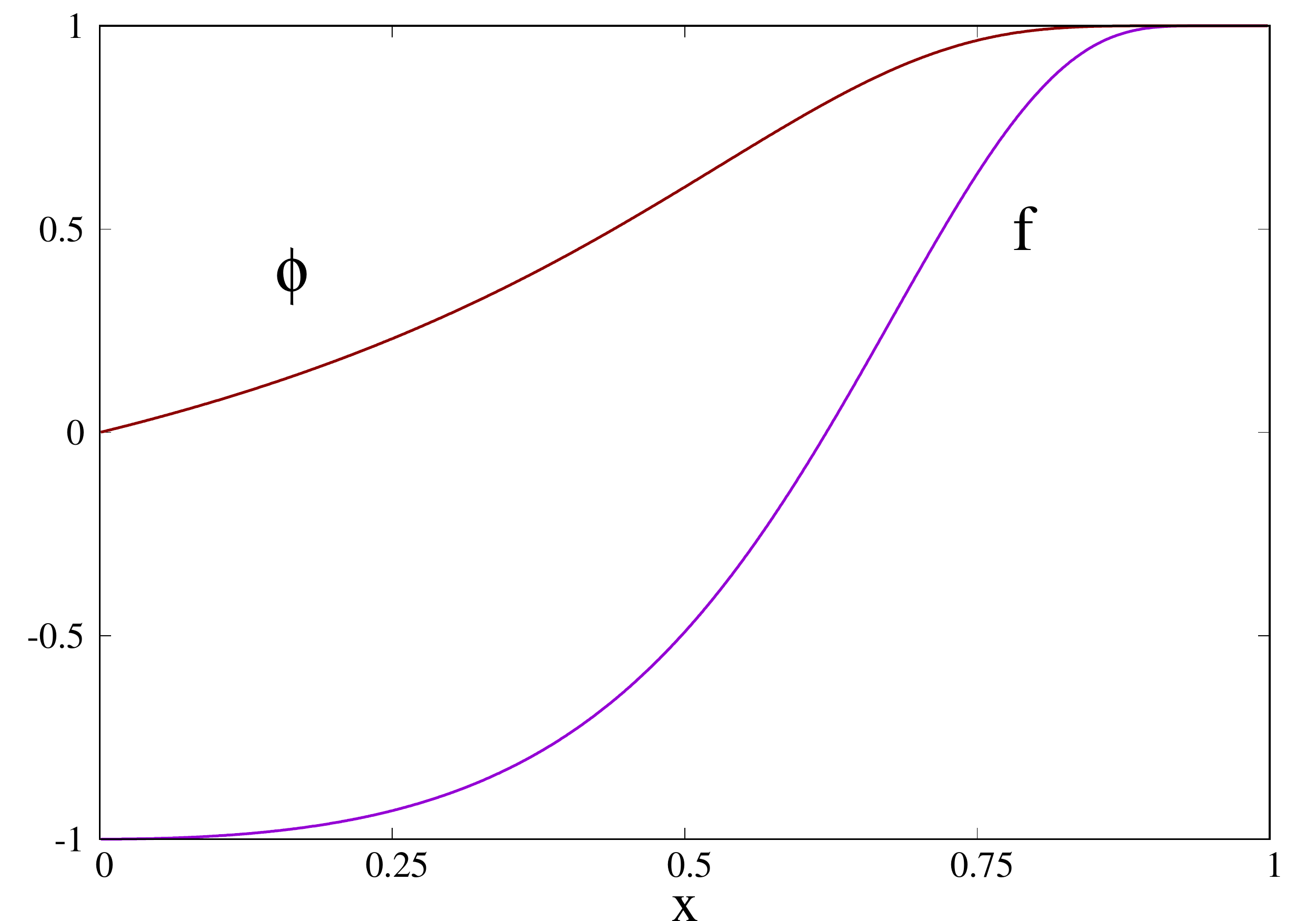}
		\includegraphics[width=5 cm,angle =-90 ]{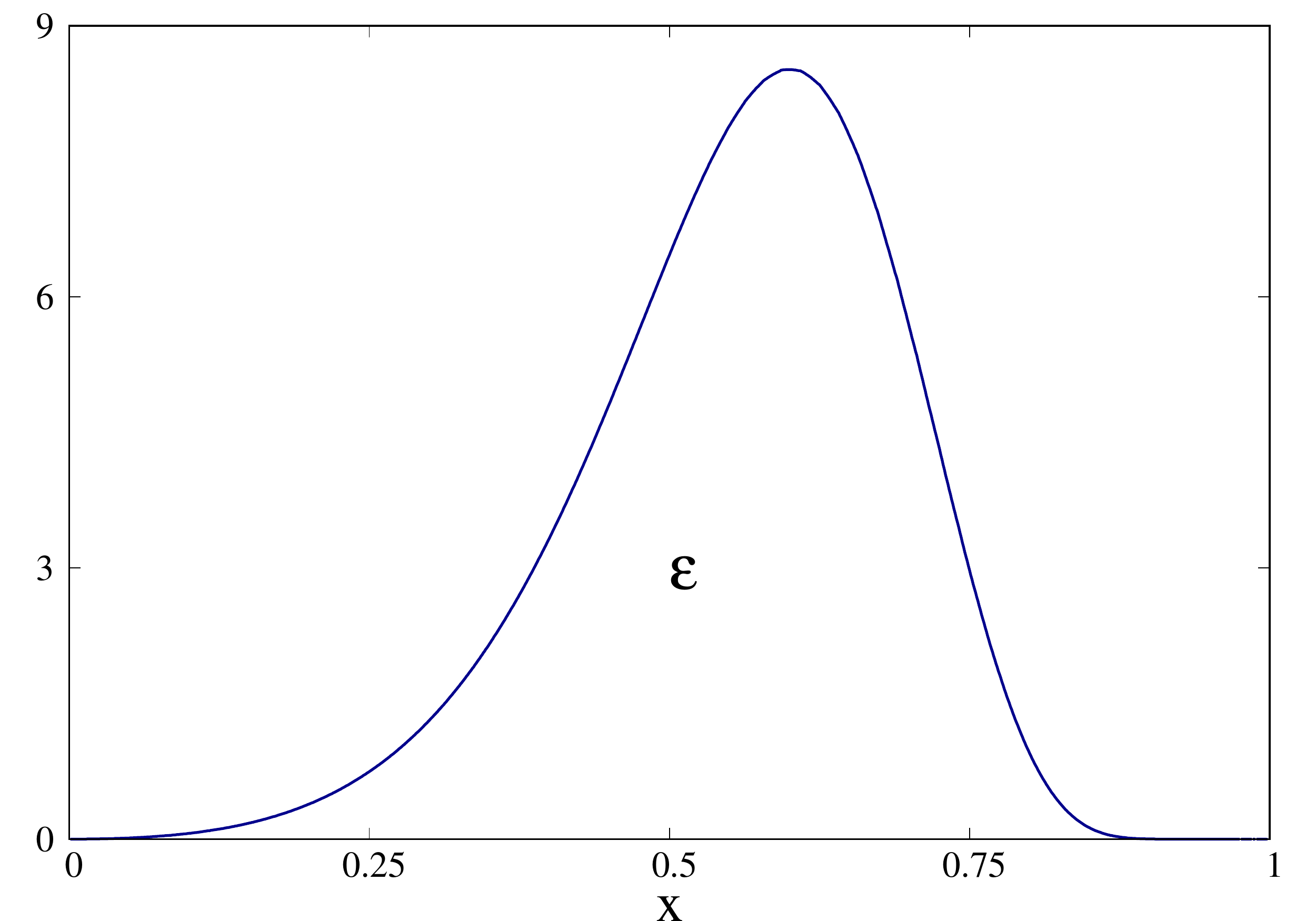}

    \caption{Profiles of $f$ and $\phi$ (left) and the energy density (right) against the compact radial coordinate $x$ for the 
  the spherically symmetric sphaleron with $g^\prime=0$.}
    \label{FIGsph}
\end{figure}

\subsection{Sphaleron}

The spherically symmetric CM monopole exists for any value of the weak mixing angle, but the sphaleron 
can be spherically symmetric only if $g^{\prime}=0$ when 
the U(1) hypercharge field decouples  \cite{Dashen:1974ck,Klinkhamer:1984di}.
The solution is obtained by setting  in \eqref{var} 
\be                 \label{sp0}
\Theta(\vartheta)=1,~~~~H_2=f(r)-1,~~~H_3=H_2\,\sin\vartheta,~~~H_4=H_2\,\cos\vartheta,~~~\phi_2=\phi(r), ~~~~
\ee
with $H_1=y=\phi_1=0$. Notice that this implies that  the U(1) field is not zero but a pure gauge, $B=\nu\, d\varphi$.
Since the U(1) gauge transformations are still allowed when $g^{\prime}=0$, the pure gauge $B$ 
 can be gauged away, but at the expense of giving the Higgs field a $\varphi$-depending phase. 
 Therefore, it is preferable  to work in the gauge \eqref{sp0} where nothing depends on $\varphi$. 
 
 Injecting \eqref{sp0}  to the equations, the angular variables 
decouple yielding 
\be            \label{eq-sphal}
f^{\prime\prime}&=&\frac{f(f^2-1)}{r^2}+\frac{1}{2}\,\phi^2 (f-1)\,, \nn \\
(r^2\phi^\prime)^\prime &=&\frac14\,(\nu^2+1) (f-1)^2\phi+\frac{\beta r^2}{4}\,(\phi^2-1)\phi\,, \nn \\
(\nu^2-1) f^\prime&=&(\nu^2-1) (f-1)\phi=0.
\ee
Only trivial solutions are possible for arbitrary $\nu$, but for $\nu^2=1$ there is  a non-trivial solution with
asymptotics  $f=-1+{\cal O}(r^2)$, $\phi={\cal O}(r)$ as $r\to 0$ and 
$f=1+{\cal O}\left(e^{-\mw r} \right)$, $\phi=1+{\cal O}\left(e^{-\mh r} \right)$ for $r\to \infty$ \cite{Dashen:1974ck,Klinkhamer:1984di}.
This solution is show in Fig.\ref{FIGsph}. 
Its total energy, 
\be               \label{Energy1}
    E&=&4\pi\int_0^\infty dr\left(f'^2+\frac{ (f^2-1)^2}{2r^2}+(r\phi')^2+\frac{1}{2}(f-1)^2\phi^2+\frac{r^2\beta}{8}(\phi^2-1)^2\right)
     \equiv 4\pi \int_0^1 \varepsilon\, dx,~~~~~~~~
\ee
is finite and evaluates to 
$
E=33.538.
$
Since $g^\prime=0$, the electromagnetic field is zero and 
the sphaleron does not support long-range fields, hence  its 
energy density $\varepsilon(x)$  approaches zero at infinity, as seen in Fig.\ref{FIGsph}.

\begin{figure}[th]
\hbox to \linewidth{ \hss
	\includegraphics[width=8 cm]{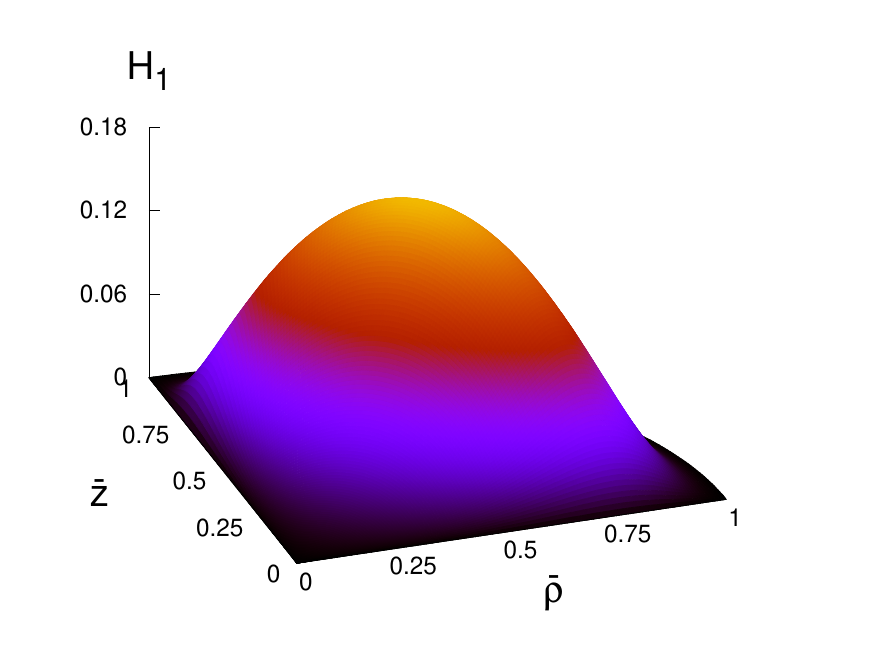}
	\includegraphics[width=8 cm]{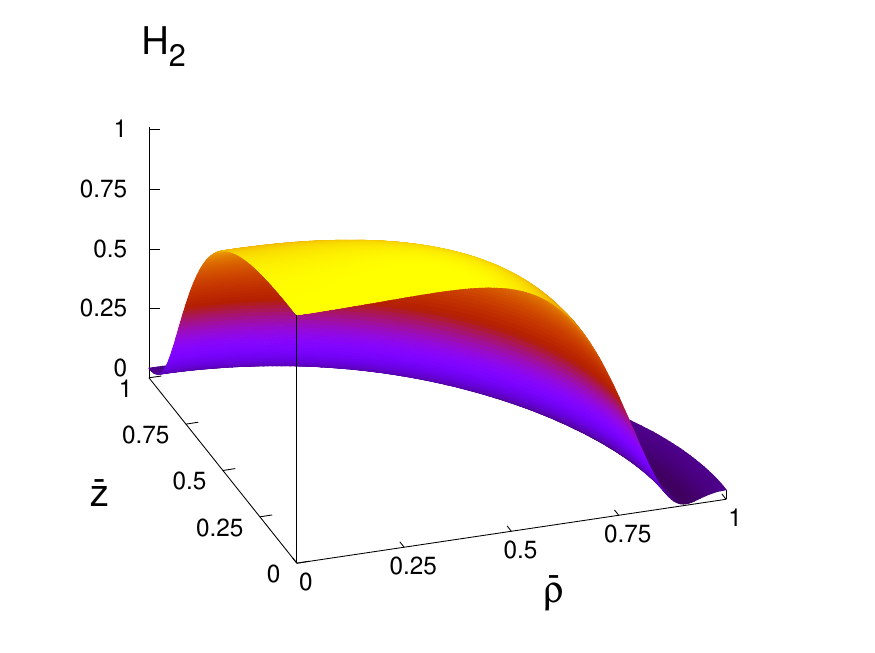}	
\hss}
\hbox to \linewidth{ \hss
	\includegraphics[width=8 cm]{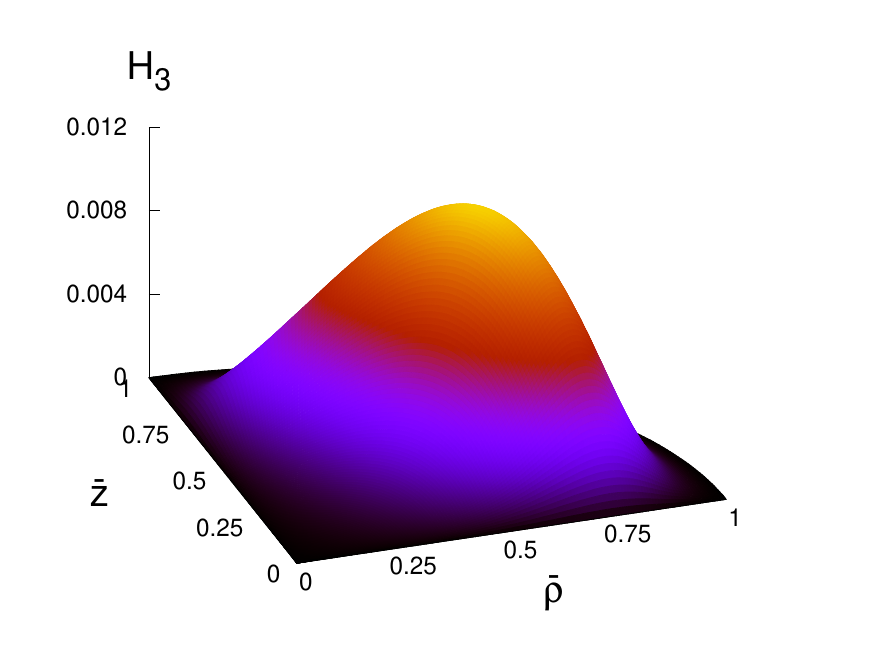}
	\includegraphics[width=8 cm]{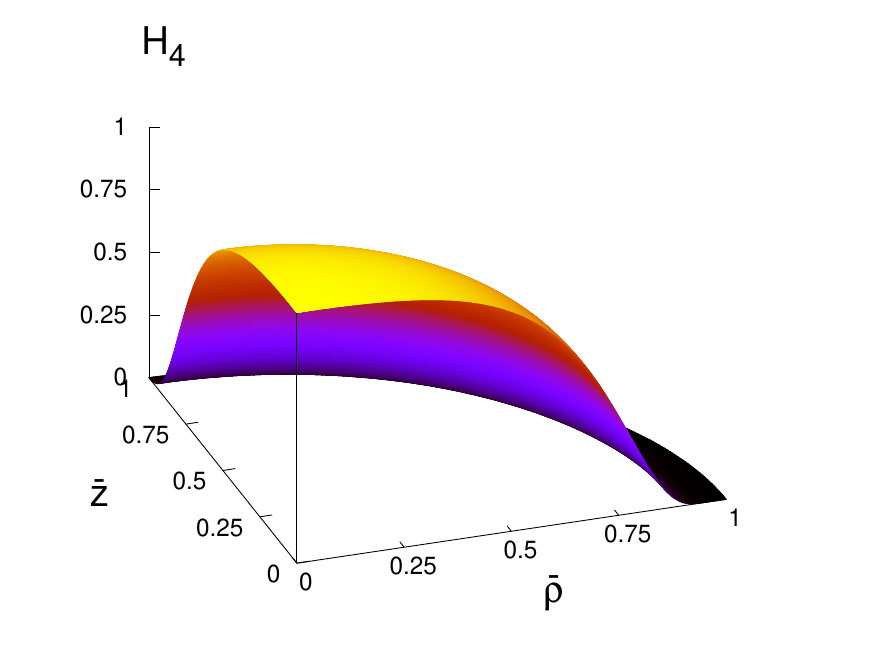}	
\hss}
\caption{The SU(2) amplitudes  for the $\nu=2$ monopole solution against 
$\bar{\rho}=x\,\sin\vartheta$ and $\bar{z}=x\,\cos\vartheta$.}
\label{Fig1}
\end{figure}

\section{AXIALLY SYMMETRIC MONOPOLES}
\setcounter{equation}{0}

The spherically symmetric monopoles of Cho-Maison exists for any $\thetaw$ but only for the magnetic charge $P=\pm 1/e$. 
In order to construct their generalizations for higher values of $|P|$, one should relax the assumption  of spherical symmetry. 
The simplest possibility is to consider axially symmetric fields discussed in Section \ref{II}.
Summarizing the discussion there, here are the 
boundary conditions for the axially symmetric monopoles: 
\be            \label{BCmon}
\text{axis~~}\underline{\vartheta=0}:&&~~~~~H_1=H_3=y=\phi_1=0,~~~~~~
\partial_\vartheta H_2=\partial_\vartheta H_4=\partial_\vartheta \phi_2=0;\nn \\
\text{equator~~}\underline{\vartheta=\pi/2}:&&~~~~~H_1=H_3=y=\phi_1=0,~~~~~~
\partial_\vartheta H_2=\partial_\vartheta H_4=\partial_\vartheta \phi_2=0; \nn \\
\text{origin~~}\underline{r=0}:&&~~~~~H_1=H_3=y=\phi_1=\phi_2=0,~~~~~~
H_2=H_4=1; \nn \\
\text{infinity~~}\underline{r=\infty}:&&~~~~~H_1=H_2=H_3=H_4=y=\phi_1=0,~~~~~\phi_2=1. 
\ee
The conditions at the symmetry axis and in the equatorial plane are determined by \eqref{eq},\eqref{axis}, while 
those at the origin and at infinity are the same as for spherically symmetric monopoles in \eqref{ssm}. 
It turns out that when  these boundary conditions are fulfilled, 
the relation $H_2=H_4$ at the axis mentioned in \eqref{axis} is also fulfilled; we checked this numerically.

Our aim is to solve the field equations with these boundary conditions to determine the components of the ``state vector"
\be       \label{state}
\Psi=[H_1,H_2,H_3,H_4,y,\phi_1,\phi_2],
\ee
which are functions of $r,\vartheta$. We solve the equations with the 
FreeFem++  numerical solver based on the finite element method  \cite{MR3043640}. 
This solver uses the weak form of  differential equations obtained by transforming them into integral equations, 
expanding with respect to basis functions obtained by triangulating  the integration domain, and handling 
the non-linearities with the Newton-Raphson procedure. The numerical procedure is stable and shows a fast convergence rate 
on 4 laptop parallel   processors.

The equations contain the parameter $\nu$, and for $\nu=1$ the solution is known -- this is the spherically symmetric CM monopole for which 
\be                  \label{sps}
\underline{\nu=1}:~~~~H_1=H_3=y=\phi_1=0,~~~~H_2=H_4=f(r),~~~~\phi_2=\phi(r),
\ee
with $f(r)$ and $\phi(r)$ shown in Fig.\ref{cho}. We use this solution as the starting point in the iterative 
procedure to change the value of $\nu$. 
Of course, $\nu$ should be integer for  the line singularities in the fields to be absent, but the equations can be solved 
any real $\nu$.  Our numerical scheme converges well for $\nu\neq 1$ and we were able to go as far as  $\nu=100$,
after which the virial relation deteriorates. The latter is defined as follows. 

\begin{figure}[th]
\hbox to \linewidth{ \hss
	\includegraphics[width=8 cm]{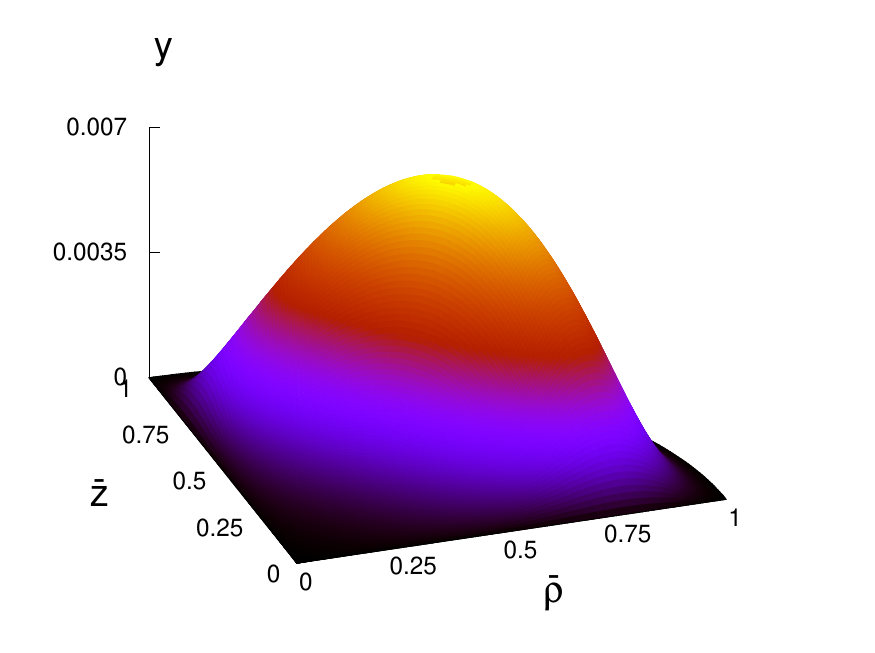}
	\includegraphics[width=8 cm]{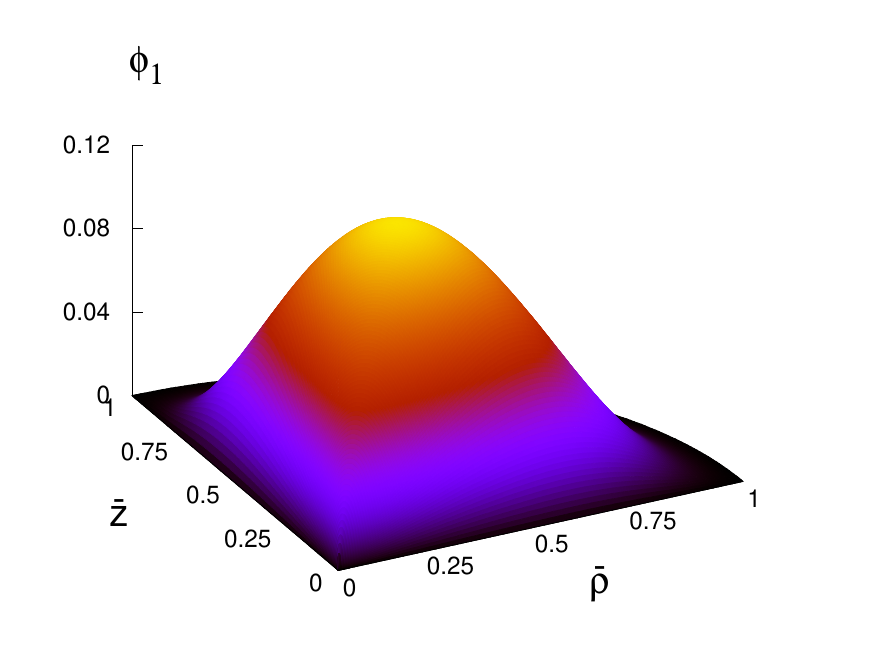}
\hss}
\hbox to \linewidth{ \hss
	\includegraphics[width=8 cm]{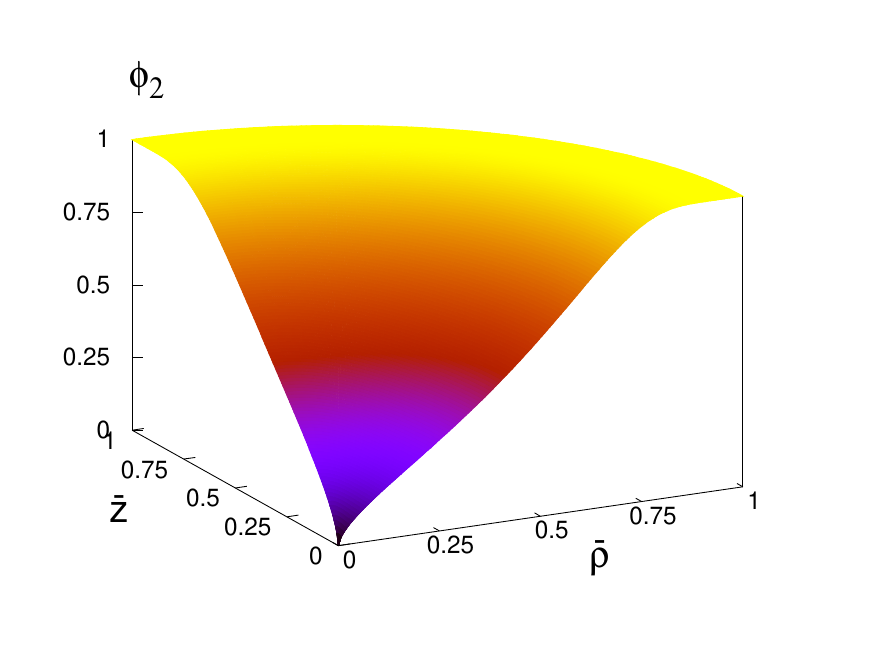}
	\includegraphics[width=8 cm]{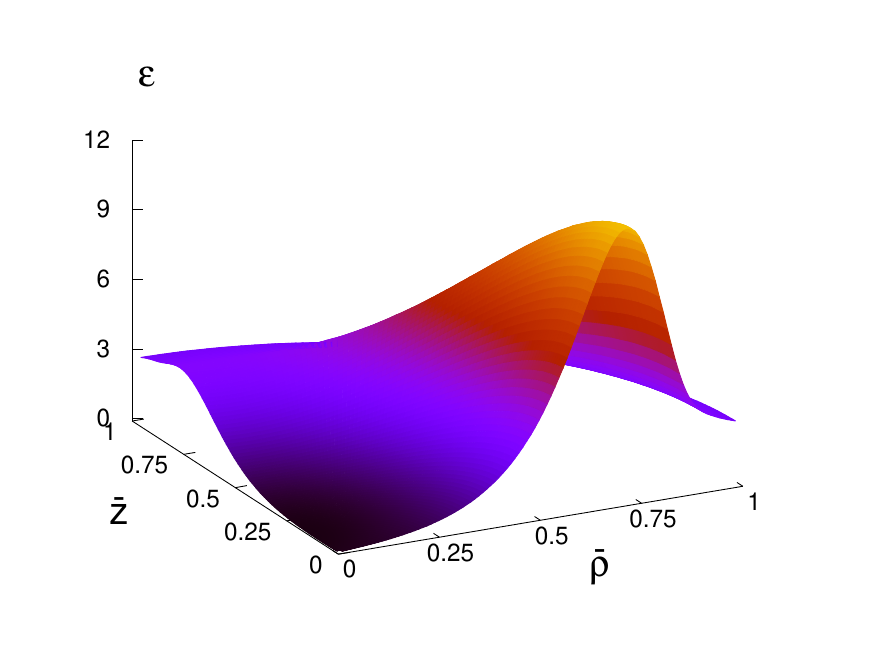}
\hss}
\caption{The  U(1)  and Higgs amplitudes  and the energy density for the $\nu=2$ monopole solution.}
\label{Fig2}
\end{figure}

\subsection{Virial relation}

The energy \eqref{EN} is infinite 
due to the contribution of the U(1) term ${\cal E}_{B}$.
Since $Y=\cos\vartheta+y\,\sin\vartheta$, one has 
\be               \label{EB}
{\cal E}_B=\left(
(\partial_r Y)^2
+\frac{1}{r^2}\,(\partial_\vartheta Y)^2
\left)\,\frac{\nu^2}{\sin\vartheta}\right.\right.\,=\frac{\nu^2}{r^2}\,\sin\vartheta+\ldots \,,
\ee
and injecting this  to \eqref{EN} yields 
\be                \label{ENo}
E=\int T_{00}\, \sqrt{-\rm g}\, d^3 x=\frac{2\pi\nu^2}{g'^2}\int_0^\infty{\frac{dr}{r^2}}+E_{\rm reg}\equiv E_{\rm U(1)}+E_{\rm reg}. 
\ee
Here the first term on the right is infinite and is the same as  $E_{\rm U(1)}$ in the energy  \eqref{01}  
of the pointlike monopole. The second term on the right, $E_{\rm reg}$, 
is finite and contains  the  finite part of the U(1) contribution,
denoted by the dots in \eqref{EB},
and also contributions of the SU(2) and Higgs fields.
In other words, $E_{\rm reg}$ is the regularized energy obtained by subtracting the divergent term $E_{\rm U(1)}$. 
It is determined by the state vector $\Psi$,
\be                           \label{rho}
E_{\rm reg}[\Psi]\equiv 2\pi \int_{0}^{\pi} \sin\vartheta\, d\vartheta \int_0^1 \varepsilon(x,\vartheta)\, dx\,,
\ee
which reduces to \eqref{Energy} with $\nu^2=1$ for the spherically symmetric field \eqref{sps}. The field equations 
determining the state vector   $\Psi$ are obtained by varying the total energy $E=E_{\rm U(1)}+E_{\rm reg}$,
but they can equally  be obtained by varying only $E_{\rm reg}$, 
\be
\frac{\delta E_{\rm reg}[\Psi]}{\delta\Psi}=0,
\ee
since $E_{\rm U(1)}$ does not depend on $\Psi$. 
If $\Psi(r,\vartheta)$ is a solution  then $E_{\rm reg}$ should be stationary with respect to the rescaling $r\to \lambda\, r$,
which leads to the virial relation, 
\be            \label{v}
v\equiv \left.\frac{d}{d\lambda} \ln E_{\rm reg} [\Psi(\lambda\, r,\vartheta)]\right|_{\lambda=1}=0.
\ee
This relation is fulfilled for all our solutions with a precision depending on the numbers of the discretization points $N_x$ 
and $N_\vartheta$ along the $x,\vartheta$ axes (these numbers determine the triangulation pattern for the FreeFem++ solver). 
Taking $N_x=100$ and  $N_\vartheta=50$ yields typically 
$v\sim 10^{-8}$ or $v\sim 10^{-7}$.

\subsection{Solutions}

The profiles of the $\nu=2$ solution are shown in Fig.\ref{Fig1} and  Fig.\ref{Fig2}. The functions $H_2,H_4$,  $\phi_2$ 
which do not vanish in the spherically symmetric limit $|\nu|=1$ 
remain essentially the same for  $|\nu|>1$ and almost do not depend on the angle $\vartheta$. 
The most notable change is that $\phi_2$ now faster approaches zero at the origin, as described by 
Eq.\eqref{lamx} below, whereas  $H_2$ is not strictly  positive. 
On the other hand, the functions $H_1,H_3,y,\phi_1$ which vanish  for 
$\nu^2=1$ no longer vanish  for $\nu^2>1$ and show a strong 
$\vartheta$-dependence. The norm of Higgs field $|\Phi|$  vanishes at the origin.

\begin{figure}
    \centering
		\includegraphics[width=5 cm,angle =-90 ]{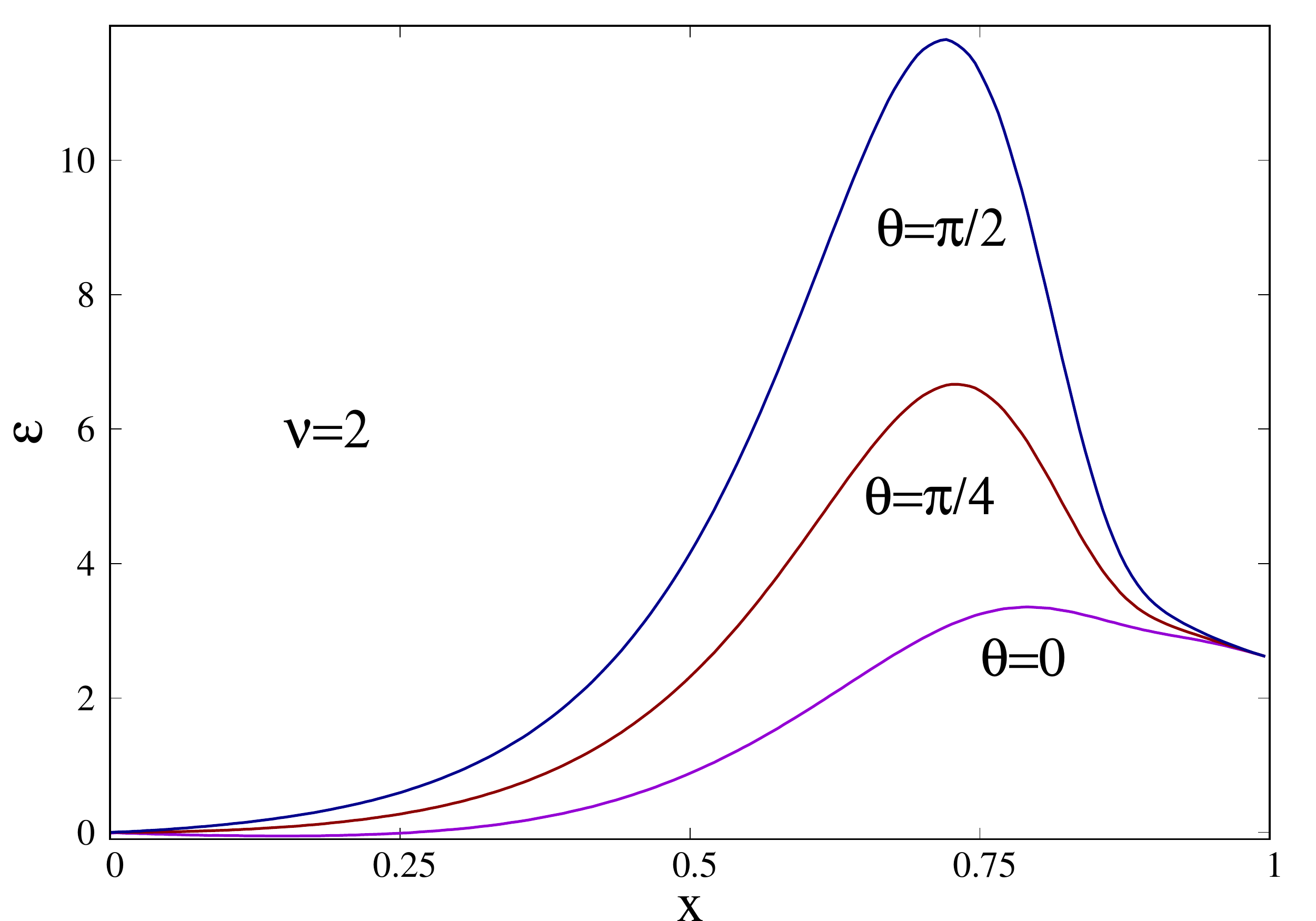}
			\includegraphics[width=5 cm,angle =-90 ]{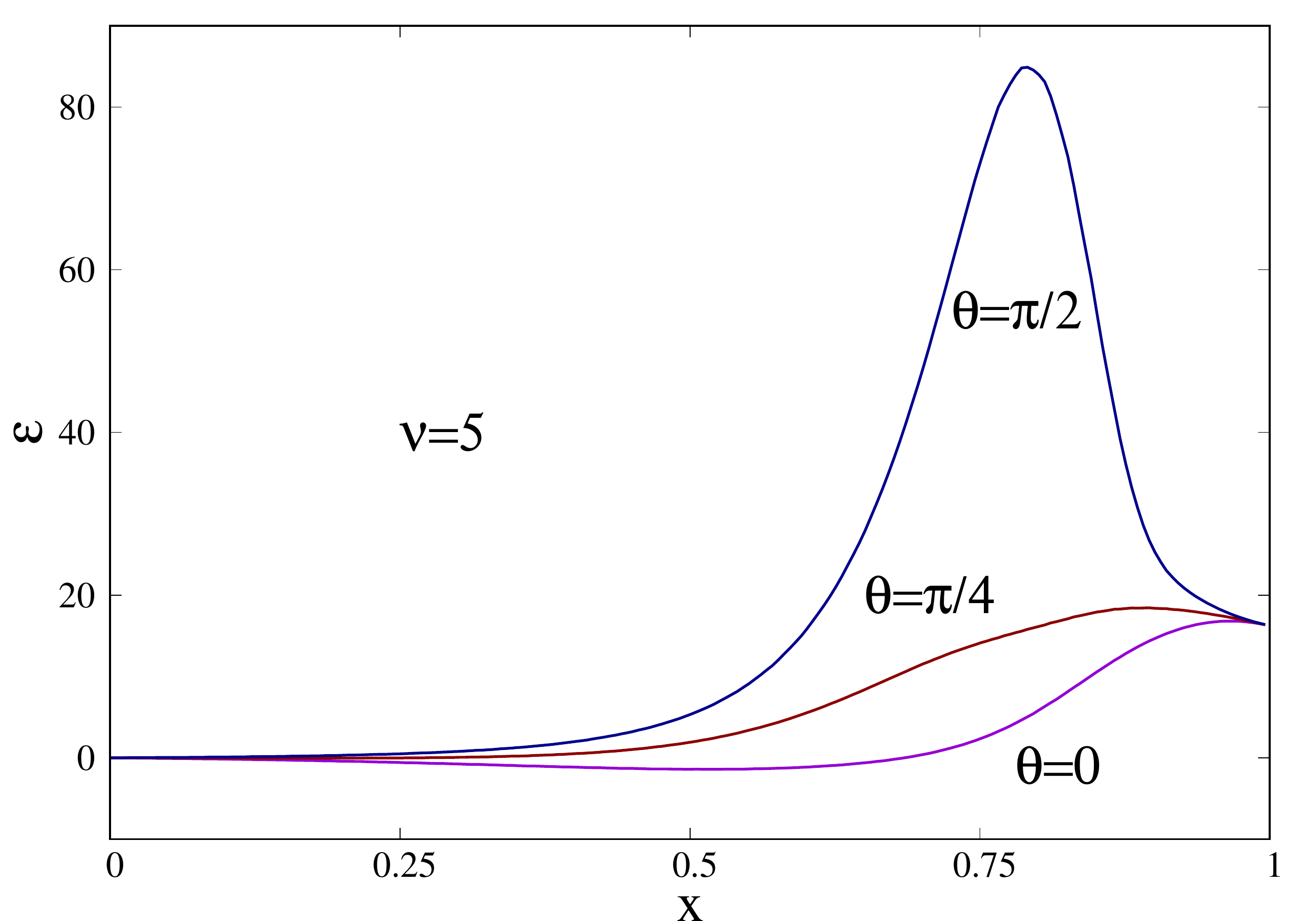}

    \caption{The energy density $\varepsilon(x,\vartheta)$ for several fixed values of $\vartheta$ for $\nu=2$ (left) and for $\nu=5$ (right).
    In the latter case the maximum is much higher.  }
    \label{en3}
\end{figure}

The energy density $\varepsilon$ defined in \eqref{rho} depends only on the radial coordinate for $\nu=\pm 1$,
but already for $\nu=2$ it shows a strong $\vartheta$-dependence with a marked maximum 
 in the vicinity of the equatorial plane, as seen in Fig.\ref{Fig2} and in Fig.\ref{en3}.
 It is interesting that $\varepsilon$ is actually not positive definite and can 
assume negative values in the central region, although the total energy density $T_{00}$ 
including the unbounded $U(1)$ contribution 
is of course always positive.

\begin{figure}
    \centering
			\includegraphics[width=15 cm,angle =0 ]{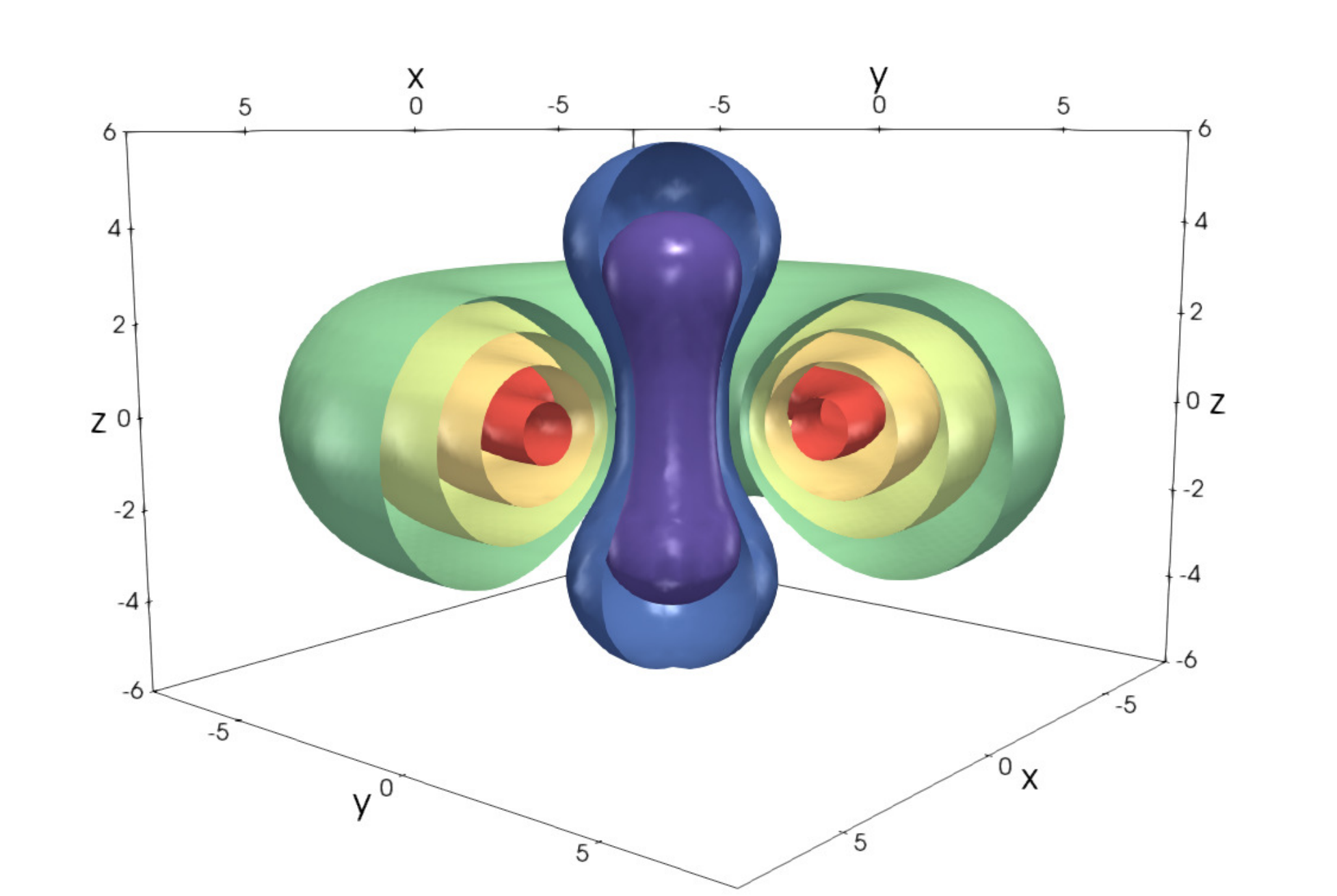}

    \caption{Surfaces of constant energy density $\varepsilon=\varepsilon_0$ 
    for the $\nu=5$ monopole solution expressed in Cartesian coordinates $x,y,z$.  
    For small values of $\varepsilon_0$ the surfaces are deformed ellipsoids but for larger $\varepsilon_0$ they become tori. 
  }
    \label{ensurf}
\end{figure}

The profiles of the energy density $\varepsilon(x,\vartheta)$ for fixed values of $\vartheta$ in Fig.\ref{en3} show that $\varepsilon$ 
is an almost monotone 
function of the radial coordinate along the symmetry axis at $\vartheta=0$, but it shows a marked maximum along the equatorial plane
for $\vartheta=\pi/2$. 
This implies that surfaces of constant energy density $\varepsilon(x,\vartheta)=\varepsilon_0$ are similar to 
ellipsoids if $\varepsilon_0$ is small, but for larger 
values of $\varepsilon_0$ they assume toroidal form, which indeed can be seen in Fig.\ref{ensurf}.

Solutions with $\nu>2$ have essentially the same structure as the $\nu=2$ solution. The functions $H_2,H_4,\phi_2$ always depend 
only weakly on the polar angle $\vartheta$ while $H_1,H_3,y,\phi_1$ show more and more pronounced extrema 
when $\nu$ increases. The Higgs field vanishes only at the  origin, and  close to the origin one has 
\be                \label{lamx}
\phi_1\sim\phi_2\sim r^\lambda~~~~\text{with}~~~~\lambda=\frac{\sqrt{1+2\nu}-1}{2}\,,
\ee
as explained in Appendix B. 
The energy density gets more and more concentrated in the equatorial region and attains higher and higher values there.
This can be seen in Fig.\ref{en3} where the density $\varepsilon$  is shown for $\nu=2$ and $\nu=5$. 
 The numerical values of the regularized energy $E_{\rm reg}$ for several values of 
the winding number $\nu$ are shown in Table I.  We include for completeness also the $\nu=1/2$ solution because it corresponds 
to the minimal value of the magnetic charge $|n|=2|\nu| =1$, but one should remember that this solution contains the line singularity.

Many technical details, as for example 
the asymptotic structure of the solutions at infinity, solutions at the origin,  are given in the two Appendices.

\begin{table}            \label{Tab1}
\begin{tabular}{|c | c | c | c | c  | c | c |}
\hline
$\nu$ &$1/2$  &  $1$  & $2$   &  $3$  & $4$  & $5$     \\
\hline
$E_{\rm reg}$ & $~6.94 ~$ & $~15.76~ $ &  $~38.12~ $ &  $~65.76~ $ &  $~97.92~ $ & $~134.13~ $  \\
\hline
$q$ & $~ -0.51~$ & $~0 ~ $ &  $~ 3.66~ $ &  $~10.61 ~ $ &  $~20.68 ~ $ & $~ 33.78~ $  \\
\hline
\end{tabular}
\caption{The energy $E_{\rm reg}$  and quadrupole moment  $q$ for several monopole solutions.}
\end{table}

\subsection{The interior structure}
The profiles functions $H_1,H_2,H_3,H_4,y,\phi_1,\phi_2$ of the solutions and the energy density are insensitive to the 
sign of $\nu$, so that for example, they are the same for $\nu=2$ and $\nu=-2$. On the other hand, the 
electromagnetic field in \eqref{Nambu} and hence the electric and magnetic currents in \eqref{cur} do depend 
on the sign of $\nu$.

\begin{figure}
    \centering

				\includegraphics[width=8 cm,angle =0 ]{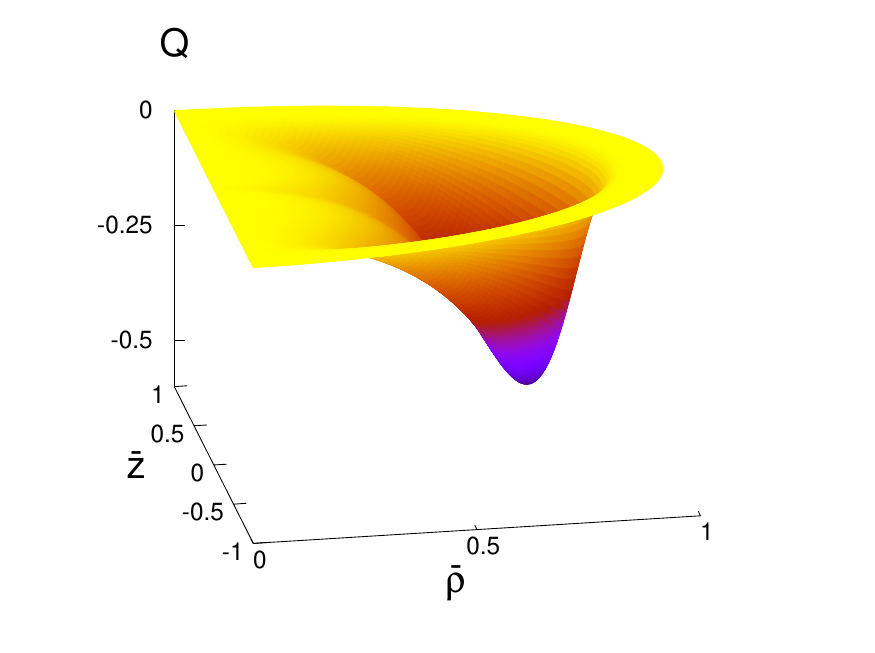}
			\includegraphics[width=8 cm,angle =0 ]{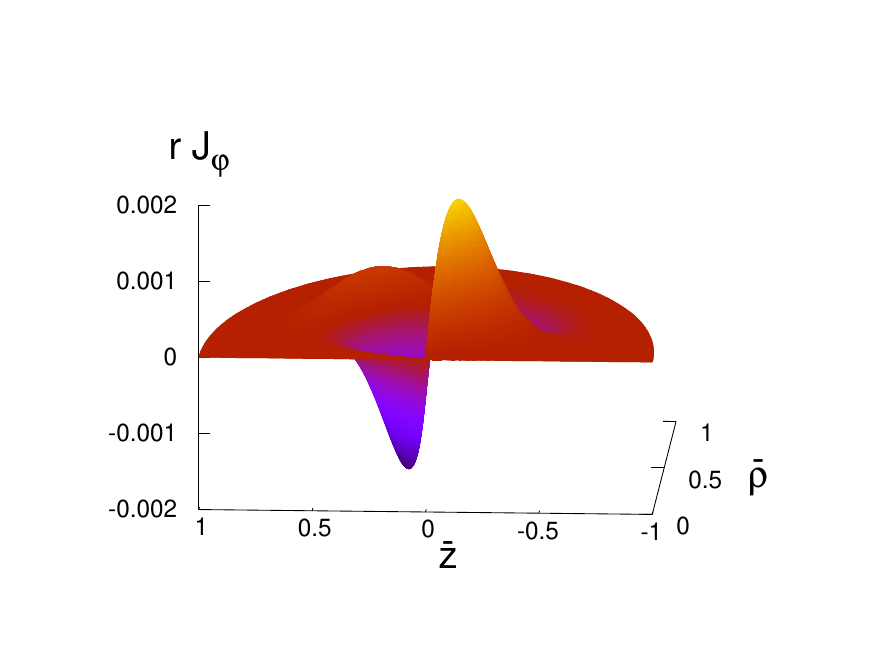}

    \caption{The magnetic charge and electric current densities  for the $\nu=2$ monopole solution.}
    \label{Q}
\end{figure}

Fig.\ref{Q} shows the magnetic charge density and the electric current density for the $\nu=2$ monopole.  
The magnetic charge splits as $P=P_{\rm U(1)}+P_{\rm SU(2)}$ 
according  to \eqref{PT}, 
 where the U(1) part  is pointlike and given by \eqref{PA}, 
while the SU(2) part is 
\be                           \label{rho0}
{P}_{\rm SU(2)} =\int \rho_{\rm SU(2)} \sqrt{-\rm g}\, d^3x\equiv 
2\pi \int_{0}^{\pi} \sin\vartheta\, d\vartheta \int_0^1 Q(x,\vartheta)\, dx\,.
\ee
This part of the charge is smoothly distributed over the space, but its value is 
the same as for the pointlike monopole, $P_{\rm SU(2)}=-g^{\prime 2}\nu/e=-\nu g^\prime/g$, and the numerical 
verification of this  is a good consistency check for our procedure. 
What is interesting is the 
profile of the charge distribution $Q(x,\vartheta)$. For $\nu=\pm 1$ when the monopole is spherically symmetric, 
comparing with \eqref{char} yields
\be
\nu=\pm 1:~~~~~~~Q=\frac{\nu g^\prime}{4\pi  g}\left( f^2 \right)^\prime_x\,,
\ee
hence $Q$ depends only on the radial coordinate.  Therefore, the SU(2) part of the charge density for the CM monopole is uniformly 
distributed over the 2-sphere. However, already for 
$\nu=2$ the charge density is not at all spherical  and shows 
a strong $\vartheta$-dependence with a profound minimum at the equatorial plane some distance away from the origin, 
as seen in Fig.\ref{Q}. This implies that the magnetic charge distribution has a toroidal shape with the maximal value 
along a ring in the equatorial plane at $z=0$.  Solutions with higher $\nu$ show a similar toroidal structure of the 
charge density. 

The electric current density $J^\mu$ vanishes for the CM monopole, but for $|\nu|>1$ 
it has a non-zero azimuthal component  $J_\varphi$. The total current through the 
$\rho-z$ half-plane is zero, $I=I_{+}+I_{-}=0$, but the currents in the $z>0$ and $z<0$ regions, 
\be                \label{I}
I_{+}=\int_0^{\infty} dz  \int_0^\infty   (\vec{J}\cdot \vec{n}_\varphi) \, \rho\,d\rho\,,~~~~~~~~
I_{-}=\int_{-\infty}^{0} dz  \int_0^\infty   (\vec{J}\cdot \vec{n}_\varphi) \, \rho\,d\rho\,,~
\ee
do not vanish. Here $\vec{n}_\varphi$ is the unit vector in the azimuthal direction. 
Therefore, the monopole contains inside two oppositely directed circular electric currents,
which can be viewed as a manifestation of the electroweak superconductivity 
 \cite{Ambjorn:1989bd,Garaud:2009uy}. 
 One has $J_\varphi\sim 1/r$ close to the origin, which does not affect the convergence of the 
integrals in \eqref{I} but complicates the graphical representation of $J_\varphi$. Therefore, we show in the plots 
the bounded product $rJ_\varphi$. 
As seen in Fig.\ref{Q}, $J_\varphi$ is antisymmetric with respect to the reflection in the equatorial plane, 
with a profound minimum in the upper hemisphere and a marked maximum in the lower hemisphere.  
This corresponds to two superconducting 
azimuthal currents  flowing 
 in opposite directions  and giving  rise to  two 
oppositely oriented magnetic moments.

\begin{figure}
    \centering

		\includegraphics[width=13 cm,angle =0 ]{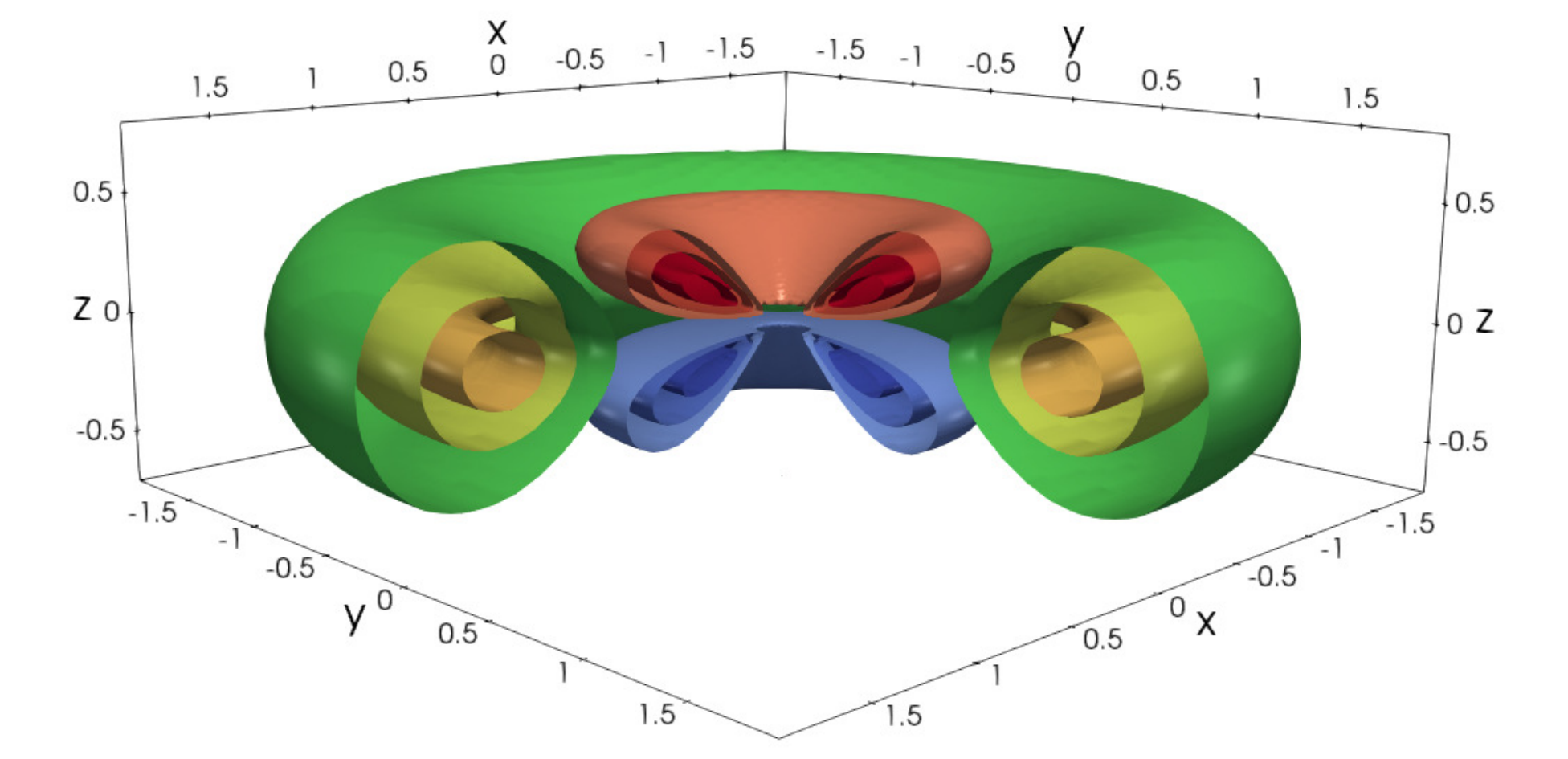}
			\includegraphics[width=17 cm,angle =0 ]{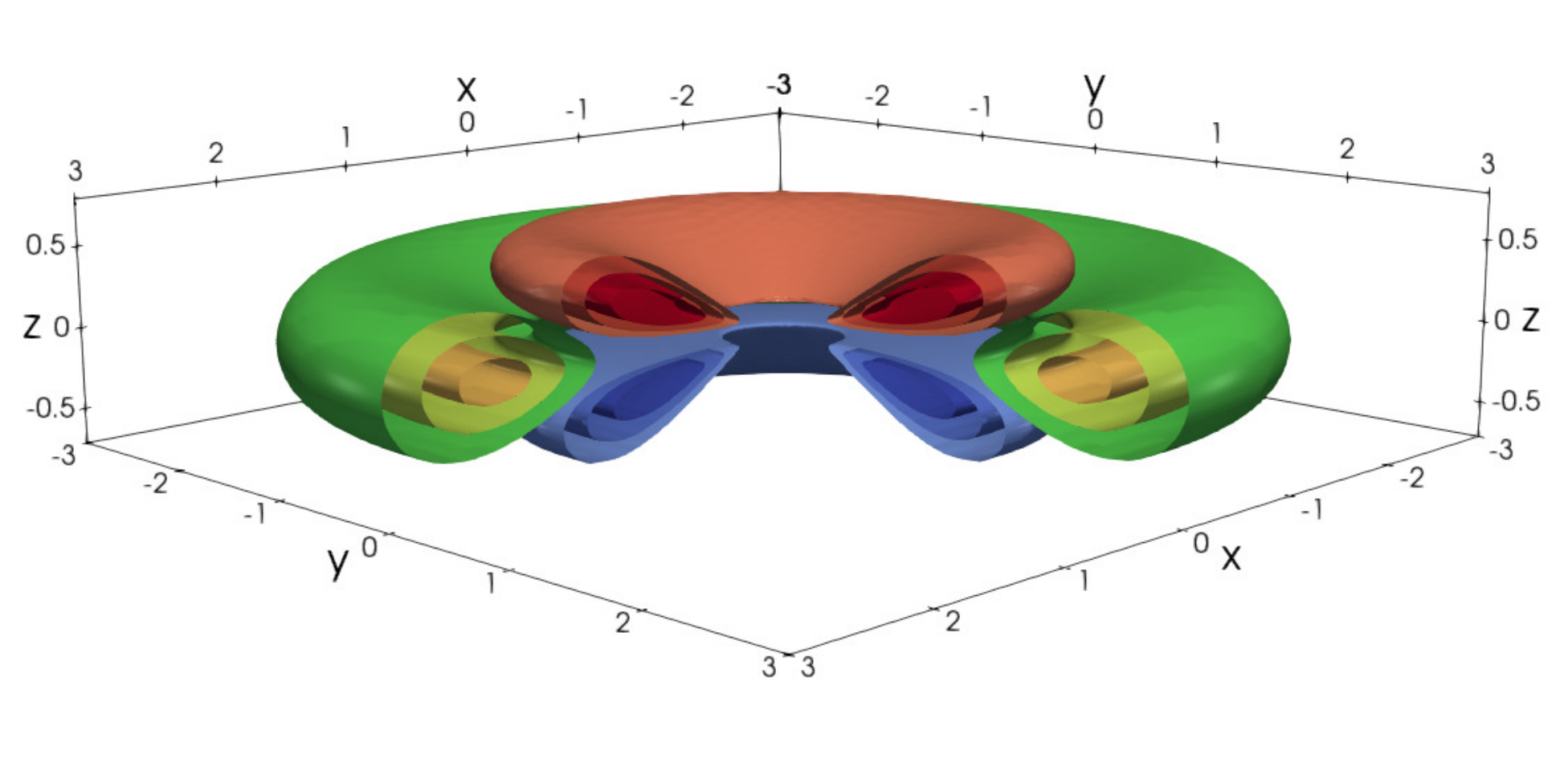}

    \caption{Surfaces of constant  magnetic charge and electric current densities for the $\nu=2$
 (upper panel) and $\nu=4$ (lower panel) monopole solutions in Cartesian coordinates of the 3-space. The ring radii in the latter case are 
 twice as large, but their thickness is the same. }
    \label{QJm}
\end{figure}

Fig.\ref{QJm} shows level surfaces for the SU(2) charge density $Q$ 
defined in \eqref{rho0} and for the current density $r J_\varphi$ 
for the $\nu=2$ and $\nu=4$ monopole solutions. The thick toroidal region containing the equatorial plane (green online)
contains  the non-Abelian magnetic charge. Although solutions with $\nu>1$ can be thought of as superpositions 
of $\nu$ Cho-Maison monopoles, these monopoles cannot be distinguished from each other 
and merge together into 
a toroidal condensate. At the same time, the Higgs field vanishes only at the origin. 
The other two tori shown in Fig.\ref{QJm} above and below the equatorial plane (red and blue online) 
correspond to two oppositely directed distributions of the  azimuthal electric current -- superconducting rings. 
As is seen in Fig.\ref{QJm}, the whole picture 
is qualitatively the same for $\nu=2$ and for $\nu=4$, and the same picture is found for other (even or odd) values of $\nu$.

All of this suggests the following qualitative description of the inner structure of the multi-monopole solutions. 
The SU(2) part of their magnetic charge is distributed over  the volume of 
a magnetically charged ring (the U(1) part of the charge is always 
located at the origin). 
The magnetic ring creates a magnetic field which is mostly anti-parallel to 
the $z$-axis 
for $z>0$ (assuming that $\nu>0$, the charge of the ring then being negative) 
and mostly parallel to the axis  in the $z<0$ region.  
This magnetic field forces the electrically charged W-bosons constituting   the condensate inside the monopole 
to Larmore orbit  in one direction for $z>0$ and in the opposite direction for  $z<0$. This
produces two circular superconducting electric currents.
These currents produce  two oppositely oriented magnetic dipole moments 
repelling each other  but  
attracted to the magnetic ring. 
Each dipole creates  
a magnetic field directed oppositely to that of the magnetic ring (Lenz's law), hence pushing 
 the individual CM monopoles (or rather their SU(2) charges) 
contained in the ring toward the 
equatorial plane.  This field overcomes the  mutual repulsion of the individual monopoles and squeezes 
them into a toroidal condensate. 

Of course, this electromagnetic analogy cannot be totally adequate  since the electromagnetic description 
applies only in the 
Higgs vacuum, whereas  the Higgs field is not in vacuum   inside the monopole.
However, the analogy is  suggestive. 

\subsection{Quadrupole moment}

The electromagnetic analogy shows that the total magnetic dipole moment of the monopole  is zero. 
Indeed, its dipole moments generated 
by the currents $I_\pm$ have opposite signs and compensate each other, while 
the magnetic charge density is everywhere sign definite. 
However, the magnetic 
quadrupole moment does not vanish. The latter is described by the traceless tensor $q_{ik}$ receiving contribution from the 
magnetic charge and electric current \cite{raab2005multipole},
\be                     \label{qik}
q_{ik}=\int \left[ 3x_i x_k-r^2\delta_{ik}\right]\,\rho_{\rm SU(2)}\, d^3 x
+\int \left[ x_i\,(\vec{r}\wedge \vec{J})_k+x_k\, (\vec{r}\wedge \vec{J})_i \right] d^3 x\,,
\ee
where $x_k=(x,y,z)$ are Cartesian coordinates.
Owing to the axial symmetry, the tensor has the structure  $q_{ik}={\rm diag}[-q/2,-q/2,q]$, where  the only independent 
component,
\be                         \label{qq}
q=q_{zz}=\int \left[ 3z^2-r^2\right]\,\rho_{\rm SU(2)}\, d^3 x + \int 2zJ_\varphi \,d^3x,
\ee
determines the deviation from the spherical symmetry. The first integral here gives the dominant contribution 
and for the {\it oblate} systems shown in Fig.\ref{QJm} one has $q>0$ since $\rho_{\rm SU(2)}$ is negative when $\nu$
is positive. We can get  the value of $q$ from our solutions as follows. 
The quadrupole moment \eqref{qik} 
determines the asymptotic form of the non-spherically symmetric part of the magnetic field  \cite{raab2005multipole}, 
\be
\delta {\cal B}_i=\frac{1}{2r^7}\left[
5\,x_i x_j x_k- r^2\left(x_i\delta_{jk}+x_j\delta_{ik}+x_k\delta_{ij}\right)\right]q_{jk}\,,
\ee
(the spherically symmetric part of the magnetic fields is the Dirac monopole \eqref{Dir}). 
In the axially symmetric case, passing to spherical coordinates, this reduces to 
\be              \label{dB}
\delta {\cal B}=\frac{3q}{4r^4}\left[ (3\cos^2\vartheta-1)\,dr+r\sin(2\vartheta)\,d\vartheta \right]. 
\ee
On the other hand, as shown by \eqref{quad} in Appendix A, the asymptotic form of the 
electromagnetic vector potential is 
\be
\delta {\cal A}=\frac{\nu}{gg^\prime}\,y_\gamma\,\sin\vartheta\, d\varphi=\frac{\nu}{gg^\prime}\,\frac{C_\gamma}{r^2}\sin^2\vartheta\cos\vartheta\, d\varphi\,,
\ee
where the value of the coefficient $C_\gamma$ is determined by  the numerics. Computing then the magnetic field 
$\delta\vec{\cal B}=\vec{\nabla}\wedge \delta\vec{\cal A}$ yields exactly the same expression as in 
\eqref{dB}, with 
\be        \label{q}
q=\frac{4\,\nu}{3\,gg^\prime}\, C_\gamma\,.
\ee
We can therefore read-off the quadrupole moment from the asymptotic form of our solutions,
and its values for the lowest $\nu$ are shown in Table I. One can see that $q$ increases with $\nu$,
which corresponds to the fact that the oblateness of the solutions increases with growing magnetic charge. 
On the other hand, $q$ becomes negative for $\nu<1$, and we checked that solutions become  {\it prolate}  in this case,
with magnetic density levels surfaces stretched along the $z$-axis.

\subsection{The limit of  large magnetic charge}

 Increasing the winding number $\nu$, we could obtain solutions up to $\nu=100$ while 
keeping small the virial $v$ in  \eqref{v}. 
Both the energy $E_{\rm reg}$ 
and quadrupole moment  $q$ always 
increase with $\nu$. 
One can use the following arguments to obtain analytical estimates. 

\begin{figure}
    \centering

		\includegraphics[width=8 cm,angle =0 ]{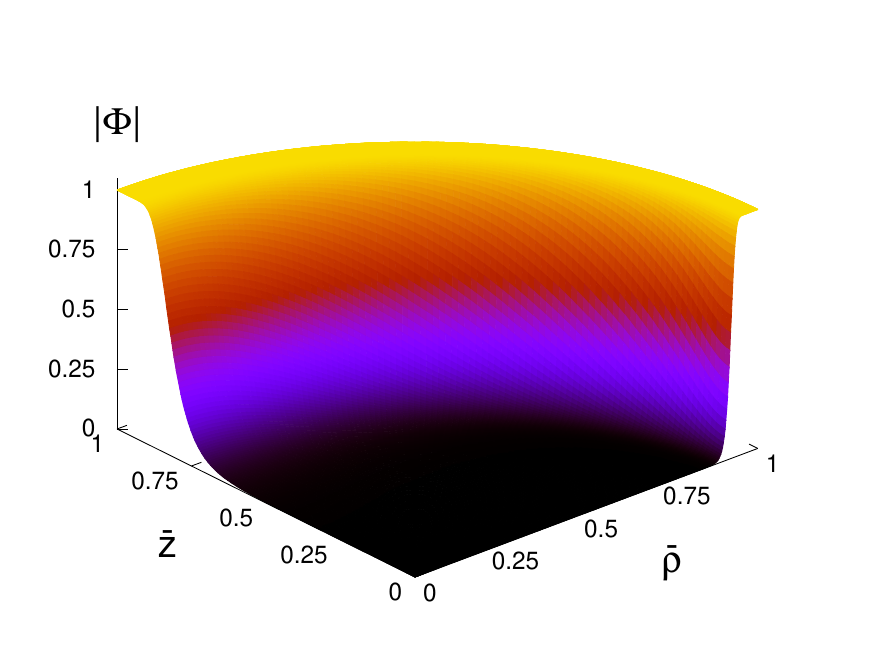}
				\includegraphics[width=8 cm,angle =0 ]{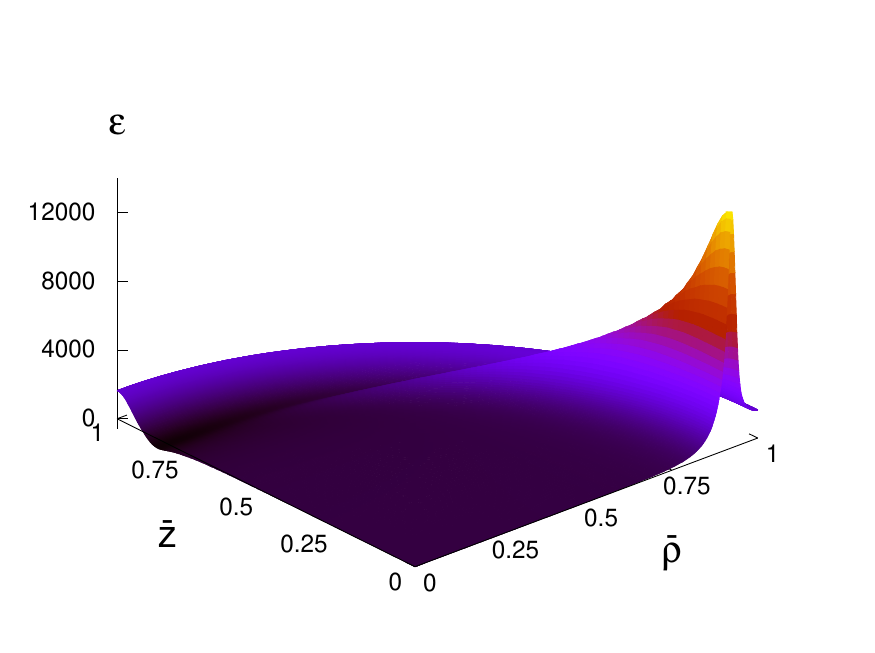}

    \caption{ The norm of the Higgs field $|\Phi|$ and the energy density $\varepsilon$ for the monopole solution with $\nu=50$.}
    \label{PHI}
\end{figure}

It is known that when the magnetic field becomes very strong, then the Higgs field 
approaches zero and the full electroweak gauge symmetry is restored \cite{Ambjorn:1988tm,Ambjorn:1989sz}. 
This can be seen in the inner  structure of the classical solutions 
\cite{Ambjorn:1989bd,Garaud:2009uy}. 
In our case, 
when the magnetic charge $P$ increases the magnetic field gets stronger, hence 
 the Higgs field in the central region of the monopole  is expected to approach zero. 
This expectation is confirmed already by the perturbative analysis    since close to the origin  one has (see Appendix B) 
\be
\phi_1\sim r^\lambda\left[\left(\sin\frac{\vartheta}{2}\right)^{\nu+1}+\left(\cos\frac{\vartheta}{2}\right)^{\nu+1}\right]\,,~~~\phi_2\sim\partial_\vartheta\phi_1\,,
\ee
with $\lambda=(\sqrt{1+2\nu}-1)/2$, hence the Higgs gets smaller when $\nu$ increases.

The numerical analysis confirms 
the expectation at the non-perturbative level  and shows 
that for large $\nu$ the monopoles develop in the central region  
a spheroidal bubble  where the norm of the Higgs field $|\Phi|=\sqrt{\phi_1^2+\phi_2^2}$ 
is very close to zero,  hence the system is in the false vacuum. This can be seen in Fig.\ref{PHI} for $\nu=50$. 
The SU(2) gauge field also vanishes in the bubble, 
since $H_1,H_3$ are very close to zero while $H_2,H_4$ are very close to unity, in which case one has $W^a_{\mu}=0$, 
as seen in \eqref{gauge2}. The $y$ amplitude is very close to zero too. 
As a result, inside the bubble there remains  only the U(1) hypercharge field,
\be             \label{in}
\underline{\text{inside:}}~~~~
B_\mu dx^\mu=\nu\,(\cos\vartheta\pm 1)d\varphi\,,
~~~~~ W^a_{\mu}=0,~~~~~~
\Phi=0.
\ee
In view of \eqref{Nambu}, this describes  the electromagnetic field 
$F_{\mu\nu}=({g}/{g^\prime})\, B_{\mu\nu}$ of 
the pointlike magnetic charge $P_{\rm U(1)}=-\nu g/g^\prime $ and 
the Z-field $Z_{\mu\nu}=B_{\mu\nu}$. Since the gauge symmetry is restored, the Z-field is massless. 

\begin{figure}
    \centering

						\includegraphics[width=17 cm,angle =0 ]{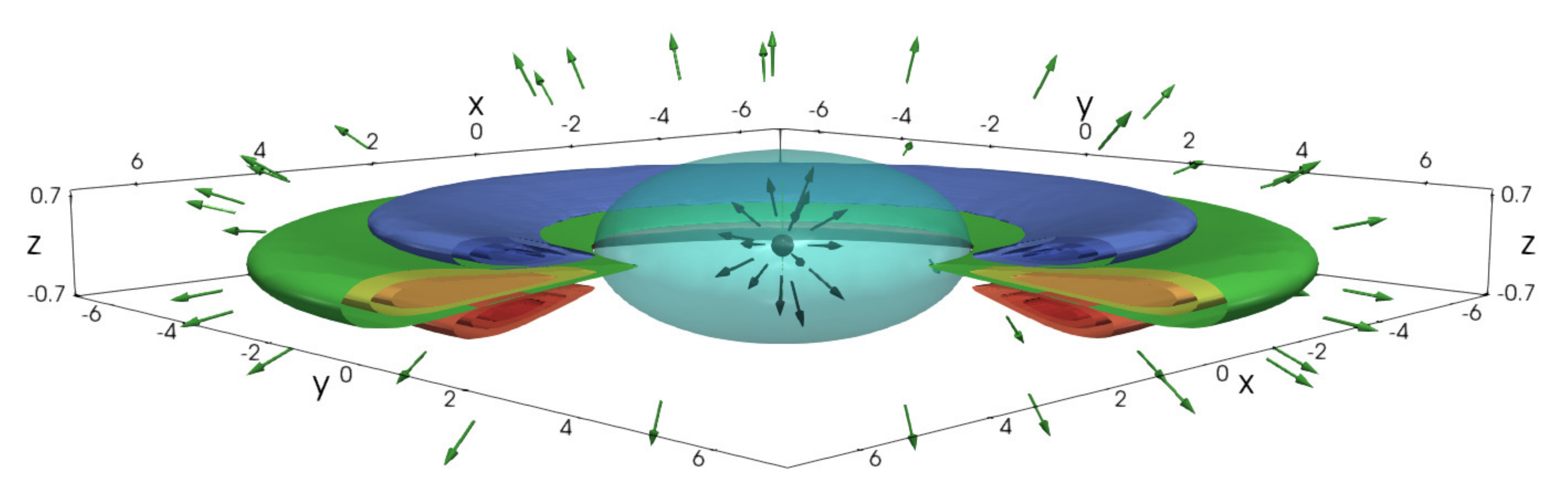}

    \caption{Profiles of the magnetic monopole solution for $\nu=-20$. The central  region is occupied by a spheroidal 
    bubble  (cyan online) containing the U(1) hypercharge field generated by a pointlike magnetic charge $P_{\rm U(1)}=-\nu g/g^\prime$ in the center. 
 This field  is strong enough to suppress  all other  fields. Outside the bubble, the non-linear  fields emerge from vacuum and 
 produce a condensate 
 forming   a ring of 
 non-Abelian magnetic charge $P_{\rm SU(2)}=-\nu g^\prime/g$  squeezed   between two superconducting rings of 
 opposite electric currents. Still farther away, the non-linear fields die away and there remains only the 
 magnetic field of the Dirac monopole of total  charge $P_{\rm U(1)}+P_{\rm SU(2)}=-\nu/e.$}
    \label{QJm1}
\end{figure}

Outside the bubble, the Higgs field approaches the vacuum value $|\Phi|=1$ generating non-zero masses for the fields, 
and being massive, the latter tend to zero at large distances exponentially fast. The monopole configuration then 
approaches that in   \eqref{Bpm},
\be             \label{Bpm1}
\underline{\text{outside:}}~~~~
B_\mu dx^\mu=\nu\,(\cos\vartheta\pm 1)\,d\varphi,~~~~\T_a W^a_{\mu}=\T_3\,B_\mu dx^{\mu},~~~~~
\Phi=
\begin{pmatrix}
0 \\
1
\end{pmatrix}. ~~~~~
\ee
This corresponds to the Dirac monopole of charge 
$P_{\rm U(1)}+P_{\rm SU(2)}=-\nu/e$. 

The Higgs field interpolates between $|\Phi|=0$ and $|\Phi|=1$ in the ``bubble crust" -- a 
transition region between the inside and outside. This region  contains a W-condensate 
in the form of rings close to the equatorial plane, as shown in Fig.\ref{QJm1} for $\nu=-20$. 
The condensate generates a magnetic charge and electric currents. 
Comparing with the similar picture in Fig.\ref{QJm}, one can see that the rings become large and strongly squashed for large $|\nu|$,
while their thickness in the $z$ direction visibly does not change. 
The total non-Abelian magnetic charge contained in the crust is 
$P_{\rm SU(2)}=-\nu g^\prime/g $.

\begin{figure}
    \centering

		\includegraphics[width=5 cm,angle =-90 ]{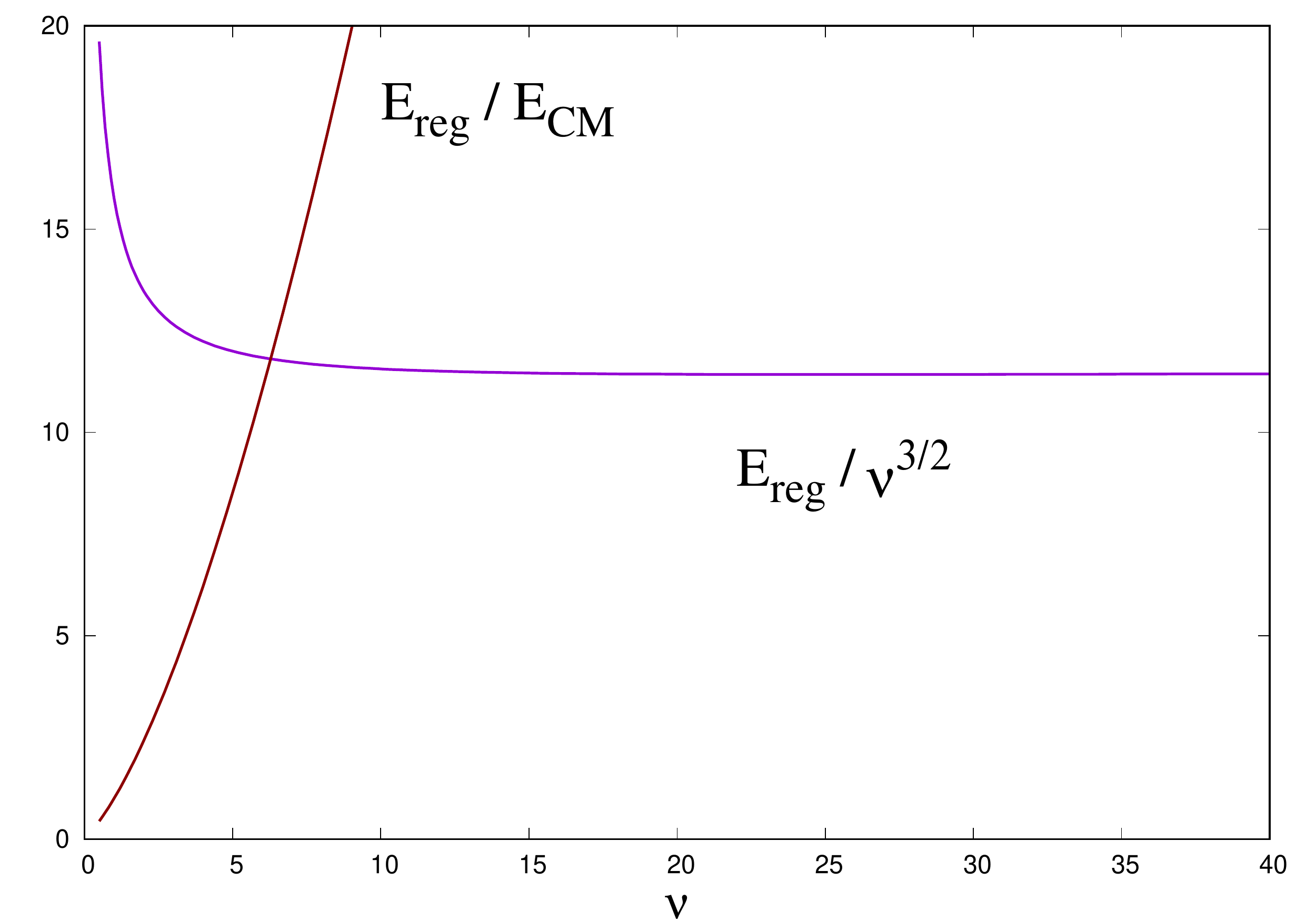}
			\includegraphics[width=5 cm,angle =-90 ]{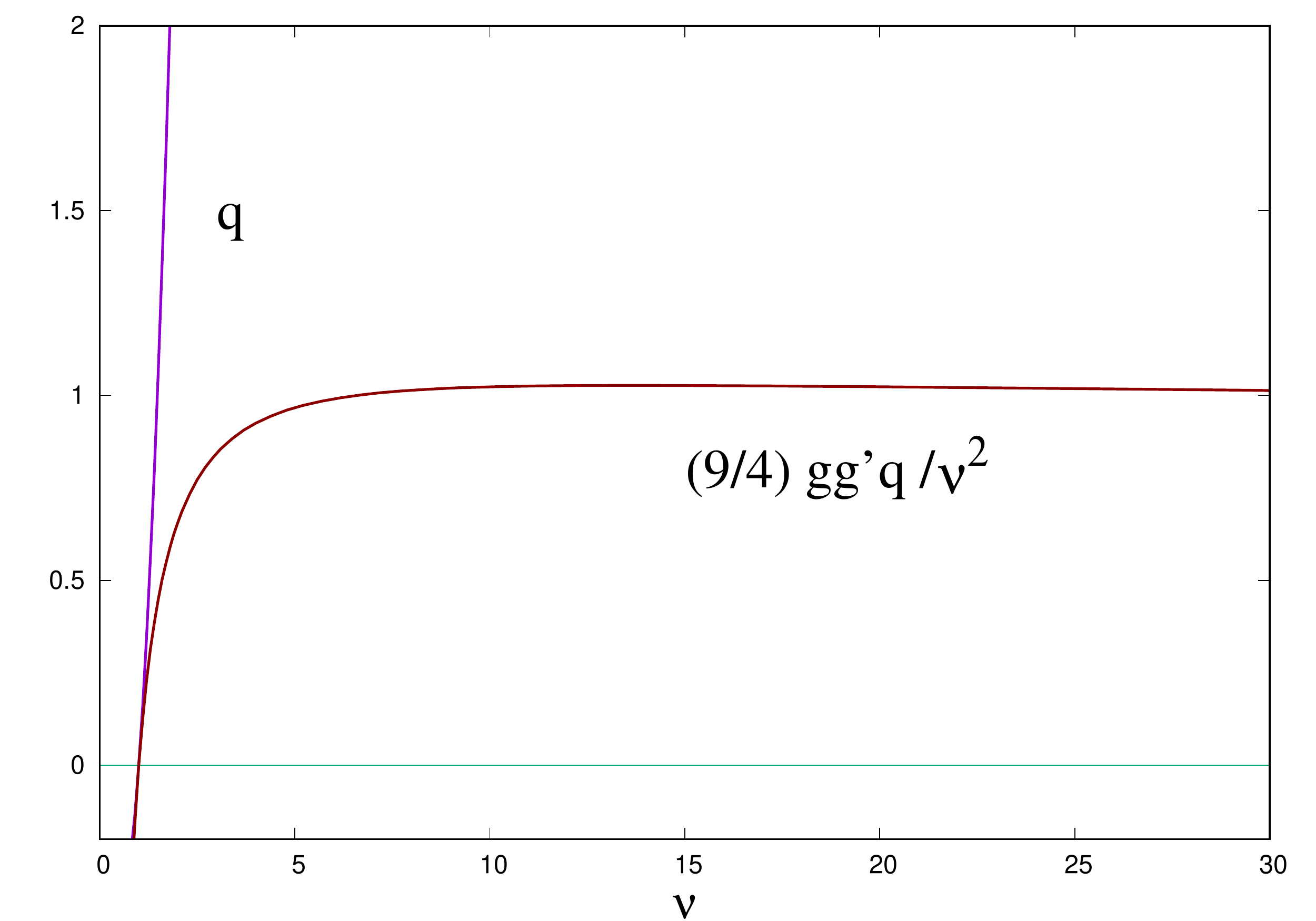}

    \caption{Left:  energy $E_{\rm reg}(\nu)$ in units of the CM monopole energy $E_{\rm CM}=E_{\rm reg}(1)$ 
    and $E_{\rm reg}(\nu)$  divided by $\nu^{3/2}$. Right:
  the quadrupole moment $q$ and also $q$ divided by $4\nu^2/(9gg^\prime)$ against the winding number $\nu$.  }
    \label{ENnu}
\end{figure}

Although the bubble is not exactly spherical (this is seen already in Fig.\ref{PHI}),  reasonable  estimates can be obtained via 
approximating the fields by the spherically symmetric expressions \eqref{ssm} with the profile functions $f(r),\phi(r)$, 
\be
f(r)=1~~\text{if}~~r<R,~~~~f(r)=0~~\text{if}~~r>R,~~~~~~\phi(r)=1-f(r).
\ee
Injecting this to Eq.\eqref{Energy} where  $\nu$ is kept arbitrary, 
yields the energy 
\be
E_{\rm reg}=E_{\rm SU(2)}=\frac{\beta}{8}\,
\frac{4\pi R^3}{3}\, 
+4\pi\,\frac{\nu^2}{2 g^2 R}\,.
\ee
Here the first term is the contribution of the constant Higgs energy density inside the bubble, and the second one 
is the non-Abelian magnetic energy  outside the bubble. 
Minimizing  with respect to $R$, yields  the following estimates for the bubble size and energy, 
\be           \label{RRR}
R=\left( \frac{4}{\beta g^2}\right)^{1/4}\sqrt{\nu}=1.29\,\sqrt{|\nu|}\,,~~~~~
~~~E_{\rm reg}=\frac{8\pi}{3}\left(\frac{\beta}{4g^2}\right)^{1/4} \nu^{3/2}=7.4\,|\nu|^{3/2}.
\ee
We can identify the bubble size  and hence the position of the bubble crust 
with the position of  the minimum of the function $Q$ shown in Fig.\ref{Q}. 
The numerically obtained values of the  bubble size are in a good agreement  with $R$ in  \eqref{RRR}. 
Moreover, as seen in Fig.\ref{ENnu}, the numerically obtained ratio 
$E_{\rm reg}/\nu^{3/2}$ indeed approaches for large $\nu$ a constant value. This value, $11.4$,  
is larger than $7.4$ suggested by formula \eqref{RRR}, but this is because the above analytical estimates 
take into account only the energy inside and outside the bubble without considering the energy in the crust. 
More accurate estimates can be obtained by introducing a finite transition region where $f(r)$ and $\phi(r)$ interpolate 
between the inside and outside values.  

Our numerics suggest that for large $\nu$ the constant in the asymptotic formula \eqref{quad} 
approaches the value $C_\gamma=\nu/3$,
hence the quadrupole moment defined by \eqref{q} is 
\be
q=\frac{4}{9\,gg^\prime}\,\nu^2\,,
\ee
which is clearly seen in Fig.\ref{ENnu}. This can be represented as 
\be
q=-\frac{4\,\nu}{9g^{\prime 2}}\,P_{\rm SU(2)}=-1.16\times  P_{\rm SU(2)}R^2\,,
\ee
with $R$ given by \eqref{RRR}. 
Therefore 
\be
q \approx -P_{\rm SU(2)}R^2\,,
\ee
which is  the quadrupole moment of a homogeneously charged  torus of radius $R$ and charge $P_{\rm SU(2)}$. 
This shows again that the above estimate for the bubble  size $R$ is sensible, because 
the quadrupole moment  in the formula \eqref{qq}  is dominated  by the magnetic charge density,
while the relative contribution of the electric current is negligible for large $|\nu|$. Specifically, the 
currents $I_\pm$ defined by \eqref{I} approach finite values
$I_\pm= \mp 0.095$  for large $\nu$. 
Since the radius $R$ of the superconducting rings is proportional to $\sqrt{\nu}$, 
the dipole moment produced by each rings scales as $\pi R^2 I_{\pm}\propto \nu$. 
The dipole moments produced by $I_{+}$ and $I_{-}$ are separated in space and 
their fields do not exactly compensate each other 
but produce a quadrupole moment, but since their separation is almost independent on $\nu$, their
 quadrupole moment  grows slower than $\nu^2$ and is sub-dominant 
as compared to that produced by the magnetic ring.

Since the  hypercharge field \eqref{in} in the monopole center is spherically symmetric, one can wonder 
why the rest of the configuration should be squashed ? Remember, however,  
that the only spherically symmetric solution for a large winding number $\nu$  is the Abelian Dirac monopole. 
All other solutions  with the same far field asymptotic are non-Abelian and non-spherically symmetric.
If they are axially symmetric, then, as shown by Eq.\eqref{A12} in Appendix A, the angular dependence 
of the W-modes in the far field zone is given in terms of the Legendre polynomials 
$P^\nu_j(\cos\vartheta)$ and $P^{\nu\pm 1}_j(\cos\vartheta)$. Since $\nu\pm 1\approx \nu$ for large $\nu$ and since 
 the leading contribution corresponds to the minimal value of $j=|\nu|$, 
the angular dependence of the W-modes is given by 
\be               \label{sq}
P^\nu_{|\nu|}(\cos\vartheta)\propto \left(\sin\vartheta\right)^{|\nu|}.
\ee 
These modes  are strongly localized around $\vartheta=\pi/2$, which agrees with the rings 
in the equatorial region shown in Fig.\ref{QJm1}. On the other hand, the angular dependence of the Z, Higgs, 
and electromagnetic modes is different. It follows that the electric currents in the two superconducting rings and the 
SU(2) magnetic charge in the central ring must be supported mainly by a condensate of W-bosons. 

It is also worth reminding that the Dirac monopole is unstable with respect to perturbations with angular momentum $j=|\nu|-1$
and the instability resides in the W-sector \cite{Gervalle:2022npx}.  
The Dirac monopole can be viewed as a superposition of two pointlike charges, $P_{\rm U(1)}$ and 
$P_{\rm SU(2)}$. It seems  plausible that the instability growth affects the SU(2)
field configuration by radiating  away all its central part, 
and what remains condenses to the rings 
squashed   according to \eqref{sq}. The total magnetic charge does not change but  its 
SU(2) part no longer remains in the center and gets distributed over the volume of the ring. 
Of course, there remains to demonstrate  that non-Abelian monopoles 
for $|\nu|>1$ are indeed stable, in which case they may be viewed as remnants of collapse of  the Dirac monopoles, 
but at least for $|\nu|=1$ the proof is available \cite{Gervalle:2022npx}. 

Although we cannot claim that monopoles with $|\nu|>1$ are stable, we believe this is indeed the case.
The stability of the $\nu=\pm 1$ CM monopoles was established via an involved partial wave analysis that 
applies only in the spherically symmetric case \cite{Gervalle:2022npx}, but  it seems that a different strategy could be 
used for multi-monopoles. Indeed, it suffices to show that the regularized 
energy functional $E_{\rm reg}[\Psi]$, or more precisely its full 3D version, admits a non-trivial minimum in the 
sector with a fixed SU(2) charge $P_{\rm SU(2)}$. 
This can probably be done via a numerical minimization 
of the energy functional in a 3D domain. However, such an analysis  requires separate studies.

\section{SPHALERONS AND THEIR INTERNAL STRUCTURE}
\setcounter{equation}{0}

Electroweak sphalerons at finite mixing angle have been much studied. These are the fundamental $\nu=1$ sphaleron 
\cite{Kleihaus:1991ks,Kunz:1992uh}, the multi-sphalerons with $|\nu|>1$ \cite{Kleihaus:1994yj,Kleihaus:1994tr},
the sphaleron-antisphaleron pairs 
\cite{Klinkhamer:1993hb,Kleihaus:2008gn}, and also spinning sphalerons \cite{Radu:2008ta,Kleihaus:2008cv,Ibadov:2010ei}. 
We have reproduced the multi-sphalerons with $\nu=1,2,\ldots$,  mainly to make sure that our 
procedure is correct, but also to compare their inner structure  with that of monopoles.

To obtain the sphaleron solutions, we use the same parameterization \eqref{var}  of the field amplitudes as for the monopoles, 
but with   $ \Theta(\vartheta)=1$ instead of $\Theta(\vartheta)=\cos\vartheta$. 
The boundary conditions at $\vartheta=\pi/2$ and at $\vartheta=0$ are provided, respectively, 
by \eqref{eq} and \eqref{axis}, while those  at $r=0,\infty$ should be the same as for 
the spherically symmetric sphaleron \eqref{sp0}:
\be               \label{sph}
\text{axis~~}\underline{\vartheta=0}:&&~~~~~H_1=H_3=y=\phi_1=0,~~~~~~
\partial_\vartheta H_2=\partial_\vartheta H_4=\partial_\vartheta \phi_2=0;\nn \\
\text{equator~~}\underline{\vartheta=\pi/2}:&&~~~~~H_1=H_4=\phi_1=0,~~~~~~
\partial_\vartheta H_2=\partial_\vartheta H_3=\partial_\vartheta y=\partial_\vartheta \phi_2=0; \nn \\
\text{origin~~}\underline{r=0}:&&~~~~~H_1=y=\phi_1=\phi_2=0,~~~~~~
H_2=-2,~~H_3=-2\sin\vartheta,~~~~H_4=-2\cos\vartheta; \nn \\
\text{infinity~~}\underline{r=\infty}:&&~~~~~H_1=H_2=H_3=H_4=y=\phi_1=0,~~~~~\phi_2=1. 
\ee
One can directly work with these boundary conditions, but
they  are singular at the origin where  $H_3,H_4$ remain $\vartheta$-dependent,
whereas $r=0$ is a single point in space where  nothing should depend on  $\vartheta$.
Alternatively, one can perform 
the gauge transformation \eqref{res} with the parameter $\chi=2\vartheta$. This 
does not affect the gauge condition \eqref{fix}, while the spherically symmetric sphaleron configuration \eqref{sp0} transforms to 
\be                 \label{sp2}
H_2=f(r)+1,~~H_3=-H_2\,\sin\vartheta,~~H_4=H_2\,\cos\vartheta,~~\phi_1=\phi(r)\sin\vartheta,~~\phi_2=\phi(r)\cos\vartheta,~~~~~
\ee
and $H_1=y=0$. Since $f(0)=-1$ and $\phi(0)=0$, all field amplitudes  now vanish at $r=0$. 
The boundary conditions for axially symmetric fields then become 
\be               \label{sph3}
\text{axis~~}\underline{\vartheta=0}:&&~~~~~H_1=H_3=y=\phi_1=0,~~~~~~
\partial_\vartheta H_2=\partial_\vartheta H_4=\partial_\vartheta \phi_2=0;\nn \\
\text{equator~~}\underline{\vartheta=\pi/2}:&&~~~~~H_1=H_4=\phi_2=0,~~~~~~
\partial_\vartheta H_2=\partial_\vartheta H_3=\partial_\vartheta y=\partial_\vartheta \phi_1=0; \nn \\
\text{origin~~}\underline{r=0}:&&~~~~~H_1=H_2=H_3=H_4=y=\phi_1=\phi_2=0; \nn \\
\text{infinity~~}\underline{r=\infty}:&&~~~~~H_1=y=0,~~
H_2=2,~~H_3=-2\,\sin\vartheta,~~H_4=2\,\cos\vartheta,~~\nn \\
&&~~~~~\phi_1=\sin\vartheta,~~\phi_2=\cos\vartheta.
\ee
This corresponds to the gauge originally used in \cite{Kleihaus:1991ks,Kunz:1992uh}. 
The $\vartheta$-dependence  is now moved to large values of $r$ where 
it causes no problems. Notice that $\phi_2$ becomes odd under the reflection $\vartheta\to\pi-\vartheta$ 
while $\phi_1$ is even. 

Using either \eqref{sph} or \eqref{sph3} with, respectively,  either \eqref{sp0} or \eqref{sp2} as the input 
configuration, our numerical scheme converges 
giving sphaleron solutions for any  $\nu$ and $\thetaw$. 
We obtain the same results as those previously reported \cite{Kleihaus:1991ks,Kunz:1992uh,Kleihaus:1994tr}, 
hence we do not show them and concentrate on the analysis of the inner sphaleron structure. 
The latter can be studied as for the monopoles  via analysing the electric and magnetic charge 
densities \eqref{cur}. In the sphaleron case there is an additional way of doing this since, unlike the 
monopoles, the fundamental sphaleron with $\nu=1$ and $\sin^2\thetaw=0.23$
is only slightly non-spherical, in which case the perturbative approach is possible. 
Specifically, the amplitudes 
$H_2,H_3,H_4,\phi_1,\phi_2$ are well described by the spherically symmetric formula 
\eqref{sp2} with $f(r),\phi(r)$ shown in Fig.\ref{FIGsph}, and the 
most notable effect of the deviation from spherical symmetry is the appearance of a non-trivial U(1) field $y$ 
which can be evaluated perturbatively  \cite{Hindmarsh:1993aw}.

 Since the current in the right hand side of the U(1) equation in \eqref{P2} is proportional to 
$g^{\prime 2}$, the U(1) amplitude $y$  is also proportional to $g^{\prime 2}$ in the lowest order, 
hence one can set 
\be                      \label{Cy}
y(r,\vartheta) =\frac{g^{\prime 2}\,r^2}{2 g^2}\,p(r)\sin\vartheta~~~~~~~\text{where}
~~~~~~~~~~~\left(r^4 p^\prime\right)^\prime =r^2(1-f)\phi^2. 
\ee
Here the differential equation for $p(r)$ is obtained by injecting $y(r,\vartheta)$ to the field equations and keeping 
only the  terms of order  $g^{\prime 2}$, whereas the $f,\phi$ amplitudes in this perturbative order are still described 
by Eqs.\eqref{eq-sphal}. 
The solution is such that for $0\leftarrow r \to \infty$ one has 
\be            \label{C}
const.\leftarrow p(r)\to \frac{C}{r^3} ~~~~~\text{with}~~~~~~C=\frac13\int_0^\infty r^2(f-1)\phi^2\,dr\,,
\ee
where the equation in \eqref{Cy} was used to evaluate $C$. 
The electromagnetic field  \eqref{Nambu} has the following non-zero components in the lowest in 
$g^\prime$ order, 
\be
F_{r\varphi}=\frac{g^\prime}{2g}\left(2f^\prime +(r^2 p)^\prime \right) \sin^2\vartheta\,,~~~~
F_{\vartheta\varphi} =\frac{g^\prime}{g}(f^2-1+r^2 p)\sin\vartheta\cos\vartheta\,.
\ee
Injecting this to \eqref{cur} determines the magnetic charge and electric current densities,
\be                  \label{rJ}
\rho_{\rm SU(2)}=\frac{1}{4\pi}\frac{2 g^\prime}{g r^2}(f-1) f^\prime\cos\vartheta,~~~~~~~
J_\varphi=-\frac{1}{4\pi}\frac{g^\prime}{g r^2} (f^2-1)(f-1)\sin^2\vartheta\,.
\ee
Notice that $p(r)$ drops out from these expressions. 
Since $f=1+{\cal O}\left( e^{-\mw r}\right)$ as $r\to\infty$, it follows that at large $r$ one has 
\be
F=d{\cal A} ~~~~\text{with}~~~~{\cal A}=\frac{g^\prime C}{2 r g}\sin^2\vartheta\, d\varphi=\frac{\vec{\mu}\wedge \vec{r}}{r^3}\,,
\ee
where the sphaleron magnetic moment is 
\be
\vec{\mu}=\vec{n}_z\,\frac{g^\prime C}{2 g}=\vec{n}_z\,\frac{g^\prime }{6g} \int_0^\infty r^2(f-1)\phi^2 dr
=\int\left( \rho_{\rm SU(2)}\, \vec{r}\, +\frac12\, \vec{r}\wedge\vec{J}\right) d^3 x\,,
\ee
with  $\vec{n}_z$ being the unit vector along the $z$-axis. Here the first integral comes from 
\eqref{C},
the second integral is the standard 
expression for the magnetic moment, and their equality 
can be checked by using \eqref{rJ} and the background equations 
\eqref{eq-sphal} \cite{Hindmarsh:1993aw}. 

Therefore, the sphaleron magnetic moment receives a contribution from the azimuthal 
electric current and also from the magnetic charge distribution. The current attains its 
maximal value in the equatorial plane whereas the magnetic charge density changes sign through 
the plane. The total magnetic charge in the $z>0$ region is 
\be
P=2\pi \int_0^{\pi/2}\sin\vartheta\, d\vartheta \int_0^\infty \rho_{\rm SU(2)} \,r^2\,dr=-\frac{g^\prime}{g}\,,
\ee
and that in the $z<0$ region is $+g^\prime/g$. As a result, the perturbative analysis indicates that 
the sphaleron contains a pair of oppositely charged magnetic monopoles with charges $\pm g^\prime/g$, encircled 
by an electric current \cite{Hindmarsh:1993aw}. 

\begin{figure}
    \centering

		\includegraphics[width=8 cm,angle =0 ]{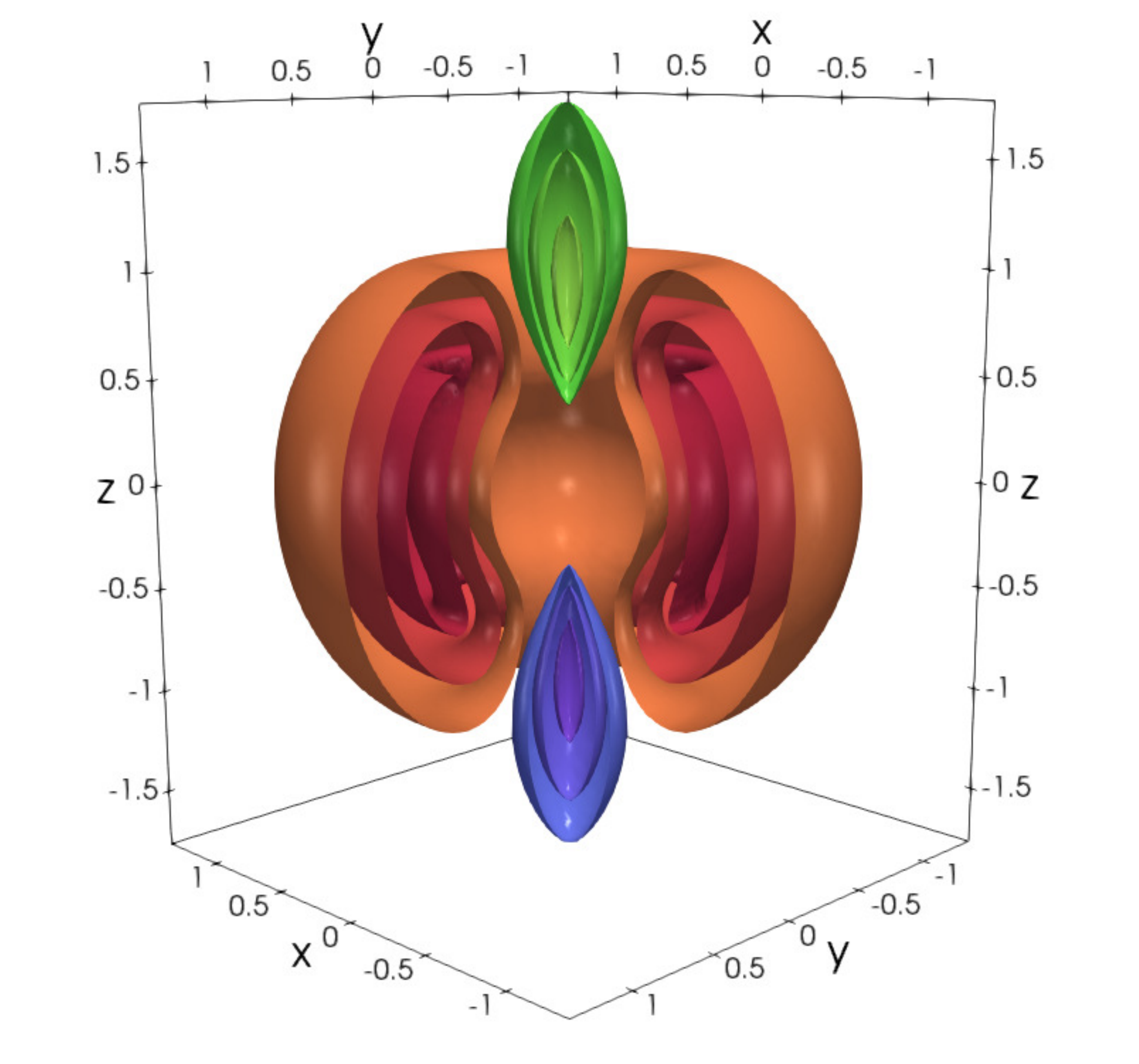}
			\includegraphics[width=8 cm,angle =0 ]{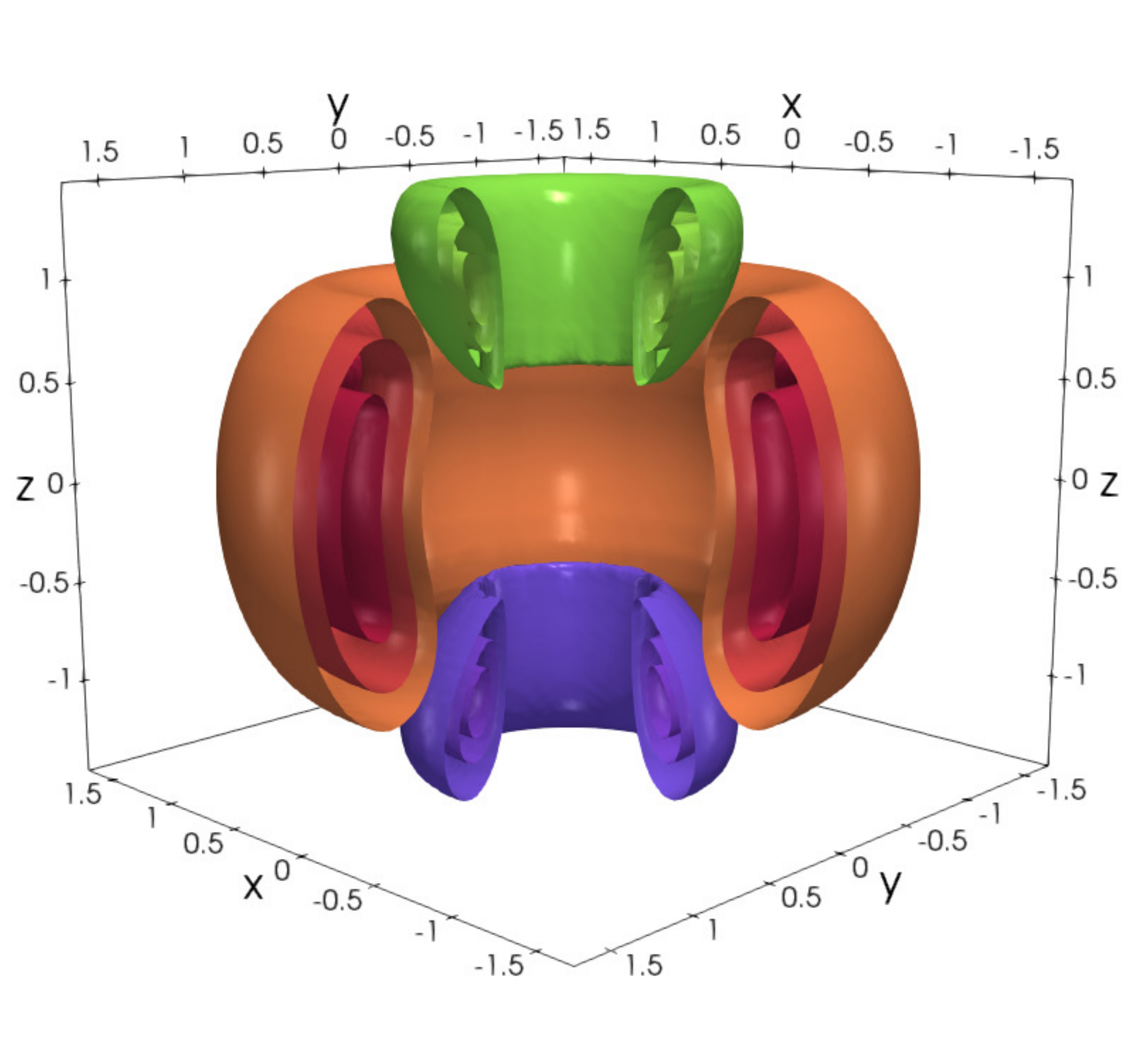}

    \caption{Surfaces of constant  magnetic charge and electric current densities for the $\nu=1$ sphaleron  
 (left) and for the $\nu=2$ sphaleron (right).}
    \label{QJ}
\end{figure}

We were able to confirm the above  considerations at the non-perturbative level by drawing 
level surfaces for the magnetic charge density and for the electric current obtained from \eqref{cur}. 
The left part of Fig.\ref{QJ} presents the result for the fundamental $\nu=1$ sphaleron (for $g^{\prime 2}=0.23$), where one can 
clearly see the thick belt representing the equatorial azimuthal current (red online) surrounding two oppositely 
charged and separated in space monopoles (green and blue online). The mutual attraction of the monopoles 
is compensated by the magnetic field created by the current, while the current itself exists because the magnetic field 
created by the monopoles forces the electric charges to Larmore orbit along the azimuthal direction. 

It is interesting that interchanging  in this picture  ``magnetic charges $\leftrightarrow$ electric currents" yields the description 
 of monopoles, because they contain inside oppositely 
 directed currents and a magnetically charged ring, instead of opposite magnetic charges and a current. 
 In this sense monopoles and sphalerons are mutually ``dual". 
 In both cases the Higgs field shows only one zero -- at the origin.

What are the monopoles inside the sphaleron ? Their charges $\pm g^\prime/g$  
may correspond either to the monopole and antimonopole of Nambu,
or to the SU(2) part of the charge of monopole and antimonopole of Cho-Maison. 
However, the total energy is finite, and in addition the distribution of the $Z$-field defined by  \eqref{Nambu} shows a $Z$-flux tube 
between the monopoles, hence they are connected through a vortex. Therefore,  these must be the Nambu monopole
and antimonopole \cite{Hindmarsh:1993aw}.   Still, the relation to the 
Cho-Maison monopoles is stunning, since comparing the regularized energy of the $\nu=2$ 
monopole solution with the energy of the $\nu=1$ sphaleron yields almost the same values: 
\be
\nu=1~~\text{sphaleron}:~~~E=38.473;~~~~~~~~~
\nu=2~~\text{monopole}:~~~E_{\rm reg}=38.119.
\ee
Therefore, the regular  part of the Cho-Maison monopole is similar to the Nambu monopole
because they both have  the same value of the 
magnetic charge and almost the same energy. 
Moreover, as seen in Fig.\ref{QJ}, for 
$\nu=2$ the sphaleron shows inside two magnetically charged rings. 
These must be the Nambu monopole and antimonopole with charges $\pm 2 g^\prime/g$.  
Therefore, for higher values of the charge the Nambu monopole contains inside a magnetic ring. 
At the same time, we know that the  $\nu=\mp 2$ generalizations of the Cho-Maison monopole also 
contains inside a magnetic ring of charge $\pm 2 g^\prime/g$, respectively,
which are the same values as for the Nambu monopole and antimonopole.

Summarizing, it seems that there exists a relation between monopoles of Nambu and monopoles of Cho-Maison. 
In some sense, the Nambu monopoles can be viewed as Cho-Maison monopoles with the 
divergent U(1) part removed.  At the same time,   the  Nambu monopole  is not  an equilibrium configuration 
of the theory because it is attached to a semi-infinite vortex pulling it. The Nambu monopole-antimonopole pair inside the sphaleron 
is static but its  total magnetic charge is zero. 
The only equilibrium 
non-Abelian configurations with a non-zero magnetic charge are the Cho-Maison monopole and its multi-charge generalizations.

\section{SUMMARY AND CONCLUDING REMARKS}

To recapitulate, we have constructed  the multi-charge generalizations for the non-Abelian electroweak monopole of Cho and Maison. 
The Cho-Maison monopole is spherically symmetric and has the 
magnetic charge $P=1/e$ or $P=-1/e$ (for the monopole and antimonopole). 
The new solutions are axially symmetric and exist for any value of  $P$,  but 
they are free of line singularities of the Dirac string type if only 
their magnetic charge  is an integer multiple of $1/e$, hence 
$P=-\nu/e=-\nu\,(g/g^\prime+g^\prime/g)\equiv P_{\rm U(1)}+P_{\rm SU(2)}$ with 
$\nu\in \mathbb{Z}$. 
Far away from the center, the solutions 
become purely electromagnetic and approach fields of the Dirac magnetic monopole of charge $P$, while
closer to the center  they contain non-linear fields and a  U(1) hypercharge field of 
Coulombian type. The latter makes an infinite contribution to the energy, but subtracting 
the Coulombian part  renders the energy finite, and  
the remaining part of the system is completely regular. 
The U(1) contribution to the magnetic charge, $P_{\rm U(1)}$, is concentrated in the monopole center,
while the SU(2) part of the charge,  $P_{\rm SU(2)}$, is smoothly  distributed over the volume of a  ring of a finite thickness.
The quantization of values of $P_{\rm SU(2)}$ is the same as for the Nambu monopole. 

The multi-monopoles  are characterized by a magnetic quadrupole moment 
that rapidly increases with growing magnetic charge. For large values of the charge, the monopoles 
are strongly squashed and their U(1) field becomes strong enough to suppress all other fields 
and restore the full gauge symmetry within a spheroidal central region -- a bubble of symmetric phase of size $R\propto \sqrt{|P|}$.
The bubble is encircled by a  belt of broken phase containing the W-condensate in the form of a magnetically 
charged ring sandwiched between two superconducting rings of oppositely directed  electric currents. 
This can be interpreted by saying that the magnetic ring creates  the circular electric currents, while  the latter 
produce  a magnetic field that squeezes the individual CM monopoles into the magnetic ring. 
The magnetic ring gives the leading contribution to the quadrupole moment $q\approx |P_{\rm SU(2)}| R^2$. 

It is interesting that exchanging   ``magnetic charges $\leftrightarrow$ electric currents" yields a 
 qualitative description of the interior  of sphalerons, so that monopoles and sphalerons are 
 mutually ``dual". It is also interesting that the structure of the regular part of the Cho-Maison monopole configuration is very similar 
 to the Nambu monopoles inside the sphalerons.

 The Cho-Maison maison  is stable with respect to any (small) perturbations,  
 hence it may be viewed  as a remnant  of decay of the Dirac monopole of the same charge \cite{Gervalle:2022npx}. 
The latter is unstable, but only with respect to spherically symmetric perturbations, hence it is conceivable that it 
radiates away a part of the energy,   while the rest 
condenses to the spherically symmetric CM monopole. The Dirac monopole with $\nu$ units of the CM magnetic charge 
is also unstable, but only with respect to perturbations with angular momentum $j=|\nu|-1$. One may  therefore conjecture 
that its instability leads to a formation of  a stable non-Abelian configuration which may have 
no symmetry at all 
or perhaps shows only discrete symmetries as for the spherical harmonics  \cite{Gervalle:2022npx}. 

Our results provide a partial confirmation of the conjecture since the spherical harmonics $Y_{jm}(\vartheta,\varphi)$ 
become axially symmetric 
for $m=0$. And indeed, we find axially symmetric non-Abelian  solutions for higher values of the magnetic charge,
although we could not yet prove that they are stable. 
However, they are presumably only a special case of more general, non-axially symmetric non-Abelian monopoles. 
In other words, the electroweak theory may admit many other not yet known non-Abelian monopole  solutions. 

It is likely that our solutions can be generalized to describe monopole-antimonopole pairs and monopole chains,
as was the case  for the t'Hooft-Polyakov monopoles  \cite{Kleihaus:2003xz,Kleihaus:2004is}.

The total energy of all electroweak monopoles is always  infinite due to the 
U(1) hypercharge field $B=\nu\,(\cos\vartheta\pm 1)d\varphi$
generated by the pointlike magnetic charge at the center, whose
 energy density $\nu^2/(2 g^{\prime 2} r^4)$ diverges at the origin. 
However, the divergence will be regularized when gravity is taken into account, which  this should impose a cutoff 
via producing an event horizon to shield the singularity at $r=0$ and render the energy finite. 
In fact, the gravitating generalization for the spherically symmetric Cho-Maison monopole is already known and is 
indeed described by a 
 black hole geometry  with a finite mass  \cite{Bai:2020ezy}. Similar black hole generalizations should exist also for axially 
 symmetric monopoles. This is almost  obvious  for large charges when the monopoles show 
 inside  the bubble of symmetric phase containing only the spherically symmetric hypercharge field $B$. 
 One can expect that switching the gravity on will  replace the underlying Minkowski geometry in the bubble by the geometry 
 of a static and spherically symmetric charged black hole, without affecting the $B$ field. 
  The minimal event horizon size will be of the order of $|\nu|/g^\prime$
 multiplied by the Planck length, which is many orders of magnitude less than the size of the bubble.  
 Therefore, the presence of a small black hole in the center should not change anything in the bubble,  
 nor should it affect the non-Abelian fields outside the bubble.  
 In other words, the inner monopole structure with the bubble and rings shown in Fig.\ref{QJm1} is expected 
 to  remain almost intact
 if the central pointlike charge is replaced by a small  black hole of the same charge. 
 Similar behaviour is known for the t'Hooft-Polyakov monopole and other solitons  
 which can incorporate a small black hole in the center without essentially changing their form \cite{Volkov:1998cc}. 
 
 Summarizing, we expect that  when coupled to gravity,  the electroweak theory should  admit magnetically charged ``hairy" black holes  
which  are either axially symmetric or have no continuous symmetries at all. The possible existence of such black holes was recently 
advocated by Maldacena \cite{Maldacena:2020skw}, and before that,  solutions of this type had been discussed at the perturbative level 
within a theory which 
 is similar although  not exactly identical to the electroweak theory \cite{Ridgway:1994sm,Ridgway:1995ke,Ridgway:1995ac}.
 However, such solutions have never been constructed explicitly. We therefore expect that taking gravity into account should promote 
  our multi-monopole solutions  to static and axially symmetric hairy black holes with a finite mass. 
 
 Before finishing, one should say that non-Abelian monopoles in the electroweak theory were reported  also in \cite{Teh:2014xva}
 (see also references therein) for the magnetic charge 
$P=1/(2e)$, which is the least possible value in the Dirac picture. The same 
axially symmetric ansatz as in our case was  used,  assuming that at infinity the U(1) field is 
$B=(1/2)\times (\cos\vartheta\pm 1)d\varphi$, whose flux through the two-sphere is $2\pi$. 
At the same time, it was assumed  that $B$ vanishes at the origin, and it was inferred from this that the energy is finite. 
However, the latter assumption is inconsistent with the former one 
since the flux of $B$  is a topological invariant that  does not depend on the size of the sphere, as seen in \eqref{flux}. 
Since its flux is conserved, $B$ cannot vanish at the origin but should diverge there  hence 
the energy should diverge as well. 
 We therefore find unclear  the status of the report.

 \section*{ACKNOWLEDGEMENTS}

The assistance of Julien Garaud in various issues  concerning the FreeFem++ numerical solver was extremely helpful for us. 

\appendix

\section{FAR FIELD ZONE}
\renewcommand{\theequation}{A.\arabic{equation}}
\setcounter{equation}{0}

In this Appendix we analyze the asymptotic behaviour of solutions at spatial infinity, both for the monopoles and sphalerons. 
This shows in particular that the sphalerons have a magnetic dipole moment, whereas for the monopoles the asymptotic expansion
starts from the quadrupole. This also shows that the gauge condition \eqref{fix} used in our calculations 
gives rise to a spurious long-range mode of a pure gauge  origin.

One has at large distances 
\be           \label{mon-far}
&&F_1=-\frac{1}{r}\,\delta H_1,~~F_2=\delta H_2,~~F_3=\Theta(\vartheta)+\delta H_3\sin\vartheta,~
~~F_4=\delta H_4\sin\vartheta, \nn \\
&&Y=\Theta(\vartheta)+\delta y\,\sin\vartheta,~~~\phi_1=\delta\phi_1,~~~~\phi_2=1+\delta\phi_2\,,
\ee
where the deviations $\delta H_1,\ldots ,\delta\phi_2$  approach  zero as 
$r\to\infty$ and where  $\Theta(\vartheta)=\cos\vartheta$ in the monopole case while $\Theta(\vartheta)=1$ in the sphaleron case. 
Since the deviations are small  in the far field zone, 
the field equations can be linearized. It is convenient to use the  original equations 
where the gauge is not fixed, then the linearized equations admit the gauge symmetry 
\be             \label{gauge-lin}
&&\delta H_1\to \delta H_1-r\partial_r\chi,~~\delta H_2\to \delta H_2+\partial_\vartheta \chi\,,~~
\delta H_4\to \delta H_4+\chi\,\frac{\Theta(\vartheta)}{\sin\vartheta}\,,~~\nn \\
&&\delta\phi_1\to \delta\phi_1+\chi/2\,,~~~~\
\delta H_3\to\delta H_3,~~~\delta y\to\delta y,~~~ \delta \phi_2\to\delta \phi_2\,,
\ee
which is 
obtained by assuming the gauge parameter $\chi$  in \eqref{res} to be small and linearizing. 

\subsection{Higgs sector}

The linearized 
equation for $\delta\phi_2$ decouples from the others,
\be
\left(\frac{\partial^2}{\partial r^2} +\frac{2}{r}\frac{\partial}{\partial r}+\frac{1}{r^2}\frac{\partial^2}{\partial \vartheta^2} 
+ \frac{\cot\vartheta}{r^2}\frac{\partial}{\partial \vartheta}-\frac{\beta}{2} \right)\delta\phi_2=0,
\ee
which is solved by 
\be
\delta\phi_2=\frac{R_H(r)}{r}\, P_j(\cos\vartheta)~~~~~~~~\text{with}~~~~~~~\left(\frac{d^2}{dr^2}-\frac{j(j+1)}{r^2}-\frac{\beta}{2}\right)R_H=0, 
\ee
where $ P_j(\cos\vartheta)$ are  the Legendre polynomials. 
 The orbital quantum number $j$ can take any value  $j=0,1,2,\ldots $, hence the general solution is
  a superposition of modes with different $j$, but the $j=0$ mode decays slower than other modes 
  hence  it is dominant at large $r$. Therefore, the 
leading contribution is described by the Yukawa potential,
\be
\delta\phi_2=\frac{C_H}{r}\,e^{-\mh \,r}\,,
\ee
where $C_H$ is an integration constant and $\mh$ is the Higgs boson mass defined in \eqref{masses}. 
This solution applies both for monopoles and sphalerons since in both cases one has  $\partial_\vartheta\phi_2=0$
for $\vartheta=0$ and for $\vartheta=\pi/2$. 

\subsection{Electromagnetic and Z sectors}

The equations for $\delta H_3$ and $\delta y$ comprise a closed system,
and setting $\delta y=y_\gamma+g^{\prime 2}\, y_Z$ and  $\delta H_3=y_\gamma-g^2 y_Z$, the system splits 
into two 
independent equations,
\be
\hat{{\cal D}}_1\,y_\gamma=0,~~~~~\left( \hat{{\cal D}}_1\,-\frac12\,\right) y_Z=0,
\ee
where the differential operator is defined by 
\be               \label{oper}
\hat{{\cal D}}_m=
\frac{\partial^2}{\partial r^2} +\frac{1}{r^2}\left(\frac{\partial^2}{\partial \vartheta^2} + \cot\vartheta\,\frac{\partial}{\partial \vartheta}
-\frac{m^2}{\sin^2\vartheta}
\right).
\ee
The eigenfunctions of the angular part of this operator are the associated Legendre polynomials $P^m_j(\cos\vartheta)$, 
the corresponding eigenvalue being  $-j(j+1)$  
with $j=|m|, |m|+1,\ldots$,  hence the solution is
\be
y_\gamma=R_\gamma(r) P^1_{j}(\cos\vartheta),~~~~~~y_Z=R_Z(r) P^1_{j}(\cos\vartheta).
\ee
Here $j=1,2,\ldots$ and 
\be
\left(\frac{d^2}{dr^2}-\frac{j(j+1)}{r^2}\right)R_\gamma=0,~~~~~~
\left(\frac{d^2}{dr^2}-\frac{j(j+1)}{r^2}-\frac{1}{2}\right)R_Z=0.
\ee
This describes the massless electromagnetic and massive $Z$ modes,
and this solution applies both to monopoles and sphalerons. 
However, the allowed values of $j$ are not the same in both cases
since the boundary conditions are different. 

For the sphalerons one should have $y=H_3=0$ at $\vartheta=0$ and $\partial_\vartheta y=\partial_\vartheta H_3=0$ 
at $\vartheta=\pi/2$, hence one can choose the minimal value of the angular momentum, 
$j=1$, which gives the dominant at infinity solution
\be
\underline{\text{sphalerons}}:~~~~~y_\gamma= \frac{C_\gamma}{r}\, \sin\vartheta\,,~~~~~
y_Z= C_Z\, e^{-\mz\, r}\,\sin \vartheta\,+ \ldots \,,
\ee
with the dots denoting subleading terms. 
The electromagnetic mode $y_\gamma$  describes the magnetic dipole  moment. 

For the monopoles one should have  $y=H_3=0$ both for $\vartheta=0$ and  $\vartheta=\pi/2$,  hence 
one cannot have $j=1$ so that the dipole moment is zero. 
The minimal possible value is 
 $j=2$, which defines the leading behaviour 
\be            \label{quad}
\underline{\text{monopoles}}:~~~~~y_\gamma= \frac{C_\gamma}{r^2}\, \sin\vartheta\cos\vartheta\,,~~~~~
y_Z= C_Z\, e^{-\mz\, r}\,\sin \vartheta\cos\vartheta\,+ \ldots \,,
\ee
and this corresponds to  the magnetic quadrupole moment. 

\subsection{W sector}

The four amplitudes $\delta H_1,\delta H_2,\delta H_4,\delta \phi_1$ 
fulfill a system of four equations admitting 
the gauge symmetry \eqref{gauge-lin}. This symmetry  can be used to impose the condition $\delta\phi_1=0$,
which corresponds to the unitary gauge. The subsequent steps 
are slightly different for monopoles and for sphalerons. 

\subsubsection{Monopoles} 
The four equations for $\delta H_1,\delta H_2,\delta H_4,\delta \phi_1$ with $\delta\phi_1=0$
are solved by setting 
\be                 \label{A12}
\delta H_1&=&\nu\,\frac{f_1(r)}{r}\,P^{\nu}_{j}(\cos\vartheta)\,,~~~\nn \\
\delta H_2&=&\nu f_3(r)P^{\nu-1}_{j}(\cos\vartheta)+ \nu f_2(r)P^{\nu+1}_{j}(\cos\vartheta)\,, \nn \\
\delta H_4&=&f_3(r)P^{\nu-1}_{j}(\cos\vartheta)- f_2(r)P^{\nu+1}_{j}(\cos\vartheta)\,.
\ee
Using the recurrence relations 
\be
(\partial_\vartheta \mp m\cot\vartheta) P^m_j(\cos\vartheta)=\lambda_\pm P^{m\pm 1}_j(\cos\vartheta)\,,
\ee
with  $\lambda_{+}=1$ and $\lambda_{-}=m(m-1)-j(j+1)$, the angular dependence separates.
The equations for $f_1(r)$ and $f_2(r)$ become 
\be          \label{W}
\left(\frac{d^2}{dr^2}+\frac{\nu^2-j(j+1)}{r^2}-\frac{g^2}{2}\right)f_1&=&0\,, \nn \\
\left(\frac{d^2}{dr^2}+\frac{\nu^2-j(j+1)}{r^2}-\frac{g^2}{2}\right)f_2&=&\frac{f_1}{r^3}\,,
\ee
and the remaining equations reduce to the constraint 
\be
f_3=f_1^\prime+(j-\nu)(j+1+\nu)f_2\,.
\ee
Denoting $C^{(1)}_W$ and $C^{(2)}_W$ two integration constants, one obtains from \eqref{W}
\be               \label{qW}
f_1=C^{(1)}_W\, e^{-\mw\,r}+\ldots \,,~~~~~
f_2=C^{(2)}_W\, e^{-\mw\,r}+\ldots\,.
\ee
This solution describes massive W boson modes. 

Summarizing, all field amplitudes approach their asymptotic values 
exponentially fast, apart from $\delta H_3$ and $\delta y$ which decay as $1/r^2$. 
This agrees with properties of the perturbative states in  the theory. 
However, this behaviour is manifest  only in the unitary gauge, while 
the gauge \eqref{fix} used for the numerical integration is not unitary. 
Solving the linearized equations in this gauge as was done above yields the 
same solutions for $\delta\phi_2$, $\delta H_3$, $\delta y$ since these amplitudes are gauge invariant, 
but the gauge-dependent amplitudes $\delta H_1,\delta H_2,\delta H_4,\delta \phi_1$  
then look completely different,
\be                \label{gauge1a}
&&\delta H_1=\frac{A}{r^2}\,\sin (2\vartheta)+\ldots,~~~
\delta H_2=\frac{A}{r^2}\,\cos (2\vartheta)+\ldots,~~~\nn \\
&&\delta H_4=\frac{A}{r^2}\,\cos^2 \vartheta+\ldots,~~~
\delta \phi_1=\frac{A}{4r^2}\,\sin (2\vartheta)+\ldots\,.
\ee
Here $A$ is  an integration constant and the dots denote subleading terms containing the exponentially small massive modes 
described by  \eqref{quad},
\eqref{qW}.  
As a result, the solution shows a second long-range tail 
in addition to the electromagnetic one. Of course, this additional mode is pure gauge 
and can be removed by the gauge transformation \eqref{gauge-lin} with the gauge parameter
\be
\chi=-2\delta\phi_1=-\frac{A}{2r^2}\sin (2\vartheta)+\ldots\,,
\ee
which is equivalent to setting $A=0$ in \eqref{gauge1a}. However, 
this mode appears in the numerical integration procedure as a result of the 
gauge condition \eqref{fix}. One might try to exclude  this spurious mode by choosing  some other gauge, 
as for example the unitary gauge. However, as shown below in Appendix B, the unitary gauge is singular at the origin,
whereas the gauge \eqref{fix} is globally regular, which is why it is preferable, even though it produces the spurious mode at infinity.

\subsubsection{Sphalerons}

Curiously, the linearized equations do not admit a complete separation of variables in this case. 
Passing to the unitary gauge  $\delta\phi_1=0$,  the four equations for $\delta H_1,\delta H_2,\delta H_4,\delta \phi_1$ 
reduce to three independent ones, of which one decouples and is solved by 
\be
\delta H_1&=&\frac{f_1(r)}{r}\,P^{\nu}_{j}(\cos\vartheta)\,,~~~\left(\frac{d^2}{dr^2}-\frac{j(j+1)}{r^2}-\frac{g^2}{2}\right)f_1=0\,.
\ee
The solution enters the equation for $\delta H_2$ as a source term,
\be
\left( \hat{{\cal D}}_\nu-\frac{g^2}{2}
\right)(\sin\vartheta \,\delta H_2)
=\frac{2}{r^2}\left(
\cot\vartheta\,\, r\, \frac{\partial}{\partial r}
+
\frac{\partial}{\partial \vartheta}
\right)\sin\vartheta\,\delta H_1\,,
\ee
while $\delta H_4$ is determined algebraically,
\be
\nu^2 \delta H_4=\partial_\vartheta(\sin\vartheta\, \delta H_2)
-\sin\vartheta\, \partial_r(r \delta H_1).
\ee
These equations admit two independent solutions decaying as $e^{-\mw\,r}$ at large $r$. 
Therefore,  the far field solution is a superposition of short-range massive modes 
and a long-range electromagnetic mode. This behaviour is manifest in the unitary gauge, while in the 
gauge \eqref{fix} used for numerical integration the gauge-dependent amplitudes 
$\delta H_1,\delta H_2,\delta H_4,\delta \phi_1$ show a long-range spurious mode similar to \eqref{gauge1a}
for the monopoles.

\section{SOLUTION AT THE ORIGIN}
\renewcommand{\theequation}{B.\arabic{equation}}
\setcounter{equation}{0}

In this Appendix we analyze the behaviour of the solutions for small $r$, close to the origin $r=0$. The complete analysis 
turns out to be rather involved, and we shall consider only  the behaviour of the Higgs field in the monopole case,
which will lead to important conclusions.

Close to the origin the monopole fields approach 
\be                \label{var1}
H_1=H_3=y=\phi_1=\phi_2=0,~~~~~~~H_2=H_4=1,
\ee
which can be called ``false vacuum".
This is an exact solution of the equations for any $r,\vartheta$, but the monopole fields approach it only for $r\to 0$. 
Therefore, for small $r$ one has 
\be
&&H_1=\delta H_1,~~~H_2=1+\delta H_2,~~~~H_3=\delta H_3,~~~~H_4=1+\delta H_4,~~~\nn \\
&&y=\delta y,~~~~~~\phi_1=\delta \phi_1,~~~~~~\phi_2=\delta \phi_2\,,
\ee
where the deviations $\delta H_1,\ldots ,\delta\phi_2$ vanish in the $r\to 0$ limit. 
Injecting this to the field equations and linearizing with respect to the deviations, it turns out that the equations for 
$\delta\phi_1$ and $\delta\phi_2$ decouple from the rest. 
One can neglect in these two equations terms proportional to the Higgs coupling $\beta$ since they are small 
as compared to the other terms if $r$ is small. After this, the equations become homogeneous in $r$ and setting 
\be
\delta\phi_1=r^\lambda\, S_1(\vartheta),~~~~~~~\delta\phi_2=r^\lambda\, S_2(\vartheta), 
\ee
the variables separate and the equations reduce to  
\be               \label{var3}
\left( \lambda(\lambda+1)+\frac{d^2}{d \vartheta^2}+\cot\vartheta\,\frac{d}{d\vartheta}
-\frac{\nu^2}{\sin^2\vartheta }+\frac{3\nu^2-1}{4}
\right)S_1
-\left(
\frac{d}{d \vartheta}+\frac{1-\nu^2}{2}\,\cot\vartheta
\right)S_2=0, \nn \\
\left( \lambda(\lambda+1)+\frac{d^2}{d \vartheta^2}+\cot\vartheta\,\frac{d}{d \vartheta}
-\frac{\nu^2+1}{4}
\right)S_2
+\left(
\frac{d}{d \vartheta}+\frac{1+\nu^2}{2}\,\cot\vartheta
\right)S_1=0.~~~~~~~~~
\ee
This defines  the eigenvalue problem to determine $\lambda$. 

If $|\nu|=1$ then setting $S_1=0$, $S_2=const.$, the equations reduce to 
\be
\lambda(\lambda+1)-\frac{1}{2}=0~~~~~\Rightarrow~~~~~\lambda=\frac{\sqrt{3}-1}{2}\,,
\ee
which reproduces the small $r$ behaviour  of the CM monopole. 
If $|\nu|\neq 1$ then the solution 
is obtained by choosing (assuming that $\nu>0$) 
\be               \label{lambb}
\lambda=\frac{\sqrt{1+2\nu}-1}{2}\,,~~~~~~S_1(\vartheta)=-\frac{2}{\nu+1}\, \frac{d}{d \vartheta}
\,S_2(\vartheta)\,.
\ee
This formula determines the rate with which the Higgs field approaches zero at the origin. 
Using this, Eqs.\eqref{var3} reduce to 
\be
\left(\frac{d^2}{d \vartheta^2}-\nu\cot\vartheta\,\frac{d}{d \vartheta}
+\frac{1-\nu^2}{4}
\right)S_2=0,
\ee
whose solution is 
\be                \label{B8}
S_2(\vartheta)=\left(\sin\frac{\vartheta}{2}\right)^{\nu+1}+\left(\cos\frac{\vartheta}{2}\right)^{\nu+1}.
\ee
Since the derivative $dS_2(\vartheta)/d\vartheta$ vanishes for $\vartheta=0$ and for $\vartheta=\pi/2$, 
the deviations $\delta\phi_1$ and $\delta\phi_2$ 
satisfy the correct boundary conditions at the symmetry axis and in the equatorial plane. 

\begin{figure}
 			
			  \includegraphics[scale=0.65]{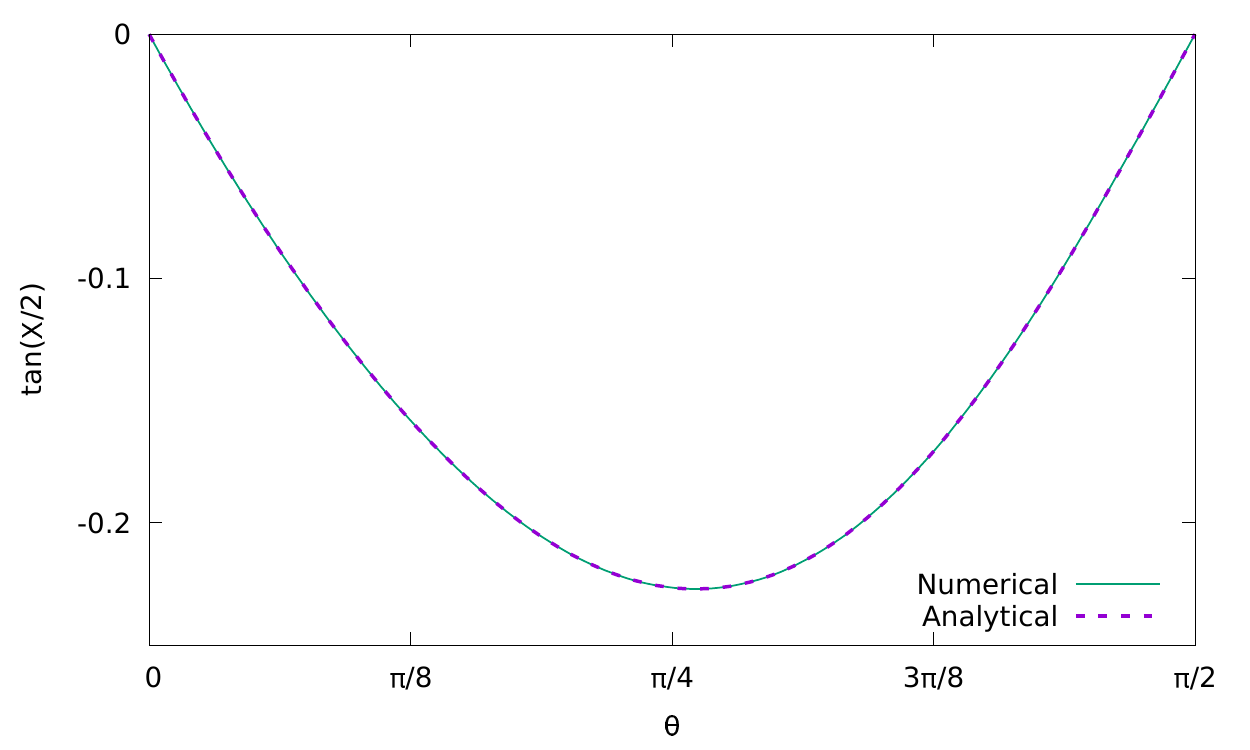}
   \includegraphics[scale=0.91]{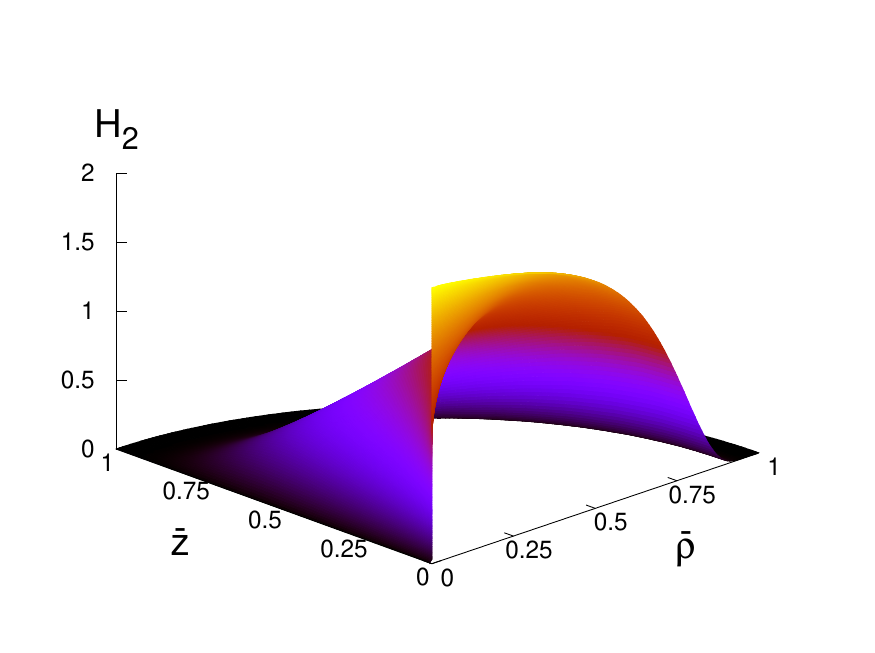}

    \caption{Left: plots of $\tan(\chi/2)$ analytically obtained from \eqref{chi} and also numerically 
    from  \eqref{chi1} for $r\to 0$. 
    The two plots exactly coincide to each other and determine the $r\to 0$ limit of the gauge transformation toward the unitary gauge. 
    Right: the $H_2$ amplitude of the $\nu=2$ solution transformed to the unitary gauge.}
    \label{vac}
\end{figure}

This result has an interesting consequence. The gauge transformation \eqref{res} changes the Higgs amplitudes as 
\be
\delta\phi_1\to \delta\tilde{\phi}_1&=& \delta\phi_1\,\cos\frac{\chi}{2}+ \delta\phi_2\,\sin\frac{\chi}{2},~~~~\nn \\
\delta\phi_2\to \delta\tilde{\phi}_2&=& \delta\phi_2\,\cos\frac{\chi}{2}- \delta\phi_1\,\sin\frac{\chi}{2},~~~~~~~
\ee
and if we require the new gauge to be unitary, $\delta\tilde{\phi}_1=0$, this implies that 
\be             \label{chi}
\tan\frac{\chi}{2}=-\frac{\delta\phi_1}{\delta\phi_2}=-\frac{S_1(\vartheta)}{S_2(\vartheta)}. 
\ee
This determines the $r\to 0$ limit of the parameter $\chi$ of the gauge transformation putting the solution to the unitary gauge. 
Notice that although $\delta{\phi}_1$ and $\delta{\phi}_2$ are small near the origin, their ratio and hence the gauge parameter $\chi$
are not small. 

This fact can be used to check the quality of our numerical solutions obtained in the gauge \eqref{fix}. In order to transform a given 
solution to the unitary gauge, one should perform the gauge transformation  \eqref{res} with the parameter 
\be             \label{chi1}
\tan\frac{\chi}{2}=-\frac{\phi_1}{\phi_2}\,,
\ee
where $\phi_1$ and $\phi_2$ are numerically obtained functions of $r,\vartheta$. This gauge parameter should agree for small $r$ 
with the one in \eqref{chi} for the procedure to be consistent, and this is indeed the case. In Fig.\ref{vac} we plot $\tan(\chi/2)$ 
given by the analytical formula \eqref{chi} and also $\tan(\chi/2)$  numerically obtained from \eqref{chi1} in the $r\to 0$ limit, 
and the two plots exactly coincide to each other so that only one curve can be  seen in Fig.\ref{vac}. Therefore, our procedure is consistent.

The same gauge transformation changes the false vacuum configuration \eqref{var1} to 
\be            \label{chi2}
&&H_1=0,~~~~~~~~H_2=1+ \frac{d\chi}{d \vartheta}\,, ~~~~~~~~y=\phi_1=\phi_2=0,\nn \\
&&H_3=(\cos\chi-1)\cot\vartheta-\sin\chi\,,~~~~~
H_4=\cos\chi+\sin\chi\cot\vartheta\,,
\ee
which  is the $r\to 0$  limit of the solution expressed in the unitary gauge. Notice however that this limit 
is $\vartheta$-dependent since $\chi$ in \eqref{chi} depends on $\vartheta$. On the other hand, 
nothing should depend on $\vartheta$ there because $r=0$ is a single point in space. 
To illustrate this, Fig.\ref{vac} shows $H_2$
for the $\nu=2$ solution, the same as in Fig.\ref{Fig1}, but transformed to the unitary gauge. 
As seen, $H_2$ does not have a definite limit at the origin $\bar{\rho}=\bar{z}=0$ but assumes there all values from the interval $[0:2]$, 
depending on the direction the origin is approached. This agrees with \eqref{chi2} since one has at the origin $H_2=1+d\chi/d\vartheta$ 
where  the derivative of $\chi(\vartheta)$ defined in \eqref{chi} varies in the interval $[-1:1]$. 

Therefore, the unitary gauge is singular at small $r$, although it is well adapted to describe the large $r$ region. 
On the other hand, the gauge \eqref{fix} is regular everywhere 
but exhibits  the spurious long-range mode \eqref{gauge1a} at large $r$.

\providecommand{\href}[2]{#2}\begingroup\raggedright\endgroup


\end{document}